# Ein stationäres Universum
# und die Grundlagen der Relativitätstheorie

*Peter Ostermann*

Es gibt ein ausgezeichnetes kosmisches Bezugssystem, in welchem die Spiralnebel statistisch ruhen. In hinreichender Entfernung von lokalen Gravitationsquellen ist die Lichtgeschwindigkeit konstant. Die entsprechenden Systemkoordinaten der allgemeinen Relativitätstheorie lassen sich verstehen als Repräsentanten des euklidischen Raums und der kosmischen Zeit. Bezüglich der eindeutig fixierten kosmischen Zeit $t^*$ aber existiert ein singularitätsfreies stationäres Linienelement, das insbesondere mit dem in den letzten Jahren gemessenen unmittelbaren Zusammenhang zwischen scheinbarer Helligkeit und Rotverschiebung von Supernovae nicht unvereinbar scheint.

Mit Unterscheidung des unserer Beobachtung zugänglichen (evolutionären) Kosmos von einem unendlichen Universum kann die physikalische Frage nur lauten, ob dieser selbst stationär oder aber Teil des stationären Universums ist.

Mit der Atomzeit, der Ephemeridenzeit und der kosmischen Zeit gibt es mindestens drei Zeitskalen, die sich ohne Bezug aufeinander definieren lassen. Nicht allein Atomuhren, sondern auch Planetenuhren stellen *natürliche* Uhren dar. Erfahrungsgemäß zeigen beide die gleiche Zeit, doch stimmt diese nicht überein mit der kosmischen Zeit. Im Unterschied zur kosmischen Zeit ist die von Atomuhren (bzw. Planetenuhren) angezeigte lokale Atomzeit (bzw. Ephemeridenzeit) aufgrund ihrer Nicht-Integrabilität ungeeignet zur Beschreibung kosmischer Abläufe. Dies vermag nur die kosmische (System-)Zeit $t^*$ zu leisten, die eindeutig fixiert ist durch die Forderung einer konstanten kosmischen Lichtgeschwindigkeit $c^* = c$. In einem räumlich euklidischen Universum ist eine entsprechende Synchronisationsbedingung bei geeigneter Koordinatenwahl immer erfüllbar.

Nach den hier entwickelten Vorstellungen unterscheidet sich das Linienelement eines stationären Universums von dem der speziellen Relativitätstheorie allein um einen kosmischen, skalaren Zeitfaktor $e^{Ht^*}$, der die Richtung der Zeit sowie die HUBBLE-Konstante als echte Naturkonstante enthält.

Bereits mit einer zunächst zur Diskussion gestellten Schar singularitätsfreier Lösungen wird versucht, solch fundamentale physikalische Schwierigkeiten zu vermeiden, wie einen zeitlichen Anfang des Universums einerseits, andererseits aber auch vollständige Materiefreiheit oder umgekehrt die Notwendigkeit einer ständigen Schöpfung aus dem Nichts.

Das *stationäre* Linienelement scheint mit wichtigen Beobachtungstatsachen so weit in Einklang zu stehen, wie man es bei Ansatz einer mittleren Energiedichte, d.h. bei Vernachlässigung aller räumlichen Inhomogenitäten (sowie gegebenenfalls einer gesamt-kosmischen Evolution), von einer Lösung der nicht-linearen EINSTEIN'schen Gleichungen realistischerweise erwarten kann. Es resultiert ein negativer kosmischer Gravitationsdruck von -1/3 der kritischen Dichte (entsprechend dem halben Betrag der mittleren phänomenologischen Massendichte, die damit nur 2/3 der kritischen Energiedichte beträgt). Dieser Druck läßt sich möglicherweise deuten als langsam veränderliche kosmologische ‚Konstante‘.

Weiterhin ergeben sich eine zeitunabhängige Rotverschiebung, die Lösung des OLBERS'schen Paradoxons, eine mittlere baryonische Strahlungsdichte in der Größenordnung einer äquivalenten 3K-Strahlung sowie andere überprüfbare Aussagen.

Eine Singularität im Sinne der heutigen Kosmologie ergibt sich aus dem stationären Linienelement erst dadurch, daß man sich auf (nicht-integrierbare) natürliche Einheiten bezieht. Demzufolge sollten sich prinzipiell alle durch Beobachtungstatsachen bestätigte Aussagen der bisherigen Kosmologie – so z.B. die Existenz der Hintergrundstrahlung als reine Schwarzkörperstrahlung – aus dem stationären Linienelement *heuristisch* dadurch gewinnen lassen, daß man *zurückblickend* die kosmischen Abläufe insgesamt auf quasi-natürliche Koordinaten bezieht und dabei dem negativen Gravitationsdruck in geeigneter Weise Rechnung trägt.

In diesem Sinne beschreibt das stationäre Linienelement (auch) einen in Bezug auf lokale natürliche Koordinaten expandierenden *Kosmos* – nun aber eingebettet in ein stationäres Universum.

# Inhalt

Die geänderten Formeln (von insgesamt 112) dieser Version v2 sind (mit einem Stern markiert):
*(55), *(57) , *(58) , *(59) , *(71) , *(72) , *(73) , *(75)

## 1. Stationarität und OLBERS'sches Paradoxon

Im folgenden verwenden wir den Begriff ‚Universum' für die Gesamtheit all dessen, was in der Vergangenheit war, in der Gegenwart ist und in der Zukunft sein wird. Demgegenüber wollen wir unter dem Begriff ‚Kosmos' die Wirklichkeit zunächst nur insoweit verstehen, wie sie der Beobachtung durch den Menschen (heute) zugänglich ist. Es ist die Frage, ob beide Begriffe synonym zu verwenden sind[1]).

Eine zeitliche Entwicklung des gesamten Universums im Sinne der heutigen Kosmologie wäre – wenn es sie so gäbe – grundsätzlich nicht-reproduzierbar. Gerade ihr Beginn würde sich einer *physikalischen* Beschreibung letztlich entziehen (ex nihilo nihil fit), jedenfalls solange man nicht bereit ist, Kausalitätsprinzip und Erhaltungssätze zumindest zeitweilig preiszugeben.

Ziel aller Bemühungen einer widerspruchsfreien Kosmologie kann es daher nur sein, die jeweiligen physikalischen Erkenntnisse und Theorien dahingehend zu erweitern, daß sie nicht nur mit allen einschlägigen Beobachtungstatsachen, sondern auch mit einem – im Sinne einer prinzipiell vollständigen Beschreibbarkeit geradezu vorgegebenen – *stationären Universum* bestmöglich in Einklang stehen. Daß sich letzteres tatsächlich erreichen läßt, ist weder beweisbar noch widerlegbar – und dennoch erstrebenswert. Das physikalische Konzept eines stationären Universums hat trotz (oder vielleicht auch wegen) der Nicht-Falsifizierbarkeit seinen Wert. Denn eine endgültige, vollständige Kosmologie kann es natürlich ebensowenig geben wie eine endgültige, allumfassende physikalische Theorie.

Nicht-stationäre Modelle des Universums aber, die einen zeitlichen Anfang implizieren, verzichten von vornherein auf die prinzipielle Möglichkeit einer vollständigen Beschreibung des Naturgeschehens[2]). Die physikalische Frage kann also nur lauten, ob der uns ‚bekannte' Kosmos selbst stationär oder aber Teil eines stationären Universums ist.

In diesem Sinne ist es durchaus denkbar, daß zwar unser Kosmos einen Anfang hätte, nicht aber das Universum (oder gar ‚Raum und Zeit') selbst.

Mit dem Abschied vom geozentrischen Weltbild hat – über das heliozentrische – eine natürliche Entwicklung zu der Anschauung geführt, daß das Universum von jedem Ort aus betrachtet und zu allen Zeiten gleich erscheinen sollte[3]). EINSTEINs ursprünglich im Sinne eines solchen *vollständigen* kosmologischen Prinzips entwickelte statische Lösung seiner Gleichungen verlangte die Einführung einer willkürlichen Konstanten [1] und erwies sich trotzdem als instabil. Die dynamische Interpretation der nur wenig später gefundenen zeitabhängigen Lösungen schien durch die Entdeckung der HUBBLE'schen Rotverschiebung [2] sehr bald eine überraschende experimentelle Bestätigung zu erfahren. Doch angesichts eines physikalisch unhaltbaren für immer materiefreien – nach DE SITTER [3] – oder aber singulären Zustands des gesamten Universums vor etwa $10^{10}$ a – nach FRIEDMANN [4] und später LEMAÎTRE [5] – der damit aus den EINSTEIN'schen Gleichungen zwangsläufig zu resultieren scheint, wurde von BONDI, GOLD [6] und HOYLE [7] versucht, die Lösung außerhalb der allgemeinen Relativitätstheorie zu suchen bzw. diese zu modifizieren. Die daraus entstandene ‚Steady-State Theory' aber hält fest an einer ‚Expansion des Universums' und behauptet deshalb die Notwendigkeit einer – physikalisch ebenfalls unhaltbaren – ständigen Schöpfung von Materie aus dem Nichts.

Wir fragen nicht, warum der Nachthimmel dunkel ist, sondern was daraus folgt, daß der Nachthimmel ist, wie er ist – und zwar in einem stationären Universum (dem insgesamt kein Bewegungszustand zugeschrieben werden kann außer dem der Ruhe).

Dabei ist allerdings von Anfang an klar, daß es hier eklatante *lokale* Verletzungen des 2. Hauptsatzes geben muß. Davon abgesehen aber wurde bereits von OLBERS [8] geschlossen, daß das Sternenlicht über hinreichend große Entfernungen absorbiert werde. Gegen diese These wurde und wird der Einwand erhoben, daß sich ein absorbierendes Medium so weit aufheizen müsse, daß es selbst Sterntemperatur erreiche. Doch stellt dieser Einwand keine stichhaltige Widerlegung dar, ganz im Gegenteil. Denn eine solche wäre nur gegeben, würde jeder Stern ewig strahlen. Dies ist natürlich nicht der Fall, da Sterne entstehen und vergehen. Gerade umgekehrt gilt: Wenn es keine entsprechende Absorption des Sternenlichts gäbe, wie könnten sich dann in einem stationären

---

[1]) Wo nicht ausdrücklich unterschieden, werden wir im folgenden davon ausgehen, daß dies der Fall sei.

[2]) Die Physik beschreibt die nach Wiederherstellung gleicher Anfangsbedingungen prinzipiell reproduzierbaren Abläufe der Natur. Es hat deshalb wenig Sinn, innerhalb der *Physik*, nach einem Beginn des gesamten Universums zu fragen.

[3]) bei geeignet gewähltem lokalen Bewegungszustand des Beobachters

Universum immer wieder neue Sterne bilden, die im Vergleich zu den früheren wieder in gleicher Helligkeit erstrahlen? – Sehr grob skizziert, ergibt sich aus dem OLBERS'schen Paradoxon in etwa folgendes:

(a) Es muß eine statistisch-stationäre Abschwächung des Sternenlichts geben – sei es durch gravitative Absorption oder auch durch relative Energieabnahme sich selbst überlassener Lichtmengen – so daß das Universum für die stationäre ca. 6000K-Strahlung unendlich vieler hinreichend weit entfernter Sonnen insgesamt undurchsichtig ist. (b) Typische Sterne, die aufgrund ihrer Energieabstrahlung eine endliche Lebensdauer von größenordnungsmäßig $10^{10}$ a haben, bilden sich wie alle anderen vergänglichen Strukturen in einem stationären Universum zwangsläufig immer wieder neu. (c) Dies geschieht im Umfeld gewisser, Licht und Materie schluckender *originärer* Gravitationszentren extremer Stärke, bei denen es sich im ursprünglichen Sinne des Wortes um Quasare[4]) oder auch um unseren ganzen Kosmos (unter anderen) handeln könnte. (d) Bei diesen gravitativen Entstehungsprozessen von Galaxien und Sternen – aus vorhandener Energie und Materie jeder Form einschließlich der jeweiligen Anteile sowohl freien als auch zuvor bereits absorbierten Sternenlichts – sollte die Entropie lokal abnehmen[5]), damit sie insgesamt stationär bleiben kann. (e) Die Bildung der verschiedenen Elemente sowie alle anderen Vorgänge – mitsamt der Fülle physikalisch relevanter Erkenntnisse und Modelle, die man heute dem ‚Urknall' oder der ‚frühen Phase des Universums' zuschreibt – sollten sich aus solchen überall im Universum wiederkehrenden Entstehungsprozessen erklären bzw. auf diese anwenden lassen. (f) Diese Prozesse wären der Grund dafür, daß sich die Temperatur eines möglichen, das Sternenlicht nur zum Teil direkt absorbierenden Mediums nicht immer weiter aufheizen muß, sondern auf einen stationären Wert[6]) einstellen kann. (g) Bestimmt man nun die mittlere kosmische Intensität der stationären 6000K-Strahlung, indem man die am Nachthimmel sichtbaren Sterne zählt und mit ihrer jeweiligen scheinbaren Helligkeit gewichtet, so findet man, daß diese gerade der Intensität einer *äquivalenten* schwarzen Strahlung im Bereich von ca. 3 K entspricht. (h) Die *zusätzlich* zu der mittleren 6000K-Strahlung tatsächlich beobachtete isotrope, schwarze 3K-Strahlung aber könnte – alternativ zur heutigen Deutung als kosmische Reliktstrahlung – möglicherweise auch ein Hinweis auf ein lichtabsorbierendes Medium mit temperaturabhängigem, selektivem Absorptionsverhalten sein. (i) Zu jeder Zeit des Universums sollte es kosmische Objekte in allen möglichen Entwicklungsstadien geben.

Angesichts der Tatsache, daß die mittlere kosmische Energiedichte offenbar nur zu einem sehr geringen Prozentsatz auf gewöhnliche Sternmaterie zurückzuführen ist – und sich ihr weitaus überwiegender Anteil damit der direkten Beobachtung entzieht – kann es keinesfalls verwundern, daß einige der genannten Prozesse heute noch ganz oder weitgehend unverstanden bzw. nicht beobachtet sind. Das hier mehr als grob skizzierte Konzept soll vorab lediglich zeigen, daß ein stationäres Universum keineswegs undenkbar ist. Vom Ansatz her ist es weitgehend unabhängig[7]) von der Gültigkeit der allgemeinen Relativitätstheorie, doch wird ein stationäres Linienelement von dieser nahegelegt.

## 2. Das kosmische Ruhsystem und die wahre Zeit

Zeigen Atomuhren notwendigerweise die *wahre* Zeit, wenn sie, an ein und demselben Ort unter denselben Bedingungen nebeneinander ruhend, immer und überall die gleiche Ganggeschwindigkeit aufweisen? – Nein.

„In der allgemeinen Relativitätstheorie können Raum- und Zeitgrößen nicht so definiert werden, daß räumliche Koordinatendifferenzen unmittelbar mit dem Einheitsmaßstab, zeitliche mit einer Normaluhr gemessen werden könnten." – (EINSTEIN 1916)

---

[4]) Die Rotverschiebung der Quasare sollte sich also kosmisch *plus* gravitativ zusammensetzen, was möglicherweise einige ansonsten unverständliche Beobachtungen bei assoziierten Galaxien erklären könnte. In diesem Zusammenhang wäre der Begriff ‚schwarzes Loch' so zu verstehen, daß in einem quasistellaren Gravitationszentrum extremer Stärke nicht nur Materie in ihrer alten Form verschwinden, sondern vor allem auch gleichzeitig Materie in neuer Form entstehen kann, wobei sich dieser – selbstverständlich *innerhalb* des absoluten Raums und der absoluten Zeit ablaufende – Vorgang selbst allerdings der Beschreibung durch die phänomenologische allgemeine Relativitätstheorie entzieht.

[5]) Gerade die Tatsache, daß die Gravitation im Unterschied zu anderen Kräften immer nur anziehend wirken kann, sehen wir in einem gewissen Widerstreit mit der zwangsläufigen Entropiezunahme bei statistischer Diffusion.

[6]) Mit Einschränkungen vergleichbar einer statistisch-stationären Temperatur der Erdatmosphäre im Sonnenlicht.

[7]) Vorausgesetzt werden vor allem Gravitationsfelder solch extremer Stärke, daß Materie und Energie jeder Form immer wieder zu Entstehungszentren von Galaxien, Nebelhaufen, Quasaren verdichtet werden können.

Mit dieser aus dem EHRENFEST'schen Paradoxon [9] – im Anschluß an KALUZAS [10] mathematisch bahnbrechende[8]) Behandlung der rotierenden Scheibe – von EINSTEIN [11] gefolgerten Notwendigkeit, in die allgemeine Relativitätstheorie eine Systemzeit $t$ einzuführen, die nicht übereinstimmt mit der Anzeige $\tau$ ruhender Atomuhren, und dazu räumliche Systemkoordinaten $x^\alpha$ (griechische Indizes $\alpha, \beta ... = 1, 2, 3$ für rein räumliche Koordinaten), die sich nicht darstellen lassen durch entsprechende von materiellen Maßstäben angezeigte Längen $L^\alpha$, sind nach unserem Verständnis bereits wenige Jahre nach Formulierung der speziellen Relativitätstheorie [12] Raum und Zeit als *absolute* Größen in die Physik zurückgekehrt.

Ganz unabhängig von der jeweils aktuellen Theorie über Zustand bzw. Entwicklung des Universums wird eine solche Auffassung von vorneherein durch folgende Tatsache nahegelegt: Mit Hilfe des DOPPLER-Effekts läßt sich statistisch immer ein ausgezeichnetes *Ruhsystem* festlegen, und zwar durch die Forderung größtmöglicher Isotropie des universalen Hintergrunds[9]). Die absoluten Geschwindigkeiten von Sonne und Erde sind auf dieser Basis bekanntlich längst ermittelt.

Die *spezielle* Relativitätstheorie zeigt nun zwar, daß in Inertialsystemen trotz Längenkontraktion und Zeitdilatation eine solche Koordinatenwahl möglich ist, bei der den Differenzen räumlicher Koordinaten unmittelbar mit Maßstäben meßbare Entfernungen, und den Differenzen der Zeitkoordinaten unmittelbar mit Lichtuhren meßbare Zeiten entsprechen. Doch die *allgemeine* Relativitätstheorie zeigt gerade, daß dies in Nicht-Inertialsystemen bzw. bei Berücksichtigung der Gravitation nicht mehr möglich ist. Angesichts der Tatsache aber, daß es abgesehen vom kosmischen Ruhsystem überhaupt nur lokale Inertialsysteme geben kann, beweist dies also die Existenz eines absoluten Raums und einer absoluten Zeit – durchaus im Sinne NEWTONs – als notwendige Voraussetzung für die Beschreibung der physikalischen Wirklichkeit[10]).

Das Problem der rotierenden Scheibe hat EINSTEIN in Verbindung mit seinem fundamentalen Äquivalenzprinzip allerdings zu der Auffassung geführt, daß der dreidimensionale Raum durch das Gravitationsfeld[11]) ‚gekrümmt' sei. Demgegenüber wurde in einer vorausgegangenen Arbeit [13] d. Verf. die Auffassung vertreten, daß jedes Linienelement der allgemeinen Relativitätstheorie am einfachsten zu verstehen ist als Aussage über die Auswirkungen von Gravitationspotential und Bewegung auf *reale Objekte*[12]) im absoluten, euklidischen Raum $x^{*\alpha}$ und in der absoluten, kosmischen Zeit $t^*$ – und nicht etwa auf ‚Raum' und ‚Zeit' selbst:

> Raum und Zeit selbst sind keine physikalischen Objekte, denen sich veränderliche Eigenschaften zuschreiben lassen. Gegenstand der physikalischen Beschreibung sind allein Veränderungen gegenüber dem, was notwendigerweise unveränderlich ist, und dessen Unveränderlichkeit keiner Erklärung bedarf.

Was nun ist die wahre Zeit? Wenn die Systemzeit eines beliebigen abgeschlossenen Teilsystems dessen interne globale Zeit darstellt, dann ist die singularitätsfreie Systemzeit des Universums $t^*$ als die wahre *kosmische* Zeit zu verstehen. In den Abschnitten 7, 8 wird sich zeigen, daß eine besonders einfache Synchronisation entsprechender im absoluten kosmischen Bezugssystem ruhender technischer Systemuhren auf Basis der EINSTEIN'schen Reflexion im Zeitmittelpunkt prinzipiell möglich wäre.

Daß die wahren Systemkoordinaten allerdings in *lokalen* Teilsystemen nicht eindeutig identifizierbar sind, kann ihre Existenz ebensowenig widerlegen, wie die Eigenschaft einer unterschiedlich temperierten Ebene, flach zu sein, bei ausschließlicher Verwendung temperaturabhängiger Maßstäbe durch die dann notwendige Bezugnahme auf weitgehend willkürlich wählbare krummlinige Koordinaten widerlegt werden kann. Die absolute,

---

[8]) Eine knappe Würdigung von KALUZAS Beitrag [10], der insgesamt eine Druckseite plus drei Halbzeilen umfaßt, wird gegeben in [13], dort insbesondere in den Abschnitten 5, 7 sowie in Anhang d, e, f.

[9]) Dies kann heute durch die Forderung größtmöglicher Isotropie der Hintergrundstrahlung definiert werden. Prinzipiell hätte man sich bereits mit HUBBLEs Entdeckung [2] auf eine maximale Isotropie der beobachtbaren statistischen Verteilung der Rotverschiebung beziehen können, vorher aber auf mittlere Sterngeschwindigkeit Null. Und selbst wenn sich der uns heute bekannte Kosmos eines Tages als Teil eines Universums aus ähnlichen (und anderen) Gebilden zeigen sollte – immer läßt sich ein ausgezeichnetes Ruhsystem finden (ein ernsthaftes Problem entstünde umgekehrt dann, wenn es *mehr* als ein einziges derartiges Ruhsystem geben sollte).

[10]) Diese Auffassung von Raum und Zeit, die hier unmittelbar zu der Möglichkeit eines im Rahmen der allgemeinen Relativitätstheorie *stationären* kosmischen Linienelements führt, wurde entwickelt in [13].

[11]) Wir unterscheiden zwischen *wahrem* Gravitationsfeld ($R^i{}_{klm} \neq 0$) und reinem Beschleunigungsfeld ($R^i{}_{klm} = 0$).

[12]) Teilchen, Felder, Körper, Gesamtheiten, Maßstäbe, Uhren

wahre Zeit *t\** aber ist zusammen mit dem absoluten, wahren Raum durch die Bedingung einer – abgesehen von lokalen Abweichungen – konstanten kosmischen Lichtgeschwindigkeit eindeutig festgelegt:

*Es gibt ein ausgezeichnetes kosmisches Bezugssystem, in welchem die Materie (bezogen auf große Skalen) annähernd homogen verteilt ist, und in welchem die Spiralnebel statistisch ruhen. In hinreichender Entfernung von lokalen Gravitationsquellen ist die Lichtgeschwindigkeit[13]) hier konstant.*

## 3. Die singularitätsfreie Lösungsschar der allgemeinen Relativitätstheorie

Könnte die wesentlich nicht-lineare allgemeine Relativitätstheorie im Mikrokosmos strenge Gültigkeit beanspruchen, so ließen sich makroskopisch wohl nur Schlüsse ziehen, die sich aus einer Näherung ergeben sollten, die dem Superpositions-Prinzip genügt. Diese Näherung müßte aber die bekannte EINSTEIN'sche Näherung[14]) für schwache Felder sein. Doch wäre eine relativistische Kosmologie auf Basis dieser Näherung von vornherein ihrer Aussagekraft beraubt. Dies deshalb, weil hier jeder Zusammenhang zwischen der 1. zeitlichen Ableitung der metrischen Koeffizienten (HUBBLE-Konstante) mit der mittleren Materiedichte verlorenginge. Es sind also die strengen EINSTEIN'schen Gleichungen anzuwenden. Eine nicht-lineare Feldtheorie kann aber kaum zugleich makroskopisch und mikroskopisch, sowohl streng als auch in linearer Näherung zutreffen. Diesen Sachverhalt interpretieren wir dahingehend: Die allgemeine Relativitätstheorie ist innerhalb gewisser Grenzen eine phänomenologisch richtige, nicht-lineare makroskopische Theorie – wie insbesondere die SCHWARZSCHILD-Lösung [14] für das Innere der Flüssigkeitskugel zeigt – die aber so, wie sie heute verstanden wird, gerade deshalb im Mikrokosmos keine strenge Gültigkeit beanspruchen kann. Nur unter dieser Voraussetzung ist es unseres Erachtens überhaupt legitim, aus der nicht-linearen allgemeinen Relativitätstheorie kosmologische Schlüsse ziehen zu wollen, indem man eine *homogene* Materieverteilung mittlerer Dichte zugrunde legt.

Nun ist ein singularitätsfreies, ja sogar ein stationäres Universum – im Unterschied zur herkömmlichen Auffassung – keineswegs unvereinbar mit den in lokalen Feldern immer wieder bestätigten EINSTEIN'schen Gravitationsgleichungen. Es ist allerdings unvereinbar mit EINSTEINs Gleichsetzung der mittels spektraler Maßstäbe und Uhren gemessenen lokalen Entfernungen und Zeitspannen mit dem wahren Raum und der wahren Zeit[15]).

EINSTEINs Bezeichnung für das, was wir hier und im folgenden auch ‚spektral' bzw. ‚atomar' nennen, lautet ‚natürlich'. Insbesondere spricht er immer wieder von ‚natürlichen' Maßstäben und Uhren. Von diesen Uhren jedoch, die wir heute zusammenfassend ‚Atomuhren'[16]) nennen, können wir zunächst – wie von allen anderen – lediglich sagen, daß je zwei Uhren gleicher Bauart, relativ zueinander ruhend, die gleiche Ganggeschwindigkeit aufweisen werden[17]).

In diesem Zusammenhang ist anzumerken, daß das Planetensystem nicht weniger eine *natürliche* Uhr[18]) darstellt als z.B. ein frequenzstabilisierter Maser. Die astronomische Erfahrung zeigt nun, daß die auf den Schwerpunkt des Sonnensystems bezogene ‚Atomzeit' und die mit der ‚Planetenuhr' gemessene Ephemeridenzeit[19]) tatsächlich synchron vergehen.

---

[13]) Bzgl. nicht-statischer lokaler Teilbereiche betrifft diese Aussage die 2-Weg-Lichtgeschwindigkeit (s. [13]).

[14]) Es sei daran erinnert, daß diese bekanntlich zu einem falschen Wert für die Periheldrehung des Merkur führt, wohingegen die strenge Theorie den richtigen Wert liefert.

[15]) Daß es durchaus einen Sinn hat, von ‚wahrer' Zeit zu sprechen, ergibt sich bereits aus der Tatsache, daß die von Atomuhren angezeigte Zeit jedenfalls *nicht* die wahre Zeit sein kann (wie das oben zitierte einfache Beispiel der rotierenden Scheibe zeigt).

[16]) einschließlich der ursprünglich gemeinten EINSTEIN'schen Lichtuhr aus Maßstab, zwei Spiegeln und Lichtsignal

[17]) Dagegen sind beliebige natürliche Uhren bezüglich ihrer *Zeitnullpunkte* in der Regel nicht synchronisiert. Dies gilt für die von EINSTEIN immer wieder herangezogenen Spektralfrequenzen von Atomen, die überhaupt keine feststellbaren Zeitnullpunkte haben, ebenso wie für ‚Zerfallsuhren', z.B. einige Klumpen Uranerz verschiedener Herkunft, die in der Regel unterschiedliche Zeitnullpunkte haben werden. – Daß es aber bezüglich der Zeitnullpunkte keine *natürliche* Synchronisation von Uhren gibt, ist letztlich auch der Grund dafür, daß es keinen eindeutigen, vom Synchronisationsverfahren unabhängigen Betrag der Einweg-Lichtgeschwindigkeit in lokalen Inertialsystemen geben kann (s. [13]).

[18]) Wenn es überhaupt Uhren gibt, welche die Bezeichnung ‚natürlich' verdienen, dann sind es, wie die historische Entwicklung zeigt, jedenfalls der Mond im Schwerefeld der Erde sowie Erde und Planeten im Schwerefeld der Sonne (abgesehen vom Sonderfall der Rotationsuhr Erde, s. Abschn. 8/f).

[19]) Wir sprechen hier von einer *Ephemeridenzeit* des Sonnensystems, bei deren Verwendung die Bewegung von Mond und Planeten richtig, d.h. im Sinne konstanter Umlaufzeiten von Perihel zu Perihel (nach Berücksichtigung wechselseitiger

Demgegenüber gibt es aber die oben definierte kosmische Zeit $t^*$. Diese kosmische Zeit ist *wahr*[20], indem sie prinzipiell eindeutig identifizierbar ist. Die kosmische Zeit ist *absolut*, indem sie – unabhängig von jedem lokalen Geschehen – gleichförmig verfließt[21]). Daraus folgt: 1. sie verfließt seit je und für immer, 2. ein Nullpunkt der kosmischen Zeit muß für jeden ihrer Abschnitte frei wählbar sein, 3. auch die absolute Einheit $T^*$ eines im Zusammenhang betrachteten kosmischen Zeitabschnitts ist jeweils frei wählbar, dann aber konstant.

Nun wird gerade das der genannten EINSTEIN'schen Gleichsetzung von *Eigenzeit*[22]) mit wahrer Zeit entsprechende Linienelement in der bekannten FRIEDMAN(N)-LEMAÎTRE(-ROBERTSON-WALKER)-Form (RW) [4], [5], [16], [17] seit langem allen gängigen relativistischen Ansätzen zur Kosmologie zugrunde gelegt. Als Zeitvariable ist dort eine zur Anzeige $\tau$ ruhender[23]) Atomuhren *lokal* äquivalente Quasi-Eigenzeit gewählt.

Doch nach unserer Auffassung zeigen Atomuhren eben nicht die wahre Zeit. Mit Unterscheidung der wahren kosmischen Zeit $t^*$ von der Eigenzeit $\tau$ machen wir einen allgemeinen kosmologischen Ansatz, der mit der Vernachlässigung aller räumlichen Inhomogenitäten[24]) allerdings auf der üblichen groben Vereinfachung beruht (und deshalb auch nicht überinterpretiert werden darf):

$$d\bar{s}_{(n)}^{*2} \;=\; \bar{g}_{ik}^* \, dx^{*i} dx^{*k} \;=\; e^{3Ht^*(1+\Pi)} c^2 dt^{*2} - e^{2Ht^*} dl^{*2}. \tag{1a}$$

Überstreichungen wie z.B. bei $\bar{g}_{ik}^*$ verwenden wir immer, um *räumliche* Mittelung anzuzeigen. Der Stern bei den Koordinaten $x^{*i}$ bezeichnet das hier verwendete kosmische Bezugssystem S* [lateinische Indizes $i, k, l ... = 0, 1, 2, 3$ werden immer unterschieden von griechischen $\alpha, \beta ... = 1, 2, 3$ (die letzteren für rein räumliche Koordinaten)]. Dabei machen wir hier und in den folgenden Abschnitten 4, 5, 6 von der Bedingung einer konstanten kosmischen Lichtgeschwindigkeit zunächst noch keinen Gebrauch. In (1a) könnte für d$l^*$ nun gelten

$$dl^{*2} \;=\; \frac{dr^{*2}}{1 - kr^{*2} e^{3Kt^*}} + r^{*2}\left(\sin^2\vartheta^* d\varphi^{*2} \;+\; d\vartheta^{*2}\right). \tag{1b}$$

In diesem allgemeinen Fall wären $H = H(t^*)$, $\Pi = \Pi(t^*)$ und $K = K(t^*)$ als Funktionen der kosmischen Zeit $t^*$, $k$ aber als eine Konstante[25]) zu verstehen, und mit der Belegung $\Pi = -1$, $K = 0$, $t^* = t'$, $H = 1/t' \cdot \ln R(t')$ wäre dann auch die bekannte RW-Form als ein Spezialfall des kosmischen Linienelements in (1a), (1b) enthalten. Doch im folgenden wollen wir uns mit

$$H = konstant, \; \Pi = konstant, \; k = 0 \tag{1c}$$

---

Störungen) beschrieben wird, weisen aber zugleich darauf hin, daß eine frühere offizielle Zeitskala gleichen Namens seit 1984 durch verschiedene andere Zeitskalen ersetzt worden ist in der Absicht, den Ergebnissen der allgemeinen Relativitätstheorie in geeigneter Weise Rechnung zu tragen (s. Abschn. 8/f).

[20]) Der wahre Raum ist dementsprechend der Raum, in welchem die Spiralnebel [möglicherweise Quasare, Cluster oder auch ‚Blasen' – letztlich die originären Gravitationszentren (s. Abschn. 1)] gleichmäßig verteilt sind und statistisch ruhen.

[21]) ISAAC NEWTON: „*Absolute, true, and mathematical time, of itself, and from its own nature, flows equably without relation to anything external ...* " (aus der englischen Ausgabe von 1729, zitiert nach [15]). In Scholium I der PHILOSOPHIAE NATURALIS PRINCIPIA MATHEMATICA, Editio Ultima, MDCCXIV (1. Ausgabe erschien 1687) heißt es vollständig: „*Tempus Absolutum, verum, & mathematicum, in se & natura sua absque relatione ad externum quodvis, aequabiliter fluit, alioque nomine dicitur Duratio: Relativum, apparens, & vulgare est sensibilis & externa quaevis Durationis per motum mensura (seu accurata seu inaequabilis) qua vulgus vice veri temporis utitur: ut Hora, Dies, Mensis, Annus.*"

[22]) Der allgemein gebräuchliche Begriff ‚Eigenzeit' ist insofern mißverständlich, als es aus unserer Sicht nur genau eine Zeit geben kann. Im Einklang mit dieser Voraussetzung ist die ‚Eigenzeit' konkret immer zu verstehen als (physikalisch beeinflußbare) *Anzeige* von Atomuhren. Die Bezeichnungen ‚$\tau$' oder ‚d$\tau$' beziehen sich dabei (wie $\bar{s}/c$ oder d$\bar{s}/c$ ) ausschließlich auf ruhende Uhren, bei bewegten Atomuhren (oder allgemein) soll es dagegen ‚$\sigma$' bzw. ‚d$\sigma$' heißen.

[23]) Nach üblicher Interpretation – die sich gerade auch dadurch von der hier vertretenen Auffassung unterscheidet – bezieht sich der Terminus ‚ruhend' in diesem Zusammenhang auf ein ‚mitbewegtes' Koordinatensystem.

[24]) Ein der allgemeinen Relativitätstheorie besser entsprechendes, realistischeres Modell sollte diese Inhomogenitäten der Materieverteilung statistisch berücksichtigen.

[25]) bei passender Wahl der Längeneinheit LE mit den möglichen Werten $k = -1, 0, +1$ LE$^{-2}$

auf die Lösungsschar des *singularitätsfreien*[26]) Linienelements konzentrieren, und zwar bei *euklidischem* 3-dimensionalen Raum, für dessen Beschreibung wir – sofern nicht ausdrücklich eine andere Form angegeben ist – kartesische Koordinaten verwenden werden. Dabei ist $\Pi$ eine zunächst noch unbestimmte konstante Zahl. $H$ wird sich sogleich als HUBBLE-Konstante erweisen. Dieses singularitätsfreie Linienelement

$$(1) = (1a) + (1b) + (1c) \tag{1}$$

hat die besondere Eigenschaft, daß sich zumindest die relativen zeitlichen Änderungen der Ablaufgeschwindigkeit atomarer Prozesse und spektraler Uhren, der Ausdehnung atomarer Körper und spektraler Maßstäbe sowie des kosmischen Koordinatenwerts der Lichtgeschwindigkeit als zeitlich konstant erweisen.

## 4. Der Energie-Impuls-Tensor der statistisch ruhenden Materie

Im folgenden werden wir von den EINSTEIN'schen Gleichungen bei räumlicher Mittelung über kosmische Bereiche ausgehen, d.h. wir setzen zunächst:

$$\overline{E}_{ik}^* \equiv \overline{R}_{ik}^* - \tfrac{1}{2}\overline{R}^* \overline{g}_{ik}^* = \kappa \overline{T}_{ik}^*. \tag{2}$$

Für den ursprünglichen EINSTEIN-Tensor – d.h. ohne Berücksichtigung eines kosmologischen Glieds – finden wir aus dem singularitätsfreien Linienelement (1):

$$\overline{E}_{ik}^{*(n)} = \frac{3H^2}{c^2} \begin{pmatrix} 1 & 0 & 0 & 0 \\ 0 & \Pi e^{-Ht^*(1+3\Pi)} & 0 & 0 \\ 0 & 0 & \Pi e^{-Ht^*(1+3\Pi)} & 0 \\ 0 & 0 & 0 & \Pi e^{-Ht^*(1+3\Pi)} \end{pmatrix}. \tag{3}$$

Daraus ergibt sich mit (2) eine zeitunabhängige phänomenologische Energiedichte $\overline{T}_{00}^*$ [27]). Die zunächst noch freie Konstante $\Pi$ stellt zur Zeit $t^* = 0$ das Verhältnis der räumlichen Durchschnittswerte von kosmischem

---

[26]) In Vorwegnahme dieser Belegung steht in (1a) der Index „(n)" bereits für ‚nicht-singulär' (obwohl für $k = 1$ LE$^{-2}$ eine Nicht-Singularität natürlich nicht gegeben wäre).

[27]) Die hier angesetzte Zuordnung der Komponente $\overline{T}_{00}^*$ zur phänomenologischen Energiedichte (s. auch Abschn. 8/a) ist keineswegs selbstverständlich, obwohl einiges dafür spricht (LANDAU-LIFSCHITZ [18] z.B. zeigen, daß $T_{00}$ für beliebige $g_{ik}$ im Unterschied zur gemischten Komponente $T_0^{\,0}$ bei phänomenologischem Ansatz positiv definit ist, wie es auch sein muß). Andererseits gelten die ursprünglichen Erhaltungssätze $\partial_k \mathbf{V}_i^{\,k} = 0$ strenggenommen nur für die gemischte Tensordichte $\mathbf{V}_i^{\,k} = \mathbf{T}_i^{\,k} + \mathbf{t}_i^{\,k}$. Dabei ist $\mathbf{t}_i^{\,k} = t_i^{\,k}\sqrt{\mathbf{g}}$ die aus dem klassischen Variationsprinzip (s. z.B. PAULI [19]) entspringende Energie-Impuls-Tensordichte des Gravitationsfeldes. Doch die entsprechende kontravariante Tensordichte $\mathbf{V}^{ik}$ ist bekanntlich i.a. nicht symmetrisch, was sie im Hinblick auf die Erhaltung des Drehimpulses nach allgemeiner Auffassung sein sollte. Trotz alternativer Ansätze (z.B. in [18]), die der zuletzt genannten Schwierigkeit Rechnung tragen sollen, bedarf die Frage der Energiedichte des Gravitationsfeldes unseres Erachtens noch der endgültigen Klärung.

Im Hinblick auf das ausgezeichnete kosmische Bezugsystem (und im Unterschied zu EINSTEINs ursprünglicher, geometrischer Auffassung) sprechen wir bei $t_i^{\,k}$ nicht von einem Pseudo-Tensor, sondern von einem echten Energie-Impuls-Tensor des Gravitationsfeldes (sowie bei $\mathbf{t}_i^{\,k}$ konsequenterweise von einer echten Tensordichte), entsprechendes gilt für $V_i^{\,k}$ und $\mathbf{V}_i^{\,k}$. Das Transformationsverhalten solcher Größen und die mathematischen Grundlagen für den Übergang von einem ausgezeichneten zu einem beliebigen anderen Bezugsystem leistet die bimetrische Formulierung der allgemeinen Relativitätstheorie durch ROSEN [20] u.a. auf Basis eines mathematischen Ansatzes von LEVI-CIVITA (s. a. Literaturangaben in [21], Bd. II §54).

Im vorliegenden Fall ergibt die Energie-Impuls-Tensordichte $\overline{\mathbf{t}}_i^{*\kappa} = \overline{t}_i^{*\kappa}\sqrt{\overline{\mathbf{g}}^*}$ des Gravitationsfeldes zusammen mit der phänomenologischen Dichte $\overline{\mathbf{T}}_i^{*\kappa} = \overline{T}_i^{*\kappa}\sqrt{\overline{\mathbf{g}}^*}$ für die *gesamte* Energiedichte $\overline{\mathbf{V}}_0^{*0}$ aus Materie und Gravitationsfeld den Wert Null gemäß:

$$\kappa \overline{\mathbf{V}}_i^{*k} = \kappa \sqrt{\overline{\mathbf{g}}^*}\,\overline{V}_i^{*k} = \kappa\left(\overline{\mathbf{T}}_i^{*k} + \overline{\mathbf{t}}_i^{*k}\right) = \frac{3H^2(1-\Pi)}{c^2} e^{\frac{3}{2}Ht^*(1-\Pi)} \begin{pmatrix} 0 & 0 & 0 & 0 \\ 0 & 1 & 0 & 0 \\ 0 & 0 & 1 & 0 \\ 0 & 0 & 0 & 1 \end{pmatrix}. \tag{F1}$$

Druck zu kosmischer Energiedichte dar. Zu diesem Druckfaktor $\Pi$ lassen sich nun verschiedene plausible Überlegungen anstellen. Wenn die Spiralnebel im euklidischen Raum statistisch ruhen, dann sollten wir nach üblichem Verständnis zunächst erwarten, daß der kosmische Druck näherungsweise gleich Null, und damit also auch $\Pi = 0$ sei. Der EINSTEIN-Tensor wäre dann insgesamt zeitunabhängig:

$$\overline{E}_{ik}^{*(n)}(\Pi{=}0) \;=\; \frac{3H^2}{c^2}\begin{pmatrix}1&0&0&0\\0&0&0&0\\0&0&0&0\\0&0&0&0\end{pmatrix}. \tag{3a}$$

Doch auch im Sonderfall $\Pi = -1/3$, den wir ausführlich in Abschnitt 8 behandeln werden, ist der gesamte kovariante EINSTEIN-Tensor $\overline{E}_{ik}^{*}$ – und damit der Energie-Impuls-Tensor $\overline{T}_{ik}^{*}$ – unabhängig von der Zeit. Dies gilt in diesem Fall zusätzlich auch für die kontravariante Tensordichte $\overline{\mathbf{T}}^{*ik} = \overline{T}^{*ik}\sqrt{\overline{\mathbf{g}}^*}$. Dabei steht $\overline{\mathbf{g}}^* = -|\overline{g}_{ik}^*| =$ $e^{3(3+\Pi)Ht^*}$ für den negativen Wert der Determinante des Fundamentaltensors $\overline{g}_{ik}^*$ und ist bezogen auf kartesische Koordinaten. Im Unterschied zu $\overline{T}_{00}^{*}$ ist aber die Komponente $\overline{\mathbf{T}}_0^{*0}$ der gemischten Tensordichte hier wie dort *nicht* mehr unabhängig von der Zeit.

## 5. Der kosmische Einfluß auf die Ganggeschwindigkeit von Atomuhren

Die in mathematischer Analogie zur Bestimmung geodätischer Linien aus der Variation des singularitätsfreien kosmischen Elements (1) fließenden Bewegungsgleichungen

$$\delta \int d\overline{s}_{(n)}^{*} \;=\; 0 \tag{4}$$

ergeben als Lösungen für die räumlichen Komponenten $\overline{u}^{*\alpha}$ der mittleren Vierer-Geschwindigkeit

$$\overline{u}^{*\alpha} \;=\; \frac{d\overline{x}^{*\alpha}}{d\overline{s}_{(n)}^{*}} \;=\; \overline{u}_{(0)}^{*\alpha}\, e^{-2Ht^*}. \tag{5}$$

Für die auf absolute Koordinaten bezogenen Komponenten der gewöhnlichen mittleren Geschwindigkeit $\overline{v}^{*\alpha} = d\overline{x}^{*\alpha}/dt^*$ folgt, daß diese als Abweichungen vom Zustand *statistischer* Ruhe für $\Pi < 1/3$ mit der Zeit abnehmen sollten gemäß

$$\overline{v}^{*\alpha} \;=\; \frac{\overline{u}^{*\alpha}}{\overline{u}^{*0}} \;=\; \frac{c\,\overline{u}_{(0)}^{*\alpha}\, e^{-\tfrac{1}{2}Ht^*(1-3\Pi)}}{\sqrt{1+\left[\overline{u}_{(0)}^{*\alpha}\right]^2 e^{-2Ht^*}}} \;, \tag{6}$$

wobei $\overline{u}_{(0)}^{*\alpha}$ die Anfangswerte der räumlichen Komponenten der mittleren Vierer-Geschwindigkeit zur Zeit $t^* = 0$ wären (im Ausdruck $[\overline{u}_{(0)}^{*\alpha}]^2$ ist über $\alpha = 1, 2, 3$ zu summieren). Damit ist gezeigt, daß sich aus den relativistischen Bewegungsgleichungen auf Basis des singularitätsfreien Linienelements kein zwingender Grund ableiten läßt, von der ursprünglich so naheliegenden Vorstellung abzugehen, daß die Spiralnebel – statistisch gemittelt – im kosmischen euklidischen Raum ruhen (dies hatten wir in den Abschn. 2, 3 vorausgesetzt, s. auch Abschn. 10).

Im Hinblick auf das Fehlen jeder Energieströmung im kosmischen Energie-Impuls-Tensor ist mit $\overline{u}_{(0)}^{*\alpha} = 0$ die richtige Lösung gemäß (6) also offenbar

$$\overline{v}^{*\alpha} \;=\; 0 \;, \tag{7}$$

---

Das Resultat $\overline{\mathbf{V}}_0^{*0} = 0$ gilt auch bei anderer Definition der $t_i^k$ (z.B. nach WEINBERG [22]). Daß demzufolge die gesamte Energiedichte aus Materie und Gravitationsfeld gleich Null sei, mag zunächst befremdlich erscheinen (unter diesem Aspekt hätten reale Objekte letztlich kosmische *Strukturen* zu verkörpern, d.h. eher die Form als den Stoff). Auf Basis integrierter Koordinaten im System K' erhielte man allerdings ein anderes Ergebnis, nämlich $t_i'^{\,k} = 0$, d.h. als gesamte ‚lokale' Energiedichte $\overline{\mathbf{V}}_0'^{\,0}$ = $\overline{\mathbf{T}}_0'^{\,0}$ (s. Abschn. 8/c). Dies gilt näherungsweise auch für das adaptierte System K″ (s. Abschn. 8/d).

d.h. diejenige für statistisch ruhende Materie[28]). In diesem Fall aber ergeben die ‚geodätischen' Bewegungsgleichungen als einzige von Null verschiedene Komponente der mittleren Vierer-Geschwindigkeit ausschließlich eine kosmische Beschleunigung der Ganggeschwindigkeit von Atomuhren:

$$\bar{u}^{*0} \;=\; \frac{\mathrm{d}t^*}{\mathrm{d}\bar{s}^*_{(n)}} \;=\; \mathrm{e}^{-\frac{3}{2}Ht^*(1+\Pi)}. \tag{8}$$

Hierdurch wird das Verhalten fernab von lokalen Gravitationsquellen ruhender Atomuhren beschrieben, die der Voraussetzung gleichmäßig verteilter – und gemäß (7) tatsächlich ruhender – Materie entsprechen.

Für die Anzeige ruhender Atomuhren ist $\mathrm{d}\bar{s}^*/c \approx \mathrm{d}\tau$, und dieses lokale Eigenzeitelement ergibt sich aus (8) einfach zu[29])

$$\mathrm{d}\tau \;\approx\; \mathrm{e}^{\frac{3}{2}Ht^*(1+\Pi)}\,\mathrm{d}t^*. \tag{9}$$

Wie aber ist dies zu verstehen? Aufgrund ihrer Nicht-Integrabilität ist es klar, daß eine beliebige ruhende Atomuhr mit ihrer ‚Eigenzeit' $\tau$ (die über größere kosmische Bereiche auch *nicht* übereinstimmt mit der Zeitvariablen der RW-Form) nicht unmittelbar die wahre Systemzeit $t^*$ anzeigen kann. Das bedeutet natürlich keineswegs, daß sie als Uhr unbrauchbar wäre (ebensowenig wie ein realer Maßstab deshalb unbrauchbar ist, weil es das Phänomen der Temperaturausdehnung gibt). Was jede Atomuhr zweifellos *richtig* anzeigt, ist die Anzahl ihrer Takte, deren Dauer sich aus unserer Sicht im Verlauf der kosmischen Zeit systematisch ändert. Die infinitesimalen Intervalle der Eigenzeit $\mathrm{d}\tau$ (und der Eigenlänge $\mathrm{d}\lambda$) sind aber grundsätzlich nur näherungsweise integrabel. Wie in Abschnitt 8/c gezeigt wird, sind die durch Integration entstehenden Näherungen für $\tau$ (und $\lambda$) allerdings sehr gut brauchbar, solange gilt $Ht^*, Hr^*/c \ll 1$.

Betrachten wir nun ganz konkret das Verhalten einer Atomuhr im Vergleich zu einer im kosmischen Bezugssystem ruhenden *Systemuhr*[30]). Die Systemuhr zeige die Anzahl $Z_{t^*}$ der unveränderlichen absoluten Zeiteinheiten ZE = $T^*$ der wahren Zeit $t^*$. Die Atomuhr dagegen zeige die Anzahl $Z_\tau$ ihrer *veränderlichen* Takte $T^{[\tau]}$. Wenn es aber nur die eine wahre Zeit $t^*$ gibt, dann muß offenbar gelten

$$Z_{t^*} T^* \;=\; t^* \;=\; \sum_{\nu=1}^{Z_\tau} T_\nu^{[\tau]}, \tag{10}$$

und für jedes hinreichend kleine Intervall $\Delta t^*$ folgt näherungsweise

$$\Delta Z_{t^*} T^* \;=\; \Delta t^* \;\approx\; \Delta Z_\tau T^{[\tau]}(t^*). \tag{11}$$

Die kosmische Zeitabhängigkeit des Takts $T^{[\tau]}(t^*)$ einer Atomuhr ergibt sich aus (9), indem wir $\mathrm{d}\tau$ und $\mathrm{d}t^*$ für den Fall hinreichend kleiner, aber endlicher Zeitintervalle dort zunächst näherungsweise durch $\Delta\tau \approx \Delta Z_\tau \cdot 1\mathrm{ZE}$ und $\Delta t^* = \Delta Z_{t^*} \cdot 1\mathrm{ZE}$ ersetzen und dann durch Herauskürzen der absoluten Zeiteinheit ZE = $T^*$ von den dimensionsbehafteten Größen zu den reinen Maßzahlen $\Delta Z_\tau$, $\Delta Z_{t^*}$ übergehen:

---

[28]) An dieser Stelle sei festgehalten, daß – obwohl im Energie-Impuls-Tensor gemäß (3) keine Energieströmung auftritt – das zugrunde liegende Linienelement (1) durchaus geeignet wäre, auch eine real expandierende, lokalisierbare Materieverteilung zu beschreiben, und zwar in einem ‚mitbewegten' Koordinatensystem. Zur Interpretation der tatsächlichen kosmischen Gegebenheiten ist diese Auffassung unseres Erachtens allerdings nicht geeignet (s. auch Abschn. 10).

[29]) Trotz der infinitesimalen Zeitspannen $\mathrm{d}\tau$ bzw. $\mathrm{d}t^*$ setzen wir hier das Zeichen ‚$\approx$' (und kein Gleichheitszeichen). Diese Unterscheidung ist als Hinweis auf die Nicht-Integrabilität der Eigenzeit $\tau$ zu verstehen (im Unterschied zum Element der wahren Zeit $\mathrm{d}t^*$ – oder auch der Quasi-Eigenzeit $\mathrm{d}t'$ – ist das Element der echten Eigenzeit $\mathrm{d}\tau$ über kosmische Zeiträume nicht integrabel, s. Abschn. 8/c). An dieser Stelle sei sogleich auch darauf hingewiesen, daß die (eingebürgerte) Variable $\tau$ als eine zur Anzeige der (eingebürgerten) ‚Eigenzeit' proportionale Größe nicht mit der auf die absolute Zeit bezogenen Schwingungsdauer $T^{[\tau]}(t^*)$ einer Spektrallinie zu verwechseln ist (s. auch Fußn.[22]). Ebenso darf das später verwendete korrespondierende $\lambda$ in diesem Zusammenhang nicht etwa als Wellenlänge verstanden werden, sondern als Variable, die proportional ist zur Maßzahl einer mit Bezug auf spektrale Einheiten gemessenen Länge.

[30]) Die kosmische Zeit kann nur von *technischen* Systemuhren richtig angezeigt werden, die durch entsprechende Eingriffe herzustellen wären.

$$\Delta Z_\tau \approx e^{\frac{3}{2}Ht^*(1+\Pi)} \Delta Z_{t^*} . \tag{12}$$

Diese Vorgehensweise ist aber nicht nur erlaubt, sondern bringt unseres Erachtens den – für das lokale Gravitationsfeld experimentell überzeugend bestätigten – Inhalt der Beziehung (9) überhaupt erst unmißverständlich zum Ausdruck:

Die *Anzahl* $Z_\tau$ der veränderlichen Takte $T^{[\tau]}(t^*)$ einer im Gravitationspotential $\overline{g}_{ik}^*$ ruhenden Atomuhr ergibt sich für ein hinreichend kurzes Intervall $\Delta t^*$ der kosmischen Zeit aus der *Anzahl* $Z_{t^*}$ der unveränderlichen Takte $T^*$ einer entsprechenden Systemuhr durch Multiplikation mit der Wurzel aus $\overline{g}_{00}^*$.

Mit Rücksicht darauf, daß für einen einzigen Takt der Atomuhr selbstverständlich $\Delta Z_\tau = 1$ zu setzen ist, folgt aus den letzten beiden Gleichungen schließlich

$$T^{[\tau]}(t^*) \approx T^* e^{-\frac{3}{2}Ht^*(1+\Pi)} , \tag{13}$$

wobei die absolute Zeiteinheit ZE = $T^*$ konstant ist. Daß aber der Takt $T^{[\tau]}(t^*)$ der Atomuhren in gleichem Maße abnehmen muß, wie umgekehrt ihre Anzeige $\tau$ – und damit auch die Frequenz $f$ eines entsprechenden Senders – gegenüber der kosmischen Zeit zunimmt, liegt auf der Hand.

Meßgeräte – indem sie vergleichen – zeigen nur Maßzahlen an. Aussagen über den veränderlichen Takt einer Uhr sind Aussagen über Zeitspannen, Aussagen über die Anzeige einer Uhr sind Aussagen über Frequenzen: Eine Aussage über die *Abnahme* der Zeitspanne also, die dem veränderlichen Takt einer Atomuhr entspricht, darf nicht verwechselt werden mit der korrespondierenden Aussage über die *Zunahme* der Anzeige einer mit dieser Uhr zu verschiedenen Zeiten wiederholt gemessenen gleichbleibenden kosmischen Zeitspanne.

Bei Bezug auf spektrale Zeiteinheiten ändern sich dagegen die Maßzahlen atomarer Zeitspannen (z.B. Schwingungsdauern von Spektrallinien am Ort ihrer Entstehung) *nicht* mit der Zeit[31]), ebensowenig wie auch die Maßzahlen atomarer Längen (z.B. Abstände benachbarter Schwingungsknoten einer stehenden Lichtwelle am Ort der Quelle) bei Bezug auf spektrale Längeneinheiten. In allen derartigen Fällen sind es lediglich die spektralen Einheiten, nicht aber ihre jeweilige Anzahl, die sich zusammen mit den zu messenden atomaren Größen ändern im Verlauf der kosmischen Zeit[32]).

## 6. Die Rotverschiebung

Zur Berechnung der kosmischen Rotverschiebung – in mathematischer Analogie zur ursprünglichen Ableitung durch LEMAÎTRE [5], s. aber Fußn.[68]) – gehen wir aus von der zuvor begründeten Vorstellung, daß die Spiralnebel, von statistischen Abweichungen abgesehen, gemäß (7) im euklidischen Raum ruhen. Wird nun im Emissionszeitpunkt $t_E^*$ der kosmischen Zeit der Wellenberg einer Lichtwelle an einem beliebigen Ort $P_E$ des euklidischen Raumes ausgesandt, so möge er zum Zeitpunkt $t_0^*$ am Ort $P_0$ eintreffen. $P_0$ habe von $P_E$ die euklidische Entfernung $l^*$. Der nächste Wellenberg werde in $P_E$ ausgesandt zum Zeitpunkt $t_E^* + \delta t_E^*$ und treffe ein in $P_0$ zum Zeitpunkt $t_0^* + \delta t_0^*$. Dann muß bei Lichtgeschwindigkeit $c^*(t^*)$ gelten

$$\int_{t_E^*}^{t_0^*} c^*(t^*)\,dt^* = l^* = \int_{t_E^*+\delta t_E^*}^{t_0^*+\delta t_0^*} c^*(t^*)\,dt^* , \tag{14}$$

woraus sich für infinitesimal kleine Intervalle $\delta t_0^*$, $\delta t_E^*$ näherungsweise ergibt:

$$\delta t_0^* \approx \delta t_E^* \frac{c^*(t_E^*)}{c^*(t_0^*)} . \tag{15}$$

---

[31]) Soweit diese Aussage die Zeit betrifft, handelt es sich bei Verwendung von Atomuhren um eine Tautologie.

[32]) Wir werden in Abschnitt 8 noch einmal darauf zurückkommen. – Unter diesem Aspekt scheint übrigens ein Nachweis von Gravitationswellen mit *beinahe* lokalen Experimenten wie z.B. GEO600 (auch abgesehen von technischen Schwierigkeiten) durchaus nicht selbstverständlich.

Bei $c^*$ handelt es sich um den Koordinatenwert der Lichtgeschwindigkeit[33], den wir als *kosmische* Lichtgeschwindigkeit bezeichnen wollen. Diese ergibt sich mit $d\bar{s}^* = 0$ unmittelbar aus dem Linienelement (1) zu

$$c^*(t^*) = \frac{dl^*}{dt^*} = c\,e^{\frac{1}{2}Ht^*(1+3\Pi)}, \tag{16}$$

und mit der Abkürzung $\Delta t^* = t_0^* - t_E^* > 0$ wird aus (15)

$$\delta t_0^* \approx \delta t_E^*\,e^{-\frac{1}{2}H\Delta t^*(1+3\Pi)}. \tag{17}$$

Ein Eigenzeitintervall $\delta\tau$ der im kosmischen Bezugssystem ruhenden Atomuhren steht mit dem entsprechenden Intervall $\delta t^*$ der kosmischen Zeit in der Beziehung (9). Für die Zeit der Aussendung $t_E^*$ bzw. die Zeit der Ankunft $t_0^*$ ergeben sich damit nun im einzelnen

$$\delta t_{0/E}^* \approx \delta\tau_{0/E}\,e^{-\frac{3}{2}Ht_{0/E}^*(1+\Pi)}. \tag{18}$$

Eingesetzt in die Beziehung (17) zwischen den Intervallen der kosmischen Zeit, folgt für die entsprechenden Intervalle der Eigenzeit schließlich einfach[34]

$$\delta\tau_0 = \delta\tau_E\,e^{H\Delta t^*}. \tag{19}$$

Zunächst steht $\delta\tau_E$ hier nur für das Eigenzeitintervall zum Zeitpunkt $t^* = t_E^*$ der Emission am Ort $P_E$. Wir wünschen aber, die Schwingungsdauer $\delta\tau_0$ einer von $P_E$ kommenden Spektrallinie am Beobachtungsort $P_0$ mit der Schwingungsdauer einer hier zur Beobachtungszeit $t^* = t_0^*$ neu entstehenden Spektrallinie desselben Typs zu vergleichen. Nun ist jedoch klar, daß die mit Atomuhren gemessene Schwingungsdauer $\delta\tau_E$ von Spektrallinien am Ort und zur Zeit ihrer Entstehung konstant ist[35]. Es gilt also $\delta\tau_E(t_0^*) = \delta\tau_E = \delta\tau_E(t_E^*)$. Gerade auf dieser Konstanz beruht ja die Konstruktion von Atomuhren (s. Fußn.[31]). Setzen wir nun wegen des in der Vergangenheit liegenden Emissionszeitpunkts $t_E^*$

$$\Delta t^* = t_0^* - t_E^* \geq 0, \tag{20}$$

so ergibt sich für den wie üblich definierten Rotverschiebungs-Parameter $z$ also

$$z \equiv \frac{\delta\tau_0}{\delta\tau_E} - 1 = e^{H\Delta t^*} - 1. \tag{21}$$

Unter der oben (implizit) benutzten Voraussetzung $H\delta t_0^*$, $H\delta t_E^* \ll 1$ hängt dieser nun also weder ab von $\Pi$ noch – wie nicht anders zu erwarten war[36] – von den absoluten Werten $t_0^*$ oder $t_E^*$, sondern alleine von der Konstanten $H$ und der *Zeitspanne* $\Delta t^*$, welche nichts anderes ist als die (positive) Laufzeit des Lichts.

Mit $t_0^* = 0$, d.h. mit Setzung des Nullpunkts der kosmischen Zeitskala in den Zeitpunkt der Beobachtung, wollen wir nun die Rotverschiebung wie üblich auch als Funktion der räumlichen Entfernung darstellen. Gemäß (16) folgt für die Entfernung $l_0^*$ der Lichtquelle in $P_E$ vom Beobachtungsort $P_0$

$$l_0^* = \int_{-\Delta t^*}^{0} c^*(t^*)\,dt^* = \frac{2c}{H(1+3\Pi)}\left[1 - e^{-\frac{1}{2}H\Delta t^*(1+3\Pi)}\right]. \qquad (\Pi \neq -1/3) \tag{22}$$

---

[33]) Ein Überblick über die verschiedenen Werte der Lichtgeschwindigkeit und deren Bedeutung findet sich in [13].

[34]) Angesichts der extrem kurzen Zeitspannen $\delta\tau_0$ bzw. $\delta\tau_E$ haben wir im Hinblick auf die gebräuchliche Schreibweise das Zeichen ‚≈' durch ‚=' ersetzt. Dies ist hier möglich, weil dabei nur Eigenzeit mit Eigenzeit verglichen wird.

[35]) Bei umgekehrter Betrachtung folgt aus (9), daß bezogen auf die *absolute* Zeit die Schwingungsdauer $\delta t_E^*(t^*)$ einer heute entstehenden Spektrallinie entsprechend *kleiner* ist als die der früher entstandenen (s. Abschn. 5).

[36]) Gerade darauf haben wir ja auch abgezielt mit unserem *exponentiellen* Ansatz des singularitätsfreien kosmischen Linienelements (1).

Das Ergebnis in (22) gilt nur unter der Voraussetzung $\Pi \neq -1/3$ (s. aber die Abschn. 7, 8) und ist zugleich mit $t_0^* = 0$ gewissermaßen *normalisiert* [37] auf die heute dementsprechend festgesetzten Einheiten von Zeit und Länge. Aus (21) und (22) ergibt sich zunächst

$$l_0^* = \frac{2c}{H(1+3\Pi)}\left[1 - (1+z)^{-\frac{1+3\Pi}{2}}\right], \qquad (\Pi \neq -1/3) \quad (23)$$

und daraus schließlich der Rotverschiebungs-Parameter $z$ in Abhängigkeit von der kosmischen Entfernung $l_0^*$, d.h. in HUBBLE-Darstellung

$$z(l_0^*) = \left[1 - \frac{H(1+3\Pi)}{2c}l_0^*\right]^{-\frac{2}{1+3\Pi}} - 1 = \frac{Hl_0^*}{c} + O^2\left(\frac{Hl_0^*}{c}\right). \qquad (\Pi \neq -1/3) \quad (24)$$

Hätte man aber bei der Berechnung des Ausdrucks (22) ohne Berücksichtigung der Normalisierung ($t_0^* = 0$) von $t_0^* - \Delta t^*$ bis $t_0^*$ integriert, so würde sich im Fall $\Pi \neq -1/3$ der Rotverschiebungsparameter $z$ für ein und denselben ruhenden Spiralnebel mit dem Zeitpunkt $t_0^*$ der Beobachtung ändern[38]. Für beliebige Belegungen von $\Pi$ kann das singularitätsfreie Linienelement (1) also nicht allgemein als stationär gelten.

Berechnet man nun die maximalen Lichtwege aus der *Vergangenheit*, so folgt für diese mit $t_0^* = 0$ und demzufolge gemäß (20) $\Delta t^* = -t_E^* := -t^*$ aus (22)

---

[37]) Das singularitätsfreie Linienelement (1) ist zu einer beliebigen Zeit $t^* = t_0^* + \Delta t^*$ einerseits gegeben als

$$d\bar{s}_{(n)}^{*2} = e^{3H(1+\Pi)(t_0^*+\Delta t^*)}c^2 dt^{*2} - e^{2H(t_0^*+\Delta t^*)}dl^{*2}, \qquad (F2)$$

andererseits aber soll es bei Berücksichtigung der Einflüsse des Gravitationspotentials auf lokale Objekte, Felder, Maßstäbe und Uhren zum Beobachtungszeitpunkt $t^* = t_0^*$ gerade mit dem Linienelement der speziellen Relativitätstheorie näherungsweise übereinstimmen. Daraus ergibt sich die Forderung

$$c^2 d\tau^2 - d\lambda^2 = ds_{(SRT)}^2 \approx e^{3Ht_0^*(1+\Pi)}c^2 dt^{*2} - e^{2Ht_0^*}dl^{*2}, \qquad (F3)$$

und die freie Wahl passender Einheiten erlaubt nun einmal für jeden Zeitzusammenhang die willkürliche ‚Normalisierung'

$$t_0^* = 0, \qquad (F4)$$

sodaß im dadurch festgesetzten Zeitnullpunkt die Einheiten der atomaren Größen mit denen der absoluten im lokalen Inertialsystem übereinstimmen: $\delta\tau(\Delta t^*=0) = \delta t^*(\Delta t^*=0)$ und $\delta\lambda(\Delta t^*=0) = \delta l^*(\Delta t^*=0)$. Es ist hier noch einmal festzuhalten, daß es in einem stationären Universum *grundsätzlich* möglich sein muß, für jeden Beobachtungszeitraum immer wieder $t_0^* = 0$ zu setzen, wobei während des jeweils betrachteten Ablaufs das zeitlich veränderliche Linienelement, in diesem Fall also

$$d\bar{s}_{(n)}^{*\,2} = e^{3H\Delta t^*(1+\Pi)}c^2 dt^{*2} - e^{2H\Delta t^*}dl^{*2}, \qquad (F5)$$

das kosmische Geschehen bestimmt. Diese Möglichkeit ist nicht so abwegig, wie es auf den ersten Blick scheinen mag. Denn ohne sprunghafte Veränderungen der spektralen Einheiten könnten schon die Abläufe in einem gewöhnlichen, mit veränderlicher Geschwindigkeit frei fallenden lokalen Inertialsystem wegen der Nicht-Integrabilität der lokalen LORENTZ-Transformation im wahren Gravitationsfeld nicht einmal näherungsweise *andauernd* mit der speziellen Relativitätstheorie in Einklang stehen.

[38]) Zunächst hätte sich für den zurückgelegten Lichtweg $l^*$ eine Abhängigkeit vom Beobachtungszeitpunkt $t_0^*$ ergeben

$$l^* = \int_{t_0^*-\Delta t^*}^{t_0^*} c^*(t^*)dt^* = e^{\frac{1}{2}Ht_0^*(1+3\Pi)}l_0^* = \frac{c^*(t_0^*)}{c}l_0^*, \qquad (\Pi \neq -1/3) \quad (F6)$$

woraus mit (21) eine Rotverschiebung resultiert hätte, die damit ihrerseits ebenfalls explizit vom Beobachtungszeitpunkt $t_0^*$ abhängig wäre. Anstelle von (24) hätten wir erhalten

$$z(t_0^*) = \left[1 - \frac{Hl_0^*(1+3\Pi)}{2c}e^{-\frac{1}{2}Ht_0^*(1+3\Pi)}\right]^{-\frac{2}{1+3\Pi}} - 1 = \frac{Hl_0^*}{c}e^{-\frac{1}{2}Ht_0^*(1+3\Pi)} + O^2\left(\frac{Hl_0^*}{c}\right). \qquad (\Pi \neq -1/3) \quad (F7)$$

Doch in einem stationären Universum sind selbstverständlich alle Aussagen sinnlos, die einen willkürlich gewählten Bezugspunkt $t_0^*$ der kosmischen Zeitskala enthalten.

$$l^*_{0\max}(t^* \to -\infty) = \frac{2c}{H(1+3\Pi)}. \qquad (\Pi > -1/3) \quad (25a)$$

Umgekehrt aber könnte das heute von einem Punkt $P_E$ ausgehende Lichtsignal mit $\Pi > -1/3$ in *Zukunft* gemäß (22) bei entsprechend langer Laufzeit $t^*$ jede Entfernung zu einem beliebigen Punkt $P_0$ des kosmischen euklidischen Raums überbrücken:

$$l^*_{0\max}(t^* \to +\infty) = \infty. \qquad (\Pi > -1/3) \quad (25b)$$

Beide Ergebnisse (25a) und (25b) gelten für $\Pi > -1/3$ (für $\Pi < -1/3$ verhielten sich die Aussagen über vergangene und zukünftige Lichtwege gerade umgekehrt). Doch dieser Sachverhalt hat einen Haken. Denn ob uns das Licht einer hinreichend weit entfernten Galaxie tatsächlich erreicht, kann in einem stationären Universum nicht davon abhängen, ob wir den Vorgang der Lichtausbreitung vom Zeitpunkt der Absorption aus rückblickend, oder umgekehrt vom Zeitpunkt der Emission aus vorausblickend beschreiben[39]).

### 7. Die allgemeine Skalarform des kosmischen Linienelements

Das zuletzt aufgetretene Dilemma hat seinen Ursprung ganz offensichtlich in der Zeitabhängigkeit (16) der kosmischen Lichtgeschwindigkeit $c^*$. Eine solche Zeitabhängigkeit aber läßt sich sehr einfach vermeiden, indem wir zu einem neuen kosmischen Koordinatensystem $S^{(c)}$ übergehen, von dem wir *fordern*, daß hier gilt

$$c^{(c)} = \frac{dl^{(c)}}{dt^{(c)}} = c, \qquad (26)$$

und dementsprechend auch

$$l^{(c)} = c\Delta t^{(c)}. \qquad (27)$$

Was nämlich könnte uns daran hindern, theoretisch eine Synchronisation beliebiger im absoluten kosmischen Raum ruhender technischer Systemuhren gemäß dem EINSTEIN'schen Prinzip der Reflexion im Zeitmittelpunkt durchzuführen? Ein stationäres kosmisches Linienelement sollte demzufolge also vom WEYL-Typus[40]) sein:

$$d\bar{s}^{(c)2} = \zeta^2 ds^{(c)2}_{(SRT)} = \zeta^2 \eta_{ik} dx^{(c)i} dx^{(c)k}, \qquad (28)$$

wobei $\eta_{ik}$ wie üblich für den Fundamentaltensor der speziellen Relativitätstheorie stehen soll. Mit (28) nämlich ist das singularitätsfreie Linienelement (1) in diejenige Form übergegangen, aus der ganz unabhängig vom Wert des Druckfaktors $\Pi$ immer eine konstante *kosmische* Lichtgeschwindigkeit $c^{(c)} = c$ resultiert. Nur in dieser allgemeinen *Skalarform* ist das kosmische Linienelement invariant gegenüber LORENTZ-Transformationen. Und allein diese Form erlaubt es, an einer (modifizierten) Festlegung des Meters auf Basis der Lichtlaufzeit (für Hin- und Rückweg) festzuhalten[41]), was gerade für die Vermessung kosmischer Entfernungen nicht nur zweckmäßig, sondern im Sinne einer überschaubaren mathematischen Behandlung nahezu unverzichtbar ist.

---

[39]) Der gleiche Einwand betrifft insbesondere auch das Linienelement der oben zitierten Steady-State Theory, das im singularitätsfreien Linienelement (1) als Sonderfall $\Pi = -1$ enthalten ist.

[40]) Ein Linienelement der Form $ds = \zeta ds_{(SRT)}$ wurde im Anschluß an NORDSTRÖM [23] von EINSTEIN-FOKKER [24] behandelt und später von WEYL [25] zu $ds = \zeta ds_{(ART)}$ erweitert. An dem ursprünglich darauf gegründeten Versuch einer Vereinheitlichung von Elektrodynamik und Gravitation hat WEYL selbst später nicht mehr festgehalten, doch hat sein darin begründetes Konzept der Eichinvarianz insbesondere für die Weiterentwicklung der Quantenmechanik bekanntlich große Bedeutung erlangt.

[41]) Siehe [13], dort wird die Möglichkeit einer internen Synchronisation allgemeiner stationärer Systeme aufgezeigt. Eine *a priori* formulierte Synchronisations-Bedingung für die Systemuhren der kosmischen Zeit auf Grundlage einer naturphilosophisch vorausgesetzten singularitätsfreien Stationarität des Universums über hinreichend große Skalen müßte offenbar lauten: Ruhende kosmische Uhren sind nur dann richtig synchronisiert, wenn sich in der von ihnen angezeigten Zeit alle kosmischen Abläufe statistisch-stationär darstellen lassen. Hier nun zeigt sich, daß die Forderung der Stationarität tatsächlich ohne weiteres auf eine konstante kosmische Lichtgeschwindigkeit $c^{(c)} = c$ führt, was gemäß der in Abschnitt 6 der zitierten Arbeit formulierten verallgemeinerten Synchronisations-Bedingung mit einer – hier allerdings nur theoretisch möglichen – EINSTEIN-Synchronisation auf Basis der Reflexion im Zeitmittelpunkt in Einklang steht.

Für den Sonderfall $\Pi = -1/3$, auf den wir in Abschnitt 8 ausführlich zurückkommen, gilt $t^{(c)} = t^*$ und wir entnehmen den kosmischen Zeitskalar $\zeta$ unmittelbar aus der singularitätsfreien Lösungsschar (1)

$$\zeta^* = e^{Ht^*}. \qquad (\Pi = -1/3) \qquad (29a)$$

Für alle anderen Werte des Druckfaktors ($\Pi \neq -1/3$) dagegen finden wir

$$\zeta^{(c)} = \left[1 + \tfrac{1}{2} H t^{(c)} (1 + 3\Pi)\right]^{\frac{2}{1+3\Pi}}, \qquad (\Pi \neq -1/3) \qquad (29b)$$

und zwar durch die Transformation vom singularitätsfreien System S* auf das System S$^{(c)}$ konstanter kosmischer Lichtgeschwindigkeit $c^{(c)} = c$ mittels der Formeln

$$t^* = \frac{1}{H} \ln\left[1 + \tfrac{1}{2} H t^{(c)} (1 + 3\Pi)\right]^{\frac{2}{(1+3\Pi)}}, \qquad (\Pi \neq -1/3) \qquad (30)$$

$$l^* = l^{(c)}.$$

Den EINSTEIN-Tensor zur allgemeinen Skalarform des kosmischen Linienelements (28) finden wir mit Rücksicht sowohl auf (29a) als auch auf (29b) für beliebige Werte von $\Pi$ zu

$$\overline{E}_{ik}^{(c)} = \frac{3H^2/c^2}{\left[1 + \tfrac{1}{2} H t^{(c)} (1+3\Pi)\right]^2} \begin{pmatrix} 1 & 0 & 0 & 0 \\ 0 & \Pi & 0 & 0 \\ 0 & 0 & \Pi & 0 \\ 0 & 0 & 0 & \Pi \end{pmatrix}, \qquad (31)$$

und im Fall $\Pi = 0$ verschwindenden kosmischen Drucks ergibt sich demzufolge

$$\overline{E}_{ik}^{(c)}(\Pi=0) = \frac{3H^2/c^2}{\left(1 + \tfrac{1}{2} H t^{(c)}\right)^2} \begin{pmatrix} 1 & 0 & 0 & 0 \\ 0 & 0 & 0 & 0 \\ 0 & 0 & 0 & 0 \\ 0 & 0 & 0 & 0 \end{pmatrix}, \qquad (31a)$$

woraus hier nun eine Singularität bei $t^{(c)} = -2/H$ resultiert. Damit zeigt sich, daß die Bedingung $c^{(c)} = c$ für ein streng stationäres Universum nach (26) zwar notwendig, aber noch nicht hinreichend ist.

Bei dieser Gelegenheit sei ausdrücklich darauf hingewiesen, daß sich prinzipiell jedes RW-Linienelement mit räumlich euklidischer Metrik ($k = 0$) in die Skalarform (28) bringen läßt[42].

---

[42]) Natürlich läßt sich umgekehrt auch das singularitätsfreie Linienelement (1) auf RW-Form bringen, und zwar durch die Transformation der kosmischen Zeit $t^*$ auf die Quasi-Eigenzeit $t'$. Dazu definieren wir zunächst den (nicht mit dem Krümmungsskalar $R$ zu verwechselnden) Skalenfaktor $R(t')$ als

$$R(t') = \left[1 + \tfrac{3}{2}(1+\Pi) H t'\right]^{\frac{2}{3(1+\Pi)}} \qquad (F8)$$

und transformieren $t^*$ gemäß

$$t^* = \frac{1}{H} \ln R(t'). \qquad (F9)$$

Daraus ergibt sich sofort

$$d\overline{s}'^2_{(RW)} = c^2 dt'^2 - R^2(t') dl^{*2}, \qquad (F10)$$

womit die Singularitätsfreiheit verloren geht. Der kovariante EINSTEIN-Tensor erhält bezogen auf die Quasi-Eigenzeit $t'$ die alternative Form

$$\overline{E}'^{(RW)}_{ik} = \frac{3H^2}{c^2 R(t')^{3(1+\Pi)}} \begin{pmatrix} 1 & 0 & 0 & 0 \\ 0 & \Pi R^2(t') & 0 & 0 \\ 0 & 0 & \Pi R^2(t') & 0 \\ 0 & 0 & 0 & \Pi R^2(t') \end{pmatrix}. \qquad (F11)$$

Als Zusammenhang des Intervalls der Eigenzeit von Atomuhren mit dem der kosmischen Zeit haben wir nun allgemein

$$d\tau \approx \zeta^{(c)} dt^{(c)}. \tag{32}$$

Der Parameter $z$ ergibt sich aus (28) gewissermaßen als reine Gravitationsrotverschiebung, und zwar analog zur Ableitung in Abschnitt 6 hier für $\Pi \neq -1/3$:

$$z = \frac{\delta\tau_0}{\delta\tau_E} - 1 = \frac{\zeta[0]}{\zeta[-\Delta t^{(c)}]} - 1 = \left[1 - \tfrac{1}{2} H\Delta t^{(c)}(1+3\Pi)\right]^{-\frac{2}{1+3\Pi}} - 1, \quad (\Pi \neq -1/3) \tag{33}$$

wobei die Normalisierung $t_0^{(c)} = 0$, $t_E^{(c)} = -\Delta t^{(c)}$ verwendet wurde. Dieses Ergebnis stimmt wegen $\Delta t^{(c)} = l_0^{(c)}/c$ näherungsweise natürlich mit dem oben erhaltenen Ergebnis $z \approx H\Delta t^{(c)} \approx Hl_0^{(c)}/c$ überein. Ohne die willkürliche Normalisierung aber wäre $z$ hier wieder explizit abhängig von $t_0^{(c)}$ (anders verhält sich dies nur im Fall $\Pi = -1/3$, das entsprechende Resultat finden wir im nächsten Abschnitt).

Es ist bemerkenswert, daß die Entwicklungen sowohl von (29a) als auch von (29b) bis einschließlich der 2. Ordnung $O^2[Ht^{(c)}]$ übereinstimmen und insbesondere in erster Näherung unabhängig sind vom Druckfaktor $\Pi$,

$$\zeta \approx 1 + Ht^{(c)} + \tfrac{1}{4}(1-3\Pi)H^2 t^{(c)2} + O^3[Ht^{(c)}], \tag{34}$$

sodaß sich alle Einflüsse des kosmischen Hintergrunds in erster Näherung $O^1(Ht^*)$ immer einfach aus einer Reihenentwicklung der stationären Lösung mit der Belegung $\zeta^* = e^{Ht^*}$ entnehmen lassen.

## 8. Die stationäre Lösung der EINSTEIN'schen Gleichungen

Wie aus der Beziehung (25b) zu ersehen war, könnte ein Lichtsignal auf Basis des singularitätsfreien Linienelements (1) bei entsprechend langer Laufzeit in Zukunft jede beliebige Entfernung des kosmischen euklidischen Raums überbrücken. Betrachtet man umgekehrt denselben Vorgang zum Zeitpunkt der Absorption des Signals im Rückblick, so ergibt sich aus (25a), daß das Signal in der Vergangenheit keinen größeren Weg als maximal $2c/[H(1+3\Pi)]$ zurückgelegt haben kann. Legt man nun aber, um diese Problematik zu vermeiden, ein kosmisches Linienelement in der Skalarform (28) zugrunde, so geht i.a. die Singularitätsfreiheit verloren. Alle diese Widersprüche verschwinden nur, wenn man mit der speziellen Belegung

$$\Pi = -\tfrac{1}{3} \tag{35}$$

am singularitätsfreien Linienelement festhält, wobei mit (16) aber nun zugleich gilt

$$c^* = \frac{dl^*}{dt^*} = c. \tag{36}$$

---

Es ist klar, daß das Linienelement (F10) für den Spezialfall $\Pi = 0$ in die bekannte EINSTEIN-DE-SITTER-Metrik übergehen muß. Beide Darstellungen (1) und (F10) des ursprünglich singularitätsfreien Linienelements wären für die Beschreibung eines kosmischen Geschehens, das nur aus dem Bewegungsablauf unveränderlicher, ausdehnungsloser Massenpunkte bestünde – welche Körper begegnen einander, und in welcher zeitlichen Reihenfolge für jeden von ihnen? – *mathematisch* natürlich äquivalent. Und doch suggerieren sie bezüglich der Frage eines Anfangs des Universums absolut gegensätzliche Auffassungen.

Denn gemäß (F8), (F10) sieht es bekanntlich so aus, als wären zur Quasi-Eigenzeit $t'_{\text{Null}} = -2/[3H(1+\Pi)]$ alle mit spektralen Maßstäben gemessenen Längen Null, alle entsprechenden Dichten unendlich, das ganze Universum also singulär gewesen. Doch ist diese Betrachtungsweise unseres Erachtens nicht legitim. Daß aber auch die echte ‚Eigenzeit' ungeeignet ist zur Beschreibung kosmischer Prozesse, ergibt sich bereits aus folgender Überlegung: Angenommen ein gemeinsamer Zeitnullpunkt sei bei zunächst geeignet positionierten, dann sich selbst überlassenen Atomuhren in allen Spiralnebeln wenigstens einmal eingestellt. Dann werden sich später immer wieder solche Uhren begegnen, die in ein und demselben Raumpunkt unterschiedliche, von ihrer Vorgeschichte abhängige Eigenzeiten anzeigen. Dem Ereignis der jeweiligen Begegnung kann sich also grundsätzlich keine eindeutige Anzeige $s$ (bzw. $\sigma$) bewegter Atomuhren zuordnen lassen. Bezieht man sich aber auf die Eigenzeit $\tau$ *ruhender* Uhren und verlangt, daß die ausgewählten Atomuhren in ihrer ursprünglichen Umgebung verbleiben, sodaß sie sich später niemals begegnen können, dann können sie nicht früher existiert haben als zum Entstehungszeitpunkt der jeweiligen Galaxis (s. a. Abschn. 8/c).

Wie bereits im allgemeineren Fall (26) entfällt demzufolge die ansonsten erforderliche Unterscheidung zwischen $l^*(t_0^*)$ und $l_0^*$ (s. Fußn.[38]). Auch ergeben sich wieder für $t^* \to \pm \infty$ gleichermaßen unendlich große maximale Lichtwege für Zukunft und Vergangenheit, und die einfache Beziehung

$$l^* = c\,\Delta t^* \tag{37}$$

erlaubt es – wie bei der Skalarform (28) – prinzipiell an einer Festlegung des Meters auf Basis der Lichtlaufzeit festzuhalten.

Für $\Pi \neq -1/3$ aber wäre entweder $c^* \neq c$, was bei unendlichem 3-dimensionalen Raum der Zukunft die Einschränkung der Vergangenheit auf ein endliches Volumen bedeuten würde. Oder aber im Energie-Impuls-Tensor träten Singularitäten auf, welche die Vergangenheit auf eine endliche Zeit begrenzen und damit einen Beginn – wenn auch nicht der Zeit selbst – so doch des Zeitraums bedeuten müßten, innerhalb dessen sich das kosmische Geschehen überhaupt physikalisch beschreiben läßt.

Nur in dem besonderen Fall $\Pi = -1/3$ haben wir ein Linienelement, das zugleich vom WEYL-Typus (s. Fußn.[40]) und singularitätsfrei ist:

$$d\bar{s}^{*2}_{(s)} = \zeta^{*2}\,ds^{*2}_{(SRT)} = e^{2Ht^*}\left\{c^2\,dt^{*2} - dl^{*2}\right\}. \tag{38}$$

Dieses Linienelement werden wir als *stationär*[43]) bezeichnen. Bereits in Abschnitt 3 haben wir darauf hingewiesen, daß es in einem stationären Universum selbstverständlich möglich sein muß, den Nullpunkt der absoluten Zeit nach Belieben zu wählen. Diese Bedingung ist auf Grundlage des Linienelements (38) bei allen physikalischen Betrachtungen tatsächlich von selbst erfüllt. Denn Messen ist Vergleichen. Wenn wir also setzen

$$t^* := t_0^* + \Delta t^*, \tag{39a}$$

dann fällt bei der *Quotientenbildung* zur Ermittlung sowohl kosmischer als auch atomarer Maßzahlen der gemeinsame konstante Faktor $e^{Ht_0^*}$ immer heraus, sodaß in diesem Sinne mit mathematisch einwandfreier Berechtigung jeweils gesetzt werden darf

$$t_0^* = 0, \quad \Delta t^* = t^*. \tag{39b}$$

In den folgenden Unterabschnitten werden wir im einzelnen belegen: Mit $\Pi = -1/3$ haben wir I) ein singularitätsfreies kosmisches Linienelement mit singularitätsfreiem Energie-Impuls-Tensor, das II) bei konstanter kosmischer Lichtgeschwindigkeit $c^* = c$ nicht nur invariant ist gegen LORENTZ-Transformationen, sondern III) die durchgängige Gültigkeit der quellenfreien MAXWELL'schen Gleichungen sowie IV) bei geeigneter Definition von Ladung und Stromstärke gemäß $\partial_i F^{ik} = j^k$ – und Kombination des üblichen elektromagnetischen Energie-Impuls-Tensors mit dem phänomenologischen Energie-Impuls-Tensor der Materie – die Gültigkeit der gesamten klassischen Elektrodynamik gewährleistet, und das sich V) darüberhinaus als stationär erweist in dem Sinne, daß bei der Quotientenbildung zum Vergleich beliebiger physikalischer Größen mit der zugehörigen Einheit (Messung) ein gemeinsamer konstanter kosmischer Zeitfaktor $e^{Ht_0^*}$ immer herausfällt (was darauf hinausläuft, daß der Bezugspunkt $t_0^*$ der kosmischen Zeitskala für die zusammenhängende Beschreibung jeder beliebigen Epoche des Naturgeschehens immer wieder gleich Null gesetzt werden darf). Außerdem werden wir die Bewegung im lokalen Gravitationsfeld betrachten und die Anzeige von Atomuhren in beliebigen Situationen gegenüber der Ephemeridenzeit klären (s. dazu insbesondere auch Anhang B).

---

[43]) In diesem Sinne werden wir die Belegung $\Pi = -1/3$ auch weiterhin mit dem Index ‚(S)' als ‚stationär' markieren. Eine besser zutreffende, aber umständliche Bezeichnung wäre ‚streng stationär', und zwar im Unterschied zum allgemeineren Linienelement (28), das trotz auftretender Singularitäten – beispielsweise bei der druckfreien Belegung $\Pi = 0$ – innerhalb geeignet definierter Anwendungsgrenzen und bei Berücksichtigung gewisser Normalisierungsprozesse mit einem stationären Universum nicht zwangsläufig unvereinbar sein muß (was in diesem Fall allerdings darauf schließen ließe, daß die allgemeine Relativitätstheorie nur zur Beschreibung eines zeitlich abgegrenzten Teilbereichs der kosmischen Realität geeignet wäre).

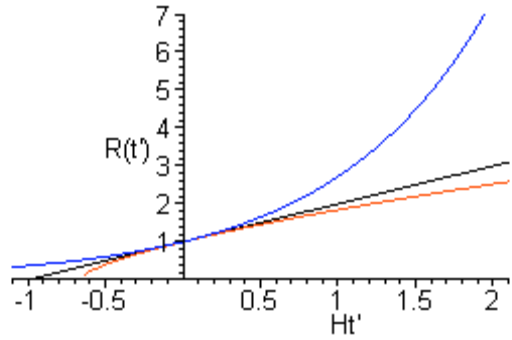

Abb. 1: Gegenüberstellung möglicher kosmischer Skalenfaktoren $R(t') = e^{Ht'}$, $1+Ht'$, $(1+3/2Ht')^{2/3}$
von oben nach unten zum Vergleich in blau, schwarz, rot
[die Variablen $t'$, $R(t')$ beziehen sich auf die RW-Form (F8)]

Astronomische Meßdaten der letzten Jahre scheinen gegen einen EINSTEIN-DE-SITTER-Skalenfaktor $(1+3/2Ht')^{2/3}$ zu sprechen, und damit eher die in den folgenden Abschnitten entwickelte Auffassung, $R(t') = 1+Ht'$, zu bestätigen (dies bedarf allerdings sorgfältiger Analysen der vorhandenen Beobachtungsdaten, und zwar auf Basis des stationären Modells in der Form (F12), die allerdings über den hier gesteckten Rahmen hinausgehen müssen, s. dazu aber die demnächst erscheinende Arbeit d. Verf. „*The Concordance Model – a Heuristic Approach from a Stationary Universe*").

### a. Stationärer Energie-Impuls-Tensor und Gravitationsdruck

Die EINSTEIN'schen Gleichungen für das stationäre Linienelement (38) finden wir aus den allgemeineren Formeln (2), (3) bzw. (31) mit $\Pi = -1/3$ zu

$$\bar{E}_{ik}^{*(s)} = \frac{3H^2}{c^2} \begin{pmatrix} 1 & 0 & 0 & 0 \\ 0 & -\frac{1}{3} & 0 & 0 \\ 0 & 0 & -\frac{1}{3} & 0 \\ 0 & 0 & 0 & -\frac{1}{3} \end{pmatrix} = \kappa \bar{T}_{ik}^*. \qquad (40)$$

Zunächst einmal kann dies nach phänomenologischer Interpretation von $\bar{T}_{ik}^*$ bei Verwendung des kinetischen Energie-Impuls-Tensors $\bar{K}_{ik}^* = \bar{\mu}_0^* c^2 \bar{u}_i^* \bar{u}_k^*$ gemäß

$$\bar{T}_{ik}^* = \bar{K}_{ik}^* - \bar{g}_{ik}^* \bar{p}^* = \bar{\mu}_0^* c^2 \bar{u}_i^* \bar{u}_k^* - e^{2Ht^*} \eta_{ik} \bar{p}^*, \qquad (41)$$

sowie aufgrund der mittleren Sterngeschwindigkeit $\bar{u}_\alpha^* = 0$ nur die Existenz eines *negativen kosmischen Gravitationsdrucks* $\bar{p}^*$ bedeuten, der bei Voraussetzung eines stationären Universums von den EINSTEIN'schen Gleichungen gewissermaßen gefordert wird. Bei Unterscheidung vom üblichen positiven Druck $p^*$ einer Flüssigkeit oder eines Gases ist ein negativer kosmischer Gravitationsdruck nicht notwendigerweise inakzeptabel – ganz im Gegenteil:

In einem großen stauberfüllten Raum werde ein Teilbereich dicht abgetrennt durch einen anfänglich offenen Kasten, anschließend werde außerhalb des Kastens aller Staub entfernt. Sollen die Verhältnisse im Inneren des Kastens die Gleichen bleiben wie zuvor, so müssen nun die Kastenwände eine einwärts gerichtete Kraft ausüben, die den Staub trotz positiven Drucks daran hindert, sich im ganzen Raum gleichmäßig zu verteilen. Denken wir uns aber im sternerfüllten Kosmos einen Teilbereich zunächst abgetrennt, anschließend außerhalb dieses Teilbereichs alle Materie entfernt, und sollen die Verhältnisse im Inneren die Gleichen bleiben wie zuvor, so müssen nun die Wände dieses Teilbereichs eine Kraft nach *außen* ausüben, welche die eingeschlossene, nahezu frei schwebende Sternmaterie daran hindert, sich trotz wechselseitiger Anziehung zu einem einzigen Körper zusammenzuklumpen. Offenbar herrscht also im Inneren ein *negativer* Druck. Dieser repräsentiert die bei statistisch gleichmäßiger Verteilung der Materie über den ganzen kosmischen Raum einander kompensierenden Gravitationskräfte. – Bei mittlerer Sterngeschwindigkeit $\bar{u}_\alpha^* = 0$ (s. Abschn. 5) ergibt sich aus $\bar{u}_i^* \bar{u}^{*i} = 1$ sofort

$$\overline{u}_0^* = \frac{1}{\overline{u}^{*0}} = e^{Ht^*}, \qquad (42)$$

und damit gemäß (41)

$$\overline{T}_{00}^* = \left(\overline{\mu}_0^* c^2 - \overline{p}^*\right) e^{2Ht^*}. \qquad (43)$$

Der negative kosmische Gravitationsdruck $\overline{p}^*$ beträgt, abgesehen vom Vorzeichen, zum Zeitpunkt $t^* = 0$ gerade ein Drittel der Dichte $\overline{T}_{00}^*$. Demzufolge haben wir zum gleichen Zeitpunkt also $\overline{\mu}_0^*(t^*= 0) = 2/3\rho_c$, wobei $\rho_c = \overline{T}_{00}^*/c^2 = 3H^2/(\kappa c^4)$ wie üblich für die kritische Dichte stehen soll. Unseres Erachtens ist es die Dichte $\overline{\mu}_0^*$ – und nicht die Komponente $\overline{T}_{00}^*$ – die der astronomischen Beobachtung mit den üblichen Verfahren zugänglich ist:

$$\begin{aligned}\overline{\mu}_0^* &= \frac{2}{3}\frac{\overline{T}_{00}^*}{c^2}e^{-2Ht^*} = \frac{2H^2}{\kappa c^4}e^{-2Ht^*}, \\ \overline{p}^* &= -\frac{\overline{T}_{00}^*}{3}e^{-2Ht^*} = -\frac{H^2}{\kappa c^2}e^{-2Ht^*}.\end{aligned} \qquad (44)$$

Es ist bemerkenswert, daß eine solche Zuordnung mit dem heutigen Wert der direkt und indirekt beobachteten kosmischen Materiedichte nicht in gleichem Maße unvereinbar ist, wie dies z.B. bei der druckfreien Lösung (3a) der Fall wäre (s. dazu Abschnitt 9/e).

Hier nun bietet es sich geradezu an, auch kurz auf die Frage der kosmologischen Konstanten einzugehen. Indem wir nämlich als *erweiterten* EINSTEIN-Tensor $\overline{\Lambda}_{ik}^*$ den um das kosmologische Glied $\Lambda \overline{g}_{ik}^*$ verminderten ursprünglichen Tensor $\overline{E}_{ik}^*$ bezeichnen, hätten wir als die dem stationären Linienelement (38) entsprechenden erweiterten EINSTEIN'schen Gleichungen,

$$\overline{\Lambda}_{ik}^{*(s)} \equiv \overline{E}_{ik}^{*(s)} - \Lambda \overline{g}_{ik}^{*(s)} = \begin{pmatrix} \frac{2H^2}{c^2} & 0 & 0 & 0 \\ 0 & 0 & 0 & 0 \\ 0 & 0 & 0 & 0 \\ 0 & 0 & 0 & 0 \end{pmatrix} + \left(\frac{H^2}{c^2} - \Lambda e^{2Ht^*}\right)\eta_{ik} = \kappa \overline{T}_{ik}^{*(\Lambda)}, \qquad (45)$$

wobei dieses Ergebnis in einer den vorausgegangenen Überlegungen angepaßten Form geschrieben ist. Setzen wir nun

$$\Lambda = \frac{H^2}{c^2} = -\kappa \overline{p}^*(t^*=0), \qquad (46)$$

so ergäbe sich für kleine Zeiten $t^*$ in der Nähe des frei wählbaren Nullpunkts $t_0^* = 0$ aus (45) näherungsweise

$$\overline{\Lambda}_{ik}^{*(s)} = \kappa \overline{T}_{ik}^{*(\Lambda)} \approx \frac{2H^2}{c^2}\begin{pmatrix} 1-Ht^* & 0 & 0 & 0 \\ 0 & Ht^* & 0 & 0 \\ 0 & 0 & Ht^* & 0 \\ 0 & 0 & 0 & Ht^* \end{pmatrix}. \qquad |Ht^*| \ll 1 \quad (45a)$$

Dies würde im Bezugszeitpunkt $t^* = 0$ also immer wieder einen druckfreien erweiterten Energie-Impuls-Tensor darstellen, wobei sich jeweils bis zur nächsten ‚Normalisierung' $t_0^* = 0$ ein positiver Druck $\overline{p}^* = 2\kappa H^3 t^*/c^2$ aufbauen sollte[44], bzw. kurz zuvor ein betragsmäßig gleicher, aber negativer Druck geherrscht hätte. Die mittlere kosmische Dichte der Materie würde auch bei dieser quasi-stationären Lösung (45) bzw. (45a) der erweiterten EINSTEIN'schen Gleichungen nur zwei Drittel der kritischen Dichte $\rho_c$ betragen, wie dies bereits oben festgestellt worden ist. Als kosmologische ‚Konstante' wäre gemäß der Belegung (46) bei jeder Normalisierung $t_0^* = 0$ immer wieder $\Lambda = H^2/c^2$ anzusetzen[45]).

---

[44]) In der Existenz des kosmischen Zeitskalars $\zeta$ bzw. $e^{Ht^*}$ könnte möglicherweise die Erklärung für kleinste Veränderungen liegen, die mit der Zeit zum *spontanen Zerfall* sich selbst überlassener, zeitweilig abgeschlossener Systeme – wie z.B. Elementarteilchen, radioaktiver Kerne, angeregter Atome – führen, die ganz offensichtlich nach ihrer jeweiligen Entstehung einem *stationären* Alterungsprozeß mit eigener Zeitkonstanten unterliegen (wobei sich raum-zeitlich lokale, d.h. vorübergehende Verletzungen der kosmischen Stationarität möglicherweise auch in einer prinzipiellen Unschärfe der physikalischen Beschreibung auswirken könnten). Das Phänomen des spontanen Zerfalls verstehen wir grundsätzlich als Hinweis darauf, daß natürliche Uhren nicht synchron zur kosmischen Zeit gehen (die Frage nach der Funktionsweise ‚biologischer Uhren' würde den Rahmen einer rein physikalischen Betrachtung weit überschreiten).

*Hält man aber an den ursprünglichen EINSTEIN'schen Gleichungen fest, so ergibt sich aus (45) mit $\Lambda = 0$ eine besondere Darstellung der Gleichung (40), in welcher der negative Gravitationsdruck selbst gewissermaßen als ‚temporäre' kosmologische Konstante auftritt.*

Der Vollständigkeit halber sei schließlich noch darauf hingewiesen, daß sich der auf die Quasi-Eigenzeit $t'$ bezogene, der RW-Form (F10) entsprechende EINSTEIN-Tensor mit $\Pi = -1/3$ unmittelbar aus (F11) ergibt[46]).

### b. Elektrodynamik im kosmischen System

Das stationäre Linienelement (38) ist ersichtlich invariant gegen LORENTZ-Transformationen, wobei allerdings nach einer solchen Transformation die absolute Geschwindigkeit $v^*$ des Inertialsystems gegenüber dem kosmischen Hintergrund im Zeitskalar auftritt. Dies läßt sich so interpretieren, daß im bewegten System die gewöhnliche HUBBLE-Konstante $H$ näherungsweise durch $H(v^*)$ zu ersetzen ist gemäß[47])

$$H(v^*) \approx \frac{H}{\sqrt{1 - \frac{v^{*2}}{c^2}}}. \tag{47}$$

Angesichts der konstanten kosmischen Lichtgeschwindigkeit $c$ und der Invarianz des stationären (38) bzw. skalarabhängigen (28) Linienelements gegenüber LORENTZ-Transformationen stellt sich nun aber die Frage: Hat die Existenz eines Zeitskalars $\zeta$ überhaupt irgendeinen Einfluß auf die Gültigkeit der MAXWELL'schen Gleichungen im kosmischen Bezugssystem? Zur Klärung dieser Frage[48]) verwenden wir die kovariante Formulierung [25] dieser Gleichungen in einem beliebigen, durch den allgemeinen Fundamentaltensor $g_{ik}$ dargestellten Gravitationsfeld und denken uns das elektromagnetische 4-Potential $A_i$ in solcher Weise vorgegeben, daß die daraus resultierenden Ladungen und Ströme gerade den realen Verhältnissen entsprechen. Im einzelnen definieren wir zunächst den kovarianten Feldtensor $F_{ik}$ sowie die kontravariante Tensordichte $\mathbf{H}^{ik}$ durch

---

[45]) Ein jederzeit druckfreier EINSTEIN-Tensor würde offensichtlich *exakt* einer Kombination des gewöhnlichen stationären Energie-Impuls-Tensors (40) und einer – allerdings anders als bei EINSTEIN hier mit dem Fundamentaltensor $\eta_{ik}$ der *speziellen* Relativitätstheorie anzukoppelnden – kosmologischen Konstanten entsprechen, wobei $\Lambda$ den gleichen Wert hätte wie oben angegeben.

[46]) Die zu dem stationären Linienelement (38) korrespondierende RW-Form ist gemäß Fußn.[42])

$$d\bar{s}'^2_{(RW)}(\Pi = -\tfrac{1}{3}) = c^2 dt'^2 - (1 + Ht')^2 dl^{*2}, \tag{F12}$$

der entsprechende Einstein-Tensor hat die Gestalt

$$\bar{E}'^{(RW)}_{ik}\left(\Pi = -\tfrac{1}{3}\right) = \frac{3H^2}{c^2} \begin{pmatrix} \frac{1}{(1+Ht')^2} & 0 & 0 & 0 \\ 0 & -\tfrac{1}{3} & 0 & 0 \\ 0 & 0 & -\tfrac{1}{3} & 0 \\ 0 & 0 & 0 & -\tfrac{1}{3} \end{pmatrix}. \tag{F13}$$

Auch für diesen in absoluten Koordinaten ursprünglich stationären Fall ergibt sich in Bezug auf die Quasi-Eigenzeit $t'$ also eine Singularität, und zwar hier zu $-t_H' = 1/H$. Wir verstehen dies als erneute Bestätigung dafür, daß sich die zeitliche Veränderung des kosmischen Linienelements – wie auch die des entsprechenden Energie-Impuls-Tensors – nicht lückenlos in der (echten) Eigenzeit $\tau$, sondern nur in der kosmischen Zeit $t^*$ beschreiben läßt (s. Abschn. 8/c).

[47]) Dies gilt strenggenommen nur in unmittelbarer Umgebung des Koordinatenursprungs, doch hätte es wenig Sinn, von irgendwelchen gegenüber dem kosmischen Hintergrund mit der Geschwindigkeit $v$ bewegten Inertialsystemen zu sprechen, deren lineare Ausmaße $\Delta x$ nicht der Bedingung $H\Delta x\, v/c^2 \ll 1$ genügen. Zwar kann man natürlich über beliebig große Raumbereiche alle Objekte gedanklich zusammenfassen, die sich unabhängig voneinander mit der gleichen Geschwindigkeit $v$ in gleicher Richtung bewegen, und diese Gesamtheit als Manifestation eines entsprechenden, unendlich ausgedehnten Inertialsystems betrachten, doch je größer die von einem solchen Inertialsystem erfaßten Raumbereiche werden, umso weniger hat es einen Sinn, auf der Gleichberechtigung dieses Systems gegenüber dem absoluten Ruhsystem des kosmischen Hintergrunds zu bestehen.

[48]) Es ist klar, daß es wenig Sinn macht, entsprechende Fragen auf Basis einer anderen Form des kosmischen Linienelements überhaupt zu stellen.

$$F_{ik} := \partial_i A_k - \partial_k A_i \,,$$
$$\mathbf{H}^{ik} := \sqrt{\mathbf{g}}\, g^{il} g^{km} F_{lm} := \sqrt{\mathbf{g}}\, H^{ik} \,, \tag{48}$$

wobei $H^{ik}$ für den mit Hilfe der allgemeinen $g^{ik}$ gebildeten kontravarianten Tensor des ursprünglich kovarianten Feldtensors $F_{ik}$ steht (im Unterschied zu $H^{ik}$ soll die Bezeichnung $F^{ik}$ dagegen dem entsprechenden Tensor der speziellen Relativitätstheorie vorbehalten bleiben). Bereits aufgrund der Definition von $F_{ik}$ ist bekanntlich das eine Paar der MAXWELL'schen Gleichungen identisch erfüllt, nämlich

$$\partial_i F_{kl} + \partial_k F_{li} + \partial_l F_{ik} \equiv 0 \,. \tag{49}$$

Dem verbleibenden Paar tragen wir Rechnung, indem wir – was angesichts vorgegebener Potentiale hier möglich ist – die Dichten von Ladung und Stromstärke *definieren* als

$$\partial_i \mathbf{H}^{ik} := \mathbf{j}^k \,. \tag{50}$$

Für den kosmologischen Fall ersetzen wir – dem stationären (38) bzw. skalarabhängigen (28) Linienelement entsprechend – die allgemeinen $g_{ik}$ nun durch die räumlich gemittelten

$$\bar{g}_{ik}^{*(s)} = \zeta^{*2} g_{ik}^{*(SRT)} = e^{2Ht^*} \eta_{ik} \tag{51}$$

und finden aufgrund der von WEYL [25] so genannten Eichinvarianz – der Skalar $\zeta$ fällt bei der Bildung von $\mathbf{H}^{ik}$ gemäß (48) einfach heraus – hier die gleichen elektromagnetischen Feldstärken wie im Inertialsystem der speziellen Relativitätstheorie:

$$\mathbf{H}^{ik} = F^{ik} \,, \tag{52}$$

woraus sich die Dichten von Ladung und Stromstärke nun wiederum bestimmen zu

$$\partial_i F^{ik} := j^k \,. \tag{53}$$

Bei vorgegebenen elektromagnetischen Potentialen $A_i$ gelten also auch bezüglich des kosmischen Ruhsystems – unabhängig vom Zeitskalar $\zeta$ – die MAXWELL'schen Gleichungen des Vakuums, und zwar die homogenen, quellenfreien in aller Strenge, die inhomogenen zumindest formal bei geeigneter Definition von Stromstärke und Ladung[49]).

Die klassische Wechselwirkung mit der Materie schließlich ergibt sich in großer Entfernung von lokalen Gravitationsquellen phänomenologisch aus dem – aufgrund unveränderter Feldstärken ebenfalls gleich bleibenden – MAXWELL'schen Energie-Impuls-Tensor

$$L_i^k := \frac{1}{4\pi}\left(F_{il} F^{lk} + \tfrac{1}{4}\delta_i^k F_{lm} F^{lm}\right) \tag{54}$$

und dem kinetischen Energie-Impuls-Tensor

$$K_i^{k\,(s)} := \mu_0 c^2 u_i u^k = \mu_0 c^2 g_{il}^{(s)} u^l u^k \,. \tag*{*(55)}$$

Für den aus diesen beiden zusammengefaßten Tensor $T_i^k$ der Dichten von Gesamtenergie und Gesamtimpuls

$$T_i^k := L_i^k + K_i^k \tag{56}$$

fordern wir wie üblich als differentielle Erhaltungssätze

$$T_{i;k}^k \overset{!}{=} 0 \,. \tag*{*(57)}$$

Mit der zusätzlich geforderten lokalen Erhaltung der Ruhmasse verknüpft,

---

[49]) Auf den keineswegs trivialen Zusammenhang von $\mathbf{j}^k$ aus (50) und $j^k$ aus (53) werden wir im Rahmen einer anderen Arbeit näher eingehen.

$$\partial_k\left(\mu_0 u^k\right) \stackrel{!}{=} 0, \qquad *(58)$$

welche bei der Bewegung geladener Teilchen im elektromagnetischen Feld – von Stoßprozessen abgesehen – tatsächlich gegeben ist, haben wir zuletzt den allgemein-relativistischen Zusammenhang zwischen LORENTZ-Kraft und geodätischer Beschleunigung:

$$F_{ik} j^k = \mu_0 \left\{ \frac{du_i}{ds} - \tfrac{1}{2} u^k u^l \partial_i g_{kl}^{(s)} \right\}. \qquad *(59)$$

Wenn es also richtig ist, im Ansatz *(55) des kinetischen Energie-Impuls-Tensors $K_i^{k(s)}$ den stationären Fundamentaltensor $g_{ik}^{(s)}$ zu verwenden, dann gilt im kosmischen System zusätzlich zu den MAXWELL'schen Gleichungen – bei hinreichender Entfernung von lokalen Gravitationsquellen – die gesamte klassische Elektrodynamik einschließlich des im Sinne der allgemeinen Relativitätstheorie modifizierten LORENTZ'schen Bewegungsgesetzes für geladene Teilchen[50].

Der oben verwendete Ansatz *(55) trägt der fundamentalen Forderung Rechnung, daß zur Ermittlung der ‚geodätischen' Bewegungsgleichungen für frei fallende Massenpunkte bei der Variation

$$\delta \int ds = 0 \qquad (60)$$

gerade dasselbe Linienelement $d\bar{s}^*$ einzusetzen sei, dessen EINSTEIN-Tensor dem mit der Gravitationskonstanten $\kappa$ multiplizierten Energie-Impuls-Tensor der Gravitationsquellen gleich sein soll. Bezogen auf kosmische Koordinaten ergibt dies im elektrodynamisch kräftefreien Fall gemäß (6) für $\Pi = -1/3$ eine kosmische Abbremsung extragalaktischer freier Teilchen. – Andererseits aber verlangt GALILEIs klassischer Trägheitssatz, daß das Linienelement (und damit auch der kinetische Energie-Impuls-Tensor $K_i^k$) in großen Entfernungen von starken Gravitationszentren die Form der speziellen Relativitätstheorie haben sollte. Dies bedeutet demzufolge: *In dem durch die Isotropie der Hintergrundstrahlung eindeutig fixierten kosmischen Ruhsystem – weit weg von allen lokalen Gravitationsquellen – kann der Trägheitssatz nur noch lokale Bedeutung haben.*

**c. Die Tauglichkeitsgrenzen der Begriffe von Eigenzeit und Eigenlänge**

Obwohl unabhängig vom Bewegungsgesetz jedenfalls die quellenfreien MAXWELL'schen Gleichungen des Vakuums auf Basis des stationären (bzw. skalarabhängigen) Linienelements ihre Gültigkeit behalten, können ganz beliebige Uhren – insbesondere auch Atomuhren – in Bezug auf die absolute Systemzeit $t^*$ [bzw. $t^{(c)}$] trotzdem nicht die gleiche Ganggeschwindigkeit aufweisen wie im Falle des Linienelements der speziellen Relativitätstheorie. Um dies zu zeigen, transformieren wir die Koordinaten $t^*$, $r^*$ des kosmischen Systems mit Hilfe der Gleichungen

$$t^* = \frac{\ln(1+Ht')}{H} \quad \Leftrightarrow \quad dt' = dt^* e^{Ht^*},$$

$$r^* = \frac{r'}{(1+Ht')} \quad \Leftrightarrow \quad dr' = dr^* e^{Ht^*} + dt^* H r^* e^{Ht^*}, \qquad (61)$$

auf die *integrierten* Koordinaten $t'$, $r'$, wobei wir die neue Zeitkoordinate $t'$ (welche derjenigen der RW-Form für die Belegung $\Pi = -1/3$ entspricht, s. Fußn.[46]) wie bereits zuvor als Quasi-Eigenzeit bezeichnen wollen. Der Grund für diese Bezeichnung tritt hier deutlich zutage. Weil das durch die Transformation entstehende Linienelement

$$d\bar{s}'^2_{(s)} = \left(1 - \frac{H^2 r'^2}{c^2(1+Ht')^2}\right) c^2 dt'^2 + \frac{2Hr'}{(1+Ht')} dt' dr' - dr'^2 - d\Sigma'^2 \qquad (62)$$

---

[50]) Es sei in diesem Zusammenhang allerdings daran erinnert, daß es notwendig ist, den Übergang des Linienelements lokaler Gravitationsquellen in dasjenige der speziellen Relativitätstheorie für große Entfernung vorauszusetzen, um Integrale für den 4-Vektor von Energie und Impuls eines abgeschlossenen Systems *einschließlich dessen Gravitationsfelds* zu erhalten, die unabhängig sind von der Koordinatenwahl (s. [22]).

mit dem Linienelement der speziellen Relativitätstheorie in der Umgebung des Koordinatenursprungs *näherungsweise* übereinstimmt – wobei $d\Sigma'^2 = r'^2(\sin^2\vartheta \, d\varphi^2 + d\vartheta^2)$ – finden wir in den mit $d\tau \approx dt'(r'=0, dr'=0)$ und $d\lambda \approx dl'(r'=0, dr'=0)$ zu (61) korrespondierenden Formeln

$$d\tau \approx dt^* e^{Ht^*},$$

$$d\lambda \approx dl^* e^{Ht^*},$$ 
(63)

tatsächlich die erwartete lokale Eigenzeit und lokale Eigenlänge, hier allerdings mitsamt ihrer Grenzen[51]. Denn im Unterschied zum stationären Linienelement (38) zeigt das Linienelement (62) einen offensichtlich lokalen Charakter, und zwar am deutlichsten dadurch, daß es trotz zuvor willkürlich gewählten Koordinatenursprungs außerhalb eines Bereichs mit Radius $r' = c(1+Ht')/H$ überhaupt nicht anwendbar ist. Betrachten wir nämlich das Linienelement (62) nicht mehr ausschließlich in unmittelbarer Umgebung des Koordinatenursprungs $r' = 0$, sondern verwenden nur noch $dr' = 0$, so finden wir unterschiedliche Ganggeschwindigkeiten der an verschiedenen Orten ruhenden Atomuhren in Bezug auf die *Systemzeit t'*, sodaß diese hier nicht mehr als Eigenzeit interpretiert werden kann. Insbesondere mit zunehmenden Werten $Ht', Hr'/c \to 1$ wird das Linienelement (62) vom Linienelement der speziellen Relativitätstheorie eklatant abweichen[52]. Damit haben wir folgenden Sachverhalt:

*Die – nach Umbenennung von $\tau$ in t' (und Setzung eines Gleichheitszeichens anstelle von '$\approx$') – durch Integration der ersten Beziehung (63) entstandene Systemvariable t' der Transformation (61) kann nur innerhalb desjenigen Bereichs als Eigenzeitanzeige $\tau$ ruhender Atomuhren aufgefaßt werden, in welchem das Linienelement $d\bar{s}'^2_{(S)}$ mit dem entsprechenden Linienelement $ds'^2_{(SRT)} = \eta_{ik} dx'^i dx'^k$ der speziellen Relativitätstheorie hinreichend gut übereinstimmt. Ein anderes Integral von $d\tau$ als die Systemvariable t' aber gibt es nicht.*

Die Eigenzeit $d\tau$ ist demzufolge nicht integrabel, alle auf einer Integrabilität der Eigenzeit beruhenden Schlüsse hinsichtlich eines Beginns des Universums (oder gar von ‚Raum und Zeit‘ selbst) sind nicht haltbar.

In (62) hat sich damit gezeigt, daß auch für kürzeste Zeitspannen $\Delta t' \to 0$ ein aus dem Linienelement (38) durch Koordinatentransformation hervorgehendes lokales Inertialsystem nicht beliebig große Bereiche des unendlichen 3-dimensionalen Raums umfassen kann. Anders als bei den absoluten kosmischen Größen, läßt sich also bei Verwendung spektraler Einheiten wegen der fehlenden Integrabilität i.a. nicht von Aussagen über lokale ‚Zeitspannen‘ auf globale ‚Zeiträume‘, von Aussagen über lokale ‚Längen‘ auf globale ‚Entfernungen‘ schließen. Dies ist der Grund, warum wir bei der Zeitvariablen der RW-Form nicht von der Eigenzeit $\tau$, sondern nur von einer Quasi-Eigenzeit $t'$ sprechen können.

Die zu (F1) korrespondierende Energie-Impuls-Tensordichte[53] $\overline{V}'^k_i$, die in diesem Fall – bezogen auf kartesische Koordinaten – wegen $\overline{g}' = 1$ und $\overline{t}'^k_i = 0$ mit $\overline{E}'^k_i$ übereinstimmt, finden wir im System der integrierten Koordinaten

$$\overline{V}'^k_i = \frac{H^2}{c^2(1+Ht')^2} \begin{pmatrix} 3 & \frac{2Hx'}{c(1+Ht')} & \frac{2Hy'}{c(1+Ht')} & \frac{2Hz'}{c(1+Ht')} \\ 0 & 1 & 0 & 0 \\ 0 & 0 & 1 & 0 \\ 0 & 0 & 0 & 1 \end{pmatrix}. \quad (\Pi = -1/3) \quad (64)$$

Im Unterschied zu (F1) ergibt sich also auf Basis der Darstellung (62) im System der integrierten Koordinaten eine nicht-verschwindende Energiedichte $\overline{V}'^0_0$ aus Materie und Gravitationsfeld. Im Sinne eines ‚expandierenden Universums‘ könnte man nun daran denken, darin ein Argument dafür zu sehen, daß eben K' das ‚richtige‘ Koordinatensystem darstelle, nicht aber K*. Denn hier hätten wir eine Energieströmung weg vom Beobachter, und zwar gerade mit der ‚passenden' Geschwindigkeit $Hx'^\alpha$ (wenn man nämlich berücksichtigt, daß die eigentliche Energiedichte wieder – wie oben erläutert – nur 2/3 der entsprechenden Komponente des Energie-Impuls-Tensors ausmachen sollte). Diese Energieströmung würde auch gerade dem zeitlichen Verlust im Inneren einer Kugelfläche mit dem Koordinatenursprung als Mittelpunkt entsprechen. Doch ist eine solche Sicht schon deshalb nicht haltbar, weil im Linienelement (62) ein spezielles, auf den jeweiligen Beobachter bezogenes Koordinatensystem im Interesse einer zweckmäßigen lokalen Anpassung willkürlich ausgezeichnet ist. So hinge die Aussage darüber, in welche Richtung eine Energieströmung – wenn es sie denn gäbe – durch eine Testfläche stattfinden sollte, beispielsweise davon ab, auf welcher Seite dieser Fläche sich der Koordinatenursprung befände. Im Hinblick auf das kosmologische Prinzip aber müßte dieser frei wählbar sein.

---

[51]) Unseres Erachtens sind es gerade diese Grenzen, die für das angemessene Verständnis eines relativistischen kosmologischen Welt-‚Bilds‘ von entscheidender Bedeutung sind.

[52]) Strenggenommen bliebe mit $dr' = 0$ eine Atomuhr einfach stehen, wenn man nämlich daran festhält, daß diese mit der EINSTEIN'schen Lichtuhr immer und überall synchron gehen sollte, die Lichtuhr nun aber keinen Raum mehr hätte für die notwendigerweise hin und her laufenden Signale. Mag andernfalls aber $dr'$ noch so wenig abweichen von Null, so machen doch die über kosmische Zeiträume $t' \to -1/H$ in der Vergangenheit aufgetretenen Abweichungen vom Linienelement der speziellen Relativitätstheorie die Interpretation von $t'$ als Eigenzeit unmöglich.

[53]) Wir sprechen hier wieder von einer echten Tensordichte und nicht von einer Pseudotensordichte (s. Fußn.[27]).

**d. Das lokal adaptierte Bezugssystem**

Auf Basis des Linienelements (62) könnten die MAXWELL'schen Gleichungen im System K′ der integrierten Koordinaten allerdings nicht einmal in erster Ordnung $Hr'/c$ Gültigkeit beanspruchen, was am Koordinatenwert der Lichtgeschwindigkeit $c'_\pm \approx Hr' \pm c$ deutlich zu erkennen ist. Deshalb wollen wir hier schließlich noch die Transformationsformeln auf *adaptierte* Koordinaten $x'''^i$ angeben

$$t^* = \frac{\ln(1+Ht'')}{H} - \frac{1}{2}\frac{Hr''^2}{c^2} \quad \Leftrightarrow \quad dt^* = \frac{dt''}{(1+Ht'')} - \frac{Hr''dr''}{c^2},$$

$$r^* = \frac{r''}{(1+Ht'')} \quad \Leftrightarrow \quad dr^* = \frac{dr''}{(1+Ht'')} - \frac{Hr''dt''}{(1+Ht'')^2}, \tag{65}$$

die das stationäre Linienelement (38) in eine solche Form $d\bar{s}''^2$ überführen, daß diese in der Umgebung des Koordinatenursprungs mit dem Linienelement der speziellen Relativitätstheorie bis auf Korrekturen zweiter und höherer Ordnung $O^2(Hr''/c, Ht'')$ übereinstimmt (66)

$$d\bar{s}''^2 = e^{-\frac{H^2 r''^2}{c^2}} \left\{ \left[1 - \frac{H^2 r''^2}{c^2(1+Ht'')^2}\right] c^2 dt''^2 - \frac{2H^2(2+Ht'')t''r''}{(1+Ht'')} dt''dr'' - \left[1 - \frac{H^2 r''^2(1+Ht'')^2}{c^2}\right] dr''^2 - d\Sigma''^2 \right\}.$$

Es ist offensichtlich, daß die Transformation (65) mit der des letzten Abschnitts (61) in der Umgebung des Koordinatenursprungs näherungsweise übereinstimmt (das Element $d\Sigma''$ ist wieder als das einer gewöhnlichen Kugelfläche zu verstehen).

Wie aus dieser bestmöglichen lokalen Anpassung deutlich zu ersehen, kann es überhaupt keine Transformation eines kosmischen Linienelements geben, die es erlauben würde, auf die Verwendung einer Systemzeit zusätzlich zur Eigenzeit zu verzichten. Umgekehrt aber darf die mathematische Möglichkeit, eine einmal gegebene kosmische Systemzeit durch Koordinatentransformation in eine andere Form überzuführen, über die Notwendigkeit der gesonderten Existenz einer absoluten Systemzeit nicht hinwegtäuschen. Allein darauf kommt es an. So läßt sich von den Variablen $t'$, $t''$ jeweils lediglich im Sinne einer Quasi-Eigenzeit sprechen, die eigentlich (integrable) *Systemzeiten* sind, und deshalb von Atomuhren nur vorübergehend und näherungsweise – sowie innerhalb eines jeweils begrenzten Raumbereichs – angezeigt werden können.

In (63) haben wir also die Auswirkungen des Gravitationspotentials auf die Anzeige von Atomuhren und spektralen Maßstäben für den stationären Fall $\Pi = -1/3$. Die bei gleichbleibenden absoluten Zeitspannen $dt^*$ von Atomuhren angezeigten Zeitspannen $d\tau$ und die bei gleichbleibenden absoluten Längen $dl^*$ spektral gemessenen Längen $d\lambda$ wachsen gegenüber dem kosmischen Bezugssystem mit der absoluten Zeit $t^*$ – wobei umgekehrt die atomaren Größen selbst mitsamt ihren spektralen Einheiten schrumpfen (s. Abschn. 5). Doch sind die mit Atomuhren gemessenen Zeitspannen für beliebige atomare Zyklen immer die gleichen. Etwas anderes dagegen ist es, kosmische Zeitspannen $\Delta t^*$ in Beziehung zu setzen zur Anzeige von Atomuhren. Dies geschieht notwendigerweise in der Astronomie beim Vergleich spektraler Meßwerte mit absoluten Bahndaten (s. Abschn. 8/f u. Anhang A).

Den mit $\Pi = -1/3$ ursprünglich singularitätsfreien, stationären EINSTEIN-Tensor (40) finden wir im lokal adaptierten Koordinatensystem nun aber weder singularitätsfrei noch stationär, sondern in einer Form, die eher der heutigen Auffassung der Kosmologie entsprechen könnte:

$$E_{ik}'' = \frac{3H^2}{c^2} \begin{pmatrix} \dfrac{1 - \dfrac{H^2 r''^2}{3c^2(1+Ht'')^2}}{(1+Ht'')^2} & -\dfrac{Hr''\left(1 - \dfrac{1}{3(1+Ht'')^2}\right)}{c(1+Ht'')} & 0 & 0 \\ -\dfrac{Hr''\left(1 - \dfrac{1}{3(1+Ht'')^2}\right)}{c(1+Ht'')} & -\dfrac{1 - \dfrac{3H^2 r''^2}{c^2}[1+Ht''(2+Ht'')]}{3(1+Ht'')^2} & 0 & 0 \\ 0 & 0 & -\dfrac{r''^2 \sin^2\vartheta}{3(1+Ht'')^2} & 0 \\ 0 & 0 & 0 & -\dfrac{r''^2}{3(1+Ht'')^2} \end{pmatrix} . (\Pi=-1/3) \quad (67)$$

Die dazu gehörende kosmische Energie-Impuls-Tensordichte $\overline{V}_i''^k$ resultiert bis auf Korrekturen der Ordnung $O^4(Ht'')$ näherungsweise in der gleichen Form (64) wie im Bezugssystem der integrierten Koordinaten. Auch im adaptierten Koordinatensystem K" ergibt sich also insgesamt eine entsprechende nicht-verschwindende kosmische Energiedichte $\overline{V}_0''^0$ aus Materie und Gravitationsfeld. Eine Entscheidung, welche der beiden Formen (40) oder (67) die richtige sein muß, fällt leicht, wenn man bedenkt, daß mit der lokal adaptierten Form ein spezielles Koordinatensystem ausgezeichnet ist. Dies ließe sich gegebenenfalls so verstehen, daß mit dem durch das Linienelement (38) beschriebenen, insgesamt stationären Universum sehr wohl unendlich viele statistisch verteilte, in Bezug auf spektrale Einheiten lokal expandierende kosmische Materieverteilungen verträglich wären.

### e.   Anzeige von Atomuhren und die Bewegung im lokalen Gravitationsfeld

EINSTEINs geometrische Begründung seiner allgemeinen Relativitätstheorie setzt voraus, daß es genau einen – gewissermaßen allumfassenden – Fundamentaltensor $g_{ik}$ gibt, aus dem sich vor allem drei elementare Zusammenhänge ableiten lassen: a) Der mit der Gravitationskonstanten $\kappa$ multiplizierte Energie-Impuls-Tensor der Materie als EINSTEIN-Tensor $E_{ik}$, dessen kovariante Divergenz aufgrund der BIANCHI-Identitäten verschwindet. b) Das Linienelement d$s$, dessen Variation die ‚geodätischen' Bewegungsgleichungen für frei fallende Testpartikel liefert. c) Die Anzeige d$\sigma$ einer Atomuhr in Abhängigkeit von der Systemzeit, wobei letztere wegen frei wählbarer Koordinaten keine unmittelbare metrische Bedeutung haben sollte. Die EINSTEIN'sche Relativitätstheorie verlangt nun, daß d$\sigma$ identisch sei mit d$s/c$.

Bei Verwendung der *lokalen* Koordinaten $x^i$, wenn es ausdrücklich um Abläufe im lokalen Gravitationsfeld geht, können wir zusammenfassend alternativ zu (60) auch schreiben

$$\delta \int m_0 c \, ds = 0, \quad (68)$$

für den Fall, daß es sich um die Bewegung von Teilchen mit von Null verschiedener Ruhemasse handelt (wir werden in Anhang B auf diese Form noch einmal zurückkommen).

Die Abweichungen des Linienelements d$s$ von dem der speziellen Relativitätstheorie d$s_{(SRT)}$ repräsentieren den Einfluß der lokalen Gravitationsquellen, sodaß ersteres mit zunehmender Entfernung in letzteres übergeht. Das Linienelement d$s$ ist aber durch die gewöhnlichen EINSTEIN'schen Gleichungen

$$E_{ik} \equiv R_{ik} - \tfrac{1}{2} R g_{ik} = \kappa T_{ik} \quad (69)$$

mit dem in diesem Fall *lokalen* phänomenologischen Energie-Impuls-Tensor

$$T_{ik} = K_{ik} - g_{ik} p = \mu_0 c^2 u_i u_k - g_{ik} p \quad (70)$$

verknüpft, der eine über kosmische Distanzen gemittelte Energiedichte naturgemäß *nicht* enthält. Dies bedeutet, daß wir die astronomischen Gegebenheiten nun gewissermaßen aus der Nähe betrachten.

Und doch liegt eine gewisse Ungereimtheit in dieser Auffassung, die an folgendem Beispiel deutlich werden mag: Angenommen, eine ideale Raumsonde wird mit hinreichender Beschleunigung von der Erde aus gestartet und nach dem endgültigen Abschalten aller Triebwerke sich selbst überlassen. Abgesehen von Reibungsverlusten durch Wechselwirkung mit interstellarer Materie, dem Austausch von Wärmestrahlung sowie dem Lichtdruck der Gestirne, die wir in diesem Gedankenexperiment allesamt vernachlässigen wollen, erfolge die Bewegung der Sonde für die späteren Zeiten gemäß den Beziehungen (68), (69), (70). Das aber bedeutet, daß die Bewegung mit dem Erreichen extragalaktischer Entfernungen in eine *gleichförmige* übergehen sollte. Dies jedoch steht im Widerspruch zum Resultat der kosmisch abgebremsten Bewegung gemäß (6) und der damit verbundenen Konsequenz, daß der GALILEI'sche Trägheitssatz bezogen auf kosmische Koordinaten nicht streng gültig sein kann. Wenn die Bewegung im lokalen Gravitationsfeld durch den Fundamentaltensor $g_{ik}$ – dessen Ableitungen gemäß (68) allein die Abweichung von der geradlinig gleichförmigen Trägheitsbewegung ergeben – bestimmt ist, diejenige im intergalaktischen Raum aber durch den räumlich gemittelten Tensor $\bar{g}_{ik}^{*(s)}$, so stellt sich die Frage, wo und wie die eine in die andere (bzw. der eine Energie-Impuls-Tensor in den anderen) übergehen kann. – Eine mögliche Lösung dieses Problems werden wir im Anhang besprechen.

Zu unserem Ansatz, daß es das Linienelement d$s$ sei, dessen Variation die Bewegungsgleichungen im lokalen Gravitationsfeld liefert, ist außerdem folgende Bemerkung angebracht: Die geodätische Beschleunigung $\delta \int ds = 0$ läßt sich bekanntlich aus dem Verschwinden der kovarianten Ableitung des lokalen kinetischen Energie-Impuls-Tensors $K^{ik}_{;k} = 0$ ableiten [indem man zugleich mit $\partial_k(\mu_0 u^k) = 0$ die (ebenfalls lokale) Erhaltung der Ruhemasse voraussetzt]. Doch in Bezug auf den kosmischen Energie-Impuls-Tensor $\bar{T}_{ik}^*$ (41) ist mit $\partial_k^*(\bar{\mu}_0^* \bar{u}^{*k}) \neq 0$ das Gesetz der lokalen Ruhmassenerhaltung bei großräumiger Betrachtung *nicht* erfüllt. Trotzdem haben wir in Abschnitt 5 bei räumlicher Mittelung über kosmische Bereiche an den ‚Bewegungsgleichungen' $\delta \int d\bar{s}^* = 0$ festgehalten. Dies ist unseres Erachtens richtig, solange aus dem kosmischen Linienelement (1) [bzw. (28), (38)], das einer homogen verteilten mittleren Dichte entspricht, nur Schlüsse gezogen werden, die eine in etwa gleichmäßig verteilte, gemäß (7) statistisch *ruhende* Materie – wie z.B. kosmische ‚Atomuhren' – betreffen. Doch sobald es um lokale Bewegungsabläufe geht, sind nach EINSTEINs Vorgehensweise die Gravitationsgleichungen $E_{ik} = 0$ des Vakuums zugrundezulegen (und nicht etwa die einer mittleren Materiedichte des Planetensystems, der Milchstraße oder irgendeines anderen – gegebenenfalls auch beliebig großen – *lokalisierbaren* Systems).

Wie aus (68) unmittelbar ersichtlich, stehen die hier entwickelten Vorstellungen selbstverständlich im Einklang mit dem Äquivalenzprinzip.

Wenn wir nun nach der Anzeige d$\sigma$ beliebig bewegter Atomuhren[54] im lokalen Gravitationsfeld fragen, so halten wir deshalb daran fest, daß diese dem Linienelement d$s/c$ der *lokalisierbaren* Gravitationsquellen entsprechen soll

$$d\sigma \approx \frac{ds}{c} = dt\sqrt{g_{ik} v^i v^k}, \qquad *(71)$$

wobei (mit $v^0 = c$) die räumlichen Komponenten $v^\alpha$ wieder als die der gewöhnlichen Geschwindigkeit zu verstehen sind[55].

In (9), (32), (63) haben wir bereits den Zusammenhang zwischen der Anzeige ruhender Atomuhren und der absoluten Zeit gefunden[56]. Schließlich aber ergibt die in lokalen Inertialsystemen gültige spezielle Relativitätstheorie zusätzlich den Einfluß der relativen Bewegung, sodaß demzufolge – alles zusammengenommen – für Atomuhren im lokalen Gravitationsfeld gerade *(71) gelten muß[57].

In großen Entfernungen von lokalen Gravitationsquellen wird *(71) übergehen in:

$$c\, d\sigma_{(m:=0)} \approx ds = \sqrt{\eta_{ik}\, dx^i dx^k}. \qquad *(72)$$

Wir abstrahieren mit $m := 0$ von allen lokalisierbaren Gravitationsquellen zugunsten einer großräumigen Betrachtung. Daraus ist zu ersehen, daß die lokalen Koordinaten $x^i$ offenbar auf spektrale Einheiten bezogen sind[58].

---

[54]) Natürlich ist die Anzeige von Atomuhren ganz unabhängig vom jeweils gewählten Koordinatensystem, sodaß gilt d$\sigma$ = d$\sigma^*$ = d$\sigma^{(c)}$ = d$\sigma'$. Worauf wir aber gegebenenfalls mit der Bezeichnung d$\sigma^*$ hinweisen wollen, ist das Koordinatensystem K*, in Bezug auf welches eine Berechnung der Anzeige gerade durchgeführt werden soll.

[55]) entfällt

[56]) Dabei sei daran erinnert, daß sich dieser Einfluß desjenigen Gravitationspotentials auf Atomuhren, dessen Gradient der Schwerebeschleunigung entspricht, bereits allein aus dem ursprünglichen Äquivalenzprinzip in Verbindung mit dem klassischen DOPPLER-Effekt ableiten läßt.

[57]) entfällt

[58]) entfällt

### f. Kosmische Zeit gegenüber Ephemeridenzeit

*Unabhängig von jeder Gravitationstheorie verstehen wir unter der Ephemeridenzeit t grundsätzlich die Zeitskala, bei deren Verwendung die Bewegung eines einzelnen idealen Planeten[59] im Feld eines idealen Zentralgestirns konstante Umlaufzeiten von Periastron zu Periastron aufweist.*

Diese Ephemeridenzeit *t* ist zu unterscheiden von der alten Zeitskala *ET* [60] gleichen Namens und stellt nichts anderes dar als die – wie wir sehen werden – richtig gewählte Systemzeit des Sonnensystems. Die reale Bestimmung der Ephemeridenzeit *t* wird in der Praxis durch Vergleich der beobachteten Bewegungsabläufe mit den auf Basis der allgemeinen Relativitätstheorie im Modell berechneten Tabellen (Ephemeriden) der jeweiligen Positionen von Mond und Planeten durchgeführt. Natürlich könnte man immer durch Koordinatentransformation zu einer anderen Systemzeit mit anderen Tabellen übergehen. Doch ist die Ephemeridenzeit – sowohl auf Basis der NEWTON'schen als auch der EINSTEIN'schen Bewegungsgesetze – im lokalen Gravitationsfeld als Systemzeit eben durch die Forderung festgelegt, daß die Umlaufzeiten von Perihel zu Perihel (bis auf nachvollziehbare Störungen) konstant sind.

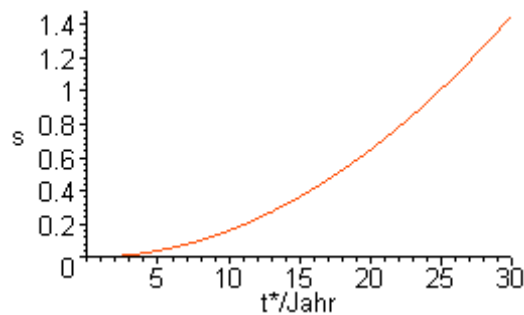

Abb. 2: Das Vorgehen einer im Schwerpunkt des Sonnensystems ruhenden Atomuhr mit der Anzeige $\sigma_A$
gegenüber der kosmischen Zeit *t\** gemäß Beziehung (74)
(hier auf Basis einer hypothetischen HUBBLE-Konstanten von 100 km/s/Mpc und bei geeigneter Festlegung der Zeiteinheit)

Für die Anzeige einer beliebigen bewegten Atomuhr im Gravitationsfeld der Sonne ergibt sich aus *(71) näherungsweise

$$\mathrm{d}\sigma \approx \frac{\mathrm{d}s}{c} \approx \mathrm{d}t\left(1+\frac{U}{c^2}-\frac{1}{2}\frac{v^2}{c^2}\right). \qquad *(73)$$

Vom absoluten kosmischen Ruhsystem führt eine GALILEI-Transformation (s. Abschn. 8/g) zu den absoluten Koordinaten des Sonnensystems, das sich bekanntlich mit einer – selbst über sehr große Zeiträume praktisch

---

[59] dessen Masse so klein sein soll, daß von den Einflüssen der Gravitationsstrahlung abgesehen werden kann

[60] Bis zur Ablösung der alten Ephemeridenzeit *ET* (Ephemeris Time) durch neuere Zeitskalen im Jahr 1984 hatte es bereits verschiedene Änderungen gegeben (sodaß bei Vergleichen gegebenenfalls die jeweils verwendete Realisierung *ET*0, *ET*1 oder *ET*2 berücksichtigt werden muß). Damals wurden TD-Zeitskalen eingeführt, die diesen Namen erhielten, weil sie (wie die alte Ephemeridenzeit *ET*) auf den nach den *dynamischen* Bewegungsgesetzen berechneten Bewegungen von Mond und Planeten beruhten. Im Unterschied zu *ET* wurden aber bei den Skalen *TDB* (Temps Dynamique Baricentrique), *TDT* (Temps Dynamique Terrestrique) die Bewegungsgleichungen der allgemeinen Relativitätstheorie – sowie der (aus unserer heutigen Sicht unvollständige, rein lokale) relativistische Einfluß auf Uhren – zugrunde gelegt. Wie die Internationale Atomzeit *TAI* (Temps Atomique International) hatte die Zeitskala *TDT* per definitionem die SI-Sekunde als Basiseinheit, wobei der Zeitpunkt 1. Januar 1977, 0:00 Uhr TAI als Zeitpunkt 1. Januar 1977, 0:00 Uhr + 32.184 Sekunden *TDT* festgesetzt wurde. Im Interesse einer Kontinuität zwischen alter und neuer Skala repräsentiert dabei der Wert 32.184 Sekunden den damaligen Unterschied zwischen *TAI* und *ET*. Trotz des Bezugs auf die gemeinsame Grundlage der SI-Sekunde aber können *TAI* als gemittelte Anzeige irdischer Atomuhren und *TDT* als eine Systemzeit im Sinne der allgemeinen Relativitätstheorie natürlich keine identischen Zeitskalen sein. Es wurde festgelegt, daß sich *TDT* von *TDB* nur um periodische Terme unterscheiden soll. Im Jahr 1991 schließlich wurden von der IAU (International Astronomical Union) als weitere Zeitskalen eingeführt: *TT* (Terrestrial Time), *TCG* (Temps Coordinate Geocentrique) und *TCB* (Temps Coordinate Baricentrique). Dabei gilt *TT* = *TDT* und bezieht sich damit auf die Oberfläche des Geoids (Erdoberfläche auf Meereshöhe).

konstanten – Geschwindigkeit (von größenordnungsmäßig $4 \cdot 10^2$ km/s) bewegt. Die absolute Zeit bleibt dabei natürlich unverändert[61]).

Für die Anzeige $\sigma_A$ einer im Schwerpunkt des Sonnensystems ruhenden ‚Atomuhr' A gilt aber in Bezug auf die kosmische Zeit $t^*$

$$\sigma_A \approx t^* + \tfrac{1}{2} H t^{*2}. \tag{74}$$

Bislang hat es keine Divergenz zwischen idealer Atomzeit und idealer Ephemeridenzeit gegeben (s. dazu auch Anhang A). Seit Einführung der SI-Sekunde (im Jahr 1969) sind mehr als 30 Jahre vergangen, doch die seither entstandene Zeitdifferenz zwischen kosmischer Zeit und Atomzeit beläuft sich auf nur ca. 1.5 s (s. Abb. 2).

Aus den Bewegungsgleichungen (68) ergeben sich nun die bekannten Lösungen für Planetenbahnen, Doppelsternsysteme u.a. mit der wichtigen Besonderheit, daß die entsprechenden Bahndaten als lokale Koordinaten zu verstehen sind.

Es ist nicht dasselbe Bezugssystem, in welchem die Spiralnebel statistisch ruhen, und die großen Halbachsen der Planetenbahnen sowie die Umlaufzeiten von Perihel zu Perihel konstant sind. Daß Planetenuhr und Atomuhr offenbar gleiche Ganggeschwindigkeiten aufweisen, ist Ausdruck der Tatsache, daß es sich bei beiden um natürliche Uhren handelt[62]).

Selbstverständlich ist es möglich, alle Bewegungsabläufe des Sonnensystems auch in das kosmische Koordinatensystem zu übertragen. Wendet man also auf das gewöhnliche Linienelement d$s$ der allgemeinen Relativitätstheorie – mit Berücksichtigung von $t = t''$ – die Umkehrtransformation zu (65) an, die zuvor das stationäre Linienelement d$\bar{s}^*$ gemäß (66) näherungsweise in das der speziellen Relativitätstheorie übergeführt hat, so folgt nun in kosmischen Koordinaten $x^{*i}$

$$\mathrm{d}s^{*2} \approx e^{2Ht^*} \mathrm{d}s^2_{(\text{ART})} \left[ t := t''(t^*, r^*),\ r := r''(t^*, r^*) \right]. \tag*{*(75)}$$

Die Bedeutung dieses Linienelements ist anhand eines einfachen Beispiels leicht zu verstehen[63]): Wie aus *(75) ersichtlich, wird der Wert der Lichtgeschwindigkeit im lokalen Gravitationsfeld durch die Koordinatentransformation nicht nennenswert beeinflußt. Erfolgt aber die Bewegung eines Planeten gemäß der üblichen Gleichungen $\delta \int \mathrm{d}s = 0$ bezüglich des Ephemeridensystems K mit konstanten großen Halbachsen $a$ und konstanten Zeitspannen $T$ von einem beliebigen Periheldurchgang zum nächsten, so erfolgt umgekehrt die gleiche Bewegung bezüglich des kosmischen Systems K* gemäß $\delta \int \mathrm{d}s^* = 0$, mit der Konsequenz, daß die großen Halbachsen $a(t^*) \approx (1-Ht^*)\, a$ und Umlaufzeiten $T(t^*) \approx (1-Ht^*)\, T$ nun bezogen auf kosmische Systemuhren und kosmische Längeneinheiten mit der Zeit schrumpfen. Dies ergibt sich durch bloße Anwendung der Koordinatentransformation auf die ursprünglichen Bahndaten, was aufgrund des ‚geodätischen'[64]) Bewegungsgesetzes (68) zulässig ist und mit diesem in Einklang steht.

---

[61]) Das ist der Grund, warum die Planetenuhr – welche z.B. auch aus zwei einander in einer Entfernung von 1 m etwa einmal wöchentlich umlaufenden Kugeln der Masse 0.8 kg bestehen könnte – eine *natürliche* Uhr darstellt. Hinsichtlich der äquivalenten Verwendung von Atomuhren ist also zu beachten, daß diese ‚nach dem Mond' gehen.

[62]) Obwohl es sich bei kosmischer Zeit und Ephemeridenzeit jeweils um Systemzeiten handelt, verlaufen beide nicht synchron. – In diesem Zusammenhang stellt sich auch die Frage, ob ein im Idealfall verlustfrei rotierender Körper gegebenenfalls als *Rotationsuhr* synchron zum Takt einer Atomuhr oder dem einer Planetenuhr gehen sollte. Seit NEWTONs Zeiten ist allerdings klar, daß die rotierende Erde – als Archetyp einer natürlichen Uhr schlechthin – starken Gezeitenkräften mit Reibungsverlusten unterliegt, was ihre Ganggeschwindigkeit gegenüber der gerade deshalb eingeführten Ephemeridenzeit (meßbar) beeinträchtigen muß. Die zur Bestimmung der (alten) Ephemeridenzeit von den Astronomen bevorzugt verwendete Planetenuhr Erde-Mond wird aber durch Gezeitenkräfte jedenfalls weit weniger beeinflußt als die Rotationsuhr ‚Erde gegenüber dem Fixsternhimmel'.

[63]) Durch bloße Koordinatentransformation läßt sich demzufolge selbst aus dem Linienelement der speziellen Relativitätstheorie ein solches erzeugen, das – bis auf Benennungen und Vorzeichen – in erster Näherung $Ht^*$ mit dem kosmischen übereinstimmt, wobei der zugehörige Energie-Impuls-Tensor natürlich aber verschwindet.

[64]) Die Bezeichnung ‚geodätisch' ist wieder – ohne konkrete geometrische Bedeutung – lediglich als Hinweis auf eine mathematische Struktur mit Extremaleigenschaften zu verstehen, welche für die Möglichkeit wesentlich ist, die Bahnen beliebiger Objekte beim Wechsel des Bezugssystems durch bloße Koordinatentransformation ineinander überzuführen.

Unabhängig vom gewählten Bezugssystem aber bleibt als koordinatenfreie Darstellung des Sachverhalts: Astronomische Uhren (bzw. Planetenuhren) gehen *lokal* synchron mit Atomuhren (bzw. EINSTEIN'schen Lichtuhren), wobei in der unterschiedlichen Ganggeschwindigkeit beider gegenüber kosmischen Systemuhren die einfache Erklärung für die Rotverschiebung ohne Bezug auf ‚Expansion des Universums' liegt.

Die beobachtbare Konsequenz einer (nicht von vorneherein undenkbaren) Übereinstimmung der kosmischen Zeit mit der Ephemeridenzeit wäre, daß eine daraus resultierende zeitliche Korrektur im Sinne einer kleinen *Zunahme* der mit Atomuhren gemessenen Umlaufzeit des 1974 von HULSE und TAYLOR [26] entdeckten binären Pulsars PSR 1913+16 betragsmäßig bei immerhin ca. 4% Prozent der tatsächlich beobachteten *Abnahme* liegen sollte, die derzeit (bis auf eine Korrektur von ca. 0.7% zur Berücksichtigung der galaktischen Bewegung) exakt dem Energieverlust durch Gravitationsstrahlung einander umlaufender *Punktmassen* zugeschrieben wird[65]). Im Hinblick auf die in [27] angegebenen Fehlertoleranzen ist damit die Möglichkeit einer solchen Übereinstimmung bereits widerlegt[66]).

**g. GALILEI-Transformation und spezielle Relativitätstheorie**

Da in der allgemeinen Relativitätstheorie jede Koordinaten-Transformation erlaubt ist, so muß natürlich auch eine GALILEI-Transformation erlaubt sein. Nach GALILEI-Transformation der Koordinaten $t=t^G$, $x=x^G+vt^G$, $y=y^G$, $z=z^G$ des kosmischen Ruhsystems K auf das bewegte Inertialsystem $S^G$ lautet das lokale Linienelement (weitab von relevanten Gravitationsquellen):

$$ds^2 = \left(1-\beta^2\right)c^2 dt^{G2} - 2\beta c\, dt^G dx^G - \left(dx^{G2} + dy^{G2} + dz^{G2}\right). \tag{76}$$

Die räumliche Entfernung, so wie sie sich bei der Ermittlung mit spektralen Einheitsmaßstäben in $S^G$ ergibt, finden wir gemäß Formel [13]/(7) und in Anlehnung an [18] zu:

$$dl = \sqrt{\frac{dx^{G2}}{1-\beta^2} + dy^{G2} + dz^{G2}}\ . \tag{76a}$$

Aus [13]/(8) ergibt sich für die Eigenzeit $\tau$ einer in $S^G$ ruhenden Atomuhr:

$$d\tau = dt^G \sqrt{1-\beta^2}\ . \tag{76b}$$

Und schließlich beträgt die 1-Weg-Lichtgeschwindigkeit für eine Ausbreitung parallel zur $x^G$-Achse im bewegten System $c_\pm^G = c/(1 \mp \beta)$, wobei aber die (lokale) Durchschnittsgeschwindigkeit für Hin- und Rückläufe nach Voraussetzung gleich $c$ bleibt. Nach diesem Muster lassen sich alle Ergebnisse der speziellen Relativitätstheorie auch auf Basis einer GALILEI-Transformation berechnen, was für *lokale* Inertialsysteme bei angemessener Wahl der Systemkoordinaten in der Regel sogar unvermeidlich ist[67]).

Angesichts der vorausgesetzten Gültigkeit des Linienelements (38) im kosmischen Bezugssystem kann aber die spezielle Relativitätstheorie auch in einem gegen den Fixsternhimmel gleichförmig bewegten Inertialsystem nur annähernde, lokale Gültigkeit haben. Demzufolge ist (auch) in Bezug auf kosmisch beeinflußte *Atomuhren* die Trägheitsbewegung über große Zeiträume keine streng gleichförmige mehr.

**h. Längenmessung und Laufzeit**

Im Fall einer zeitlich veränderlichen Lichtgeschwindigkeit wäre eine Definition des Meters auf Basis der Lichtgeschwindigkeit für kosmische Entfernungen praktisch unbrauchbar. Dies betrifft offensichtlich auch jedes

---

[65]) Wir beziehen uns mit diesen Angaben auf die übersichtliche Zusammenstellung der einschlägigen Ergebnisse bei Will [27]. Dort finden sich auch Literaturangaben zu weiteren, von TAYLOR in Zusammenarbeit mit verschiedenen Autoren publizierten Originalarbeiten (s. insbesondere auch Kapitel 14 *"An Update"*).

[66]) Bei der Auswertung von Meßdaten eines Pulsars könnte allerdings das Problem der Rotationsuhr (s.o.) eine Rolle spielen.

[67]) Bereits die Behandlung der rotierenden Scheibe im Rahmen der allgemeinen Relativitätstheorie [18] benutzt eine (lokale) GALILEI-, und nicht etwa eine (lokale) LORENTZ-Transformation (s. [13]/Abschn. 9). Letztere ist nicht integrabel.

Linienelement in RW-Form[68]). Wegen der demgegenüber beim stationären (bzw. skalarabhängigen) Linienelement konstanten kosmischen Lichtgeschwindigkeit bereitet eine solche Meterdefinition hier dagegen keinerlei prinzipielle Schwierigkeiten[69]). Wie lassen sich nun astronomische oder gar kosmische Entfernungen in Metern messen?

Ein elektromagnetisches Signal benötigt für Hin- und Rücklauf zwischen zwei festen Punkten des euklidischen Raums immer die gleiche kosmische Zeitspanne $\Delta t^*$. Doch die gemessene Anzahl spektraler Zeiteinheiten (Takte einer Atomuhr) bliebe bei verschiedenen Beobachtungszeiten nicht die gleiche, weil sich die spektralen Einheiten gegenüber den absoluten ändern. Wenn die kosmische Laufzeit eines Lichtsignals beispielsweise für Hin- und Rückweg insgesamt $\Delta t^* = 2\,l^*/c$ beträgt, dann ist die mit einer ruhenden Atomuhr gemessene Laufzeit:

$$\tau_l \;=\; \int_{t_0^*}^{t_0^*+\Delta t^*} \mathbf{e}^{Ht^*}\mathrm{d}t^* \;=\; \frac{\mathbf{e}^{Ht_0^*}}{H}\left(\mathbf{e}^{H\Delta t^*} - 1\right) \;\approx\; \Delta t^*\mathbf{e}^{Ht_0^*}\left(1 + \frac{1}{2}H\Delta t^*\right) + O^3(H\Delta t^*). \qquad (77)$$

Wird aber der Lichtweg zur wahren Entfernungsmessung in $n$ hinreichend kurze Teilstrecken unterteilt, die allesamt *gleichzeitig* zu durchlaufen sind – was prinzipiell der Messung mit entsprechenden Meterstäben zum Zeitpunkt $t_0^*$ entspricht – so erhält man

$$n\,\tau_n \;=\; n\int_{t_0^*}^{t_0^*+\frac{\Delta t^*}{n}} \mathbf{e}^{Ht^*}\mathrm{d}t^* \;=\; \frac{n\,\mathbf{e}^{Ht_0^*}}{H}\left(\mathbf{e}^{\frac{H\Delta t^*}{n}} - 1\right) \;\approx\; \Delta t^*\mathbf{e}^{Ht_0^*}\left(1 + \frac{1}{2n}H\Delta t^*\right) + O^3\!\left(\frac{H\Delta t^*}{n}\right). \qquad (78)$$

Daraus ergibt sich für die auf spektrale Längeneinheiten bezogene Entfernung $l$ durch Grenzübergang

$$l \;=\; c\lim_{n\to\infty} n\tau_n \;=\; c\Delta t^*\mathbf{e}^{Ht_0^*} \;=\; l^*\mathbf{e}^{Ht_0^*}, \qquad (79)$$

was natürlich zu erwarten war. Mit der weiter oben begründeten möglichen Festsetzung $t_0 = 0$ hat man also hier $l = l^*$. Wollte man aber aus den mit Atomuhren gemessenen Laufzeiten einfach durch Multiplikation mit $c$ auf die zurückgelegte Entfernung $l$ schließen, so erhielte man trotz konstanter Lichtgeschwindigkeit unterschiedliche Ergebnisse, und zwar im Verhältnis

$$\frac{c\tau_l}{l} \;\approx\; 1 + \frac{1}{2}H\Delta t^* + O^2(H\Delta t^*), \qquad (80)$$

sodaß eine *unmittelbar* auf die Laufzeitmessung mittels Atomuhren reduzierte Entfernungsmessung im Sinne der heutigen Meterdefinition für kosmische Entfernungen nicht möglich ist. Eine entsprechende Festlegung der Basiseinheit 1 m jedoch bleibt natürlich trotzdem möglich (s. Fußn.[69]), die allerdings – wegen der universellen zeitlichen Abnahme der spektralen Einheiten – auf einen fixierten Zeitpunkt bezogen sein müßte.

---

[68]) So läßt sich aus einem Linienelement der RW-Form, auf dem die heutige Kosmologie basiert, unseres Erachtens keineswegs eine reale Expansion ableiten. Daß insbesondere weder eine Flucht der Spiralnebel noch eine Expansion des ‚Raums' erforderlich ist, um die kosmische Rotverschiebung (s. Abschn. 6) aus der RW-Form zu erklären, läßt bereits ein unvoreingenommener Blick auf LEMAÎTRES [5] allererste – mit der Verbindung von EINSTEINS [1], DE SITTERS [3], FRIEDMANNS [4] und HUBBLES [2] Werk die gesamte nachfolgende Entwicklung prägende – Ableitung deutlich erkennen (dessen übersichtliche Vorgehensweise findet sich z.B. auch bei WEINBERG [22]). Was dort unseres Erachtens alleine benutzt wird, ist ein zeitabhängiger Koordinatenwert der Lichtgeschwindigkeit bezüglich der im kosmischen System *ruhenden* Spiralnebel.

[69]) Im Unterschied zur heutigen Definition ist dies widerspruchsfrei nur auf Basis der – lokal zwar immer, global aber nur bei hinreichender Entfernung von lokalen Gravitationsquellen im absoluten kosmischen Bezugssystem gegebenen – Konstanz der *2-Weg*-Lichtgeschwindigkeit (s. [13]) möglich, die Laufzeit könnte dabei mit *einer* Atomuhr gemessen werden, deren Anzeige – veränderliche (heutige) SI-Sekunden – konsequenterweise in die kosmische Zeit (konstante ‚kosmische' Sekunden) umzurechnen wäre.

Da sich für kosmische Distanzen selbstverständlich keine 2-Weg-Laufzeiten messen lassen, ist man hier darauf angewiesen, aus der indirekt ermittelten einfachen Laufzeit von der Quelle zum Beobachter auf die Entfernung

$$l^* = \frac{c}{H}\ln\left(1 + H\tau_l\, \mathrm{e}^{-Ht_0^*}\right) \tag{81}$$

zu schließen, was sich aus (77), hier aber mit $\Delta t^* = l^*/c$ ergibt. Daß nun $l$ mit der Beobachtungszeit $t_0^*$ gemäß (79) anwächst, könnte isoliert betrachtet durchaus eine Fehldeutung als ‚Expansion' (mit einer ‚Geschwindigkeit' $Hl$ und einer ‚Beschleunigung' $H^2l$) nahelegen, hat in Wirklichkeit aber seinen Grund im beschleunigten Gang – oder umgekehrt im schrumpfenden Takt – der Atomuhren[70]. Dieses Problem löst sich ganz und gar auf, wenn man sich bei der Laufzeitmessung auf die Rotverschiebung $z$ (bzw. auf technische Systemuhren, s. Fußn.[61]) bezieht.

Die Stationarität kosmischer Abläufe wird aber dadurch keinesfalls aufgehoben, daß man bei gleichbleibenden kosmischen Entfernungen in Bezug auf die Anzeige ‚natürlicher' Atomuhren zeitlich zunehmende Laufzeiten konstatieren wird. Denn beim Vergleich zweier zu verschiedenen Beobachtungszeiten ermittelter Werte hinge die relative Zunahme als einzig relevante Aussage über die aufgetretene Veränderung wieder nur von der *Zeitdifferenz* zwischen beiden Beobachtungen ab, nicht aber von einem beliebig gewählten Bezugspunkt der kosmischen Zeitskala[71].

*Im Falle des stationären Linienelements (38) fällt der Bezugszeitpunkt $t_0^*$ beim Vergleich spektralbasierter Meßdaten sowohl mit atomaren als auch mit absoluten Größen letztlich immer heraus!*

Mit der Belegung $\Pi = -1/3$ ändert sich nichts am ursprünglichen Ergebnis $z = \mathrm{e}^{H\Delta t^*} - 1$ gemäß (21) über die Rotverschiebung in Abhängigkeit von der Laufzeit des Lichts. Doch mit $\Delta t^* = l^*/c$ folgt nun einfach

$$z(l^*) = -1 + \mathrm{e}^{\frac{Hl^*}{c}}, \qquad (\Pi = -1/3) \tag{82}$$

und daraus unmittelbar

$$l^*(z) = \frac{c}{H}\ln(1+z) \qquad (\Pi = -1/3) \tag{83}$$

anstelle von (23), (24). Die entsprechenden Beziehungen werden wir im nächsten Abschnitt für einige vorläufige Abschätzungen verwenden.

## 9. Einige vorläufige Abschätzungen

Im folgenden sollen einige – wenn auch nur vorläufige – Abschätzungen angegeben werden, die sich in Abhängigkeit vom Druckfaktor $\Pi$ auf einige außerordentlich wichtige kosmologische Beobachtungstatsachen beziehen, welche über die bereits in Abschnitt 6 behandelte fundamentale Rotverschiebung hinausgehen. Dabei wird sich zeigen, daß nur für $\Pi = -1/3$ mit (82), (83) auch alle entsprechenden Ergebnisse (89), (90), (93), (95) und (100) streng unabhängig sind vom Beobachtungszeitpunkt $t_0^*$, sodaß in diesem Sinne mit uneingeschränkter Berechtigung tatsächlich von einem stationären Linienelement gesprochen werden kann.

---

[70]) Was geschieht, wenn zur Messung der Laufzeit eines in gleichbleibender (großer) absoluter Entfernung reflektierten Lichtsignals nun keine Atomuhr, sondern eine (kleine) lokale EINSTEIN'sche Lichtuhr verwendet wird? Während das Meßsignal unterwegs ist, läuft das Uhrensignal zwischen den Spiegeln eines Interferometer-Arms hin und her. Das Meßergebnis besteht aus einer gewissen Anzahl von ‚Takten' der Lichtuhr zwischen Aussendung und Rückkehr des Meßsignals. Die gleiche Messung werde nun zu einem späteren Zeitpunkt wiederholt. Der absolute Lichtweg des Meßsignals ist nach Voraussetzung unverändert geblieben. Auch der Koordinatenwert der Lichtgeschwindigkeit ist nach wie vor der gleiche. Doch hätte die Länge $L^{[\lambda]}(t^*)$ des realen Interferometer-Arms der Lichtuhr, der ja zugleich die spektrale Längeneinheit repräsentiert, in Bezug auf absolute Einheiten inzwischen abgenommen (ebenso wie die atomare Zeiteinheit $T^{[\tau]}(t^*)$, s. Abschn. 5), sodaß bei Berücksichtigung der lokal konstanten Lichtgeschwindigkeit $c$ die Anzahl der ‚Takte' der Lichtuhr nun tatsächlich größer wäre als zuvor. Die Zunahme der auf diese Weise spektral ermittelten Laufzeit des betrachteten Meßsignals aber ginge ganz auf das Konto der abnehmenden atomaren Zeiteinheit $T^{[\tau]}(t^*)$. Atomuhr und Lichtuhr weisen die gleiche Ganggeschwindigkeit auf.

[71]) entfällt

**a. Zur Häufigkeitsverteilung von Galaxien**

Für die Häufigkeit der Spiralnebel und Quasare läßt sich in Abhängigkeit von $z$ eine – allerdings sehr grobe – Abschätzung angeben. Zunächst folgt für $\Pi \neq -1/3$ aus (23) das Differential $dl_0^*$ der Entfernung im euklidischen Raum zu

$$dl_0^* = \frac{c}{H}(1+z)^{-\frac{3}{2}(1+\Pi)} dz. \qquad (\Pi \neq -1/3) \qquad (84)$$

Mit einer mittleren konstanten Objektzahldichte $n^*$ ergibt sich für die Anzahl $dN$ der Spiralnebel im Volumenelement einer Kugelschale mit einem euklidischen Radius zwischen $l^*$ und $l^* + dl^*$

$$dN = n^* dV^* = 4\pi n^* l^{*2} dl^*, \qquad (85)$$

und daraus schließlich mit $l^* := l_0^*$ gemäß (23), (84)

$$\frac{dN}{dz} = 16\pi n^* \frac{c^3}{H^3(1+3\Pi)^2}(1+z)^{-\frac{3}{2}(1+\Pi)}\left[1-(1+z)^{-\frac{1}{2}(1+3\Pi)}\right]^2, \quad (\Pi \neq -1/3) \qquad (86)$$

mit einem Maximum der Verteilung bei

$$z_{\max} = -1 + \left(\frac{\frac{5}{3}+3\Pi}{1+\Pi}\right)^{\frac{2}{1+3\Pi}}. \qquad (\Pi \neq -\tfrac{1}{3} \wedge \Pi > -\tfrac{5}{9}) \qquad (87)$$

Die Gesamtzahl $N$ der theoretisch sichtbaren Galaxien ergibt sich aus (86) durch Integration über $z$ (von 0 bis $\infty$) zu

$$N = \frac{32\pi c^3 n^*}{3H^3(1+3\Pi)^3}. \qquad (88)$$

Angesichts der nach Voraussetzung gleichmäßig in einem unendlichen euklidischen Raum verteilten Galaxien ist dieser für $\Pi \neq -1/3$ *endliche* Wert auf den ersten Blick ein paradoxes Ergebnis. Beachtet man aber die oben erhaltene Beziehung (25a), so entspricht das Ergebnis (88) einfach der Anzahl $N$ von Galaxien, die sich in einer Kugel mit dem Radius des maximalen Lichtwegs aus der Vergangenheit befinden.

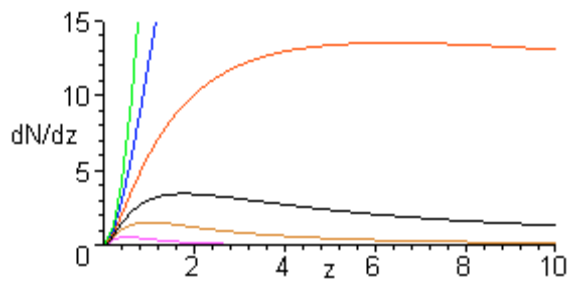

Abb. 3: Die theoretische Verteilung der Galaxien in Abhängigkeit von $z$
(von oben nach unten):
für $\Pi$ = -1 (grün), -2/3 (blau), -1/3 (rot), 0 (schwarz), +1/3 (gold), +1 (magenta)
ohne Berücksichtigung von Absorptions-, Selektions- und/oder Evolutionseffekten.

Im stationären Fall $\Pi = -1/3$ dagegen ergibt sich mit (83) anstelle von (86) analog zur obigen Ableitung

$$\frac{dN}{dz} = 4\pi n^* \frac{c^3}{H^3} \frac{\ln^2(1+z)}{(1+z)}, \qquad (\Pi = -1/3) \qquad (89)$$

und die Gesamtzahl der Galaxien folgt durch Integration hier zu $N = \infty$, ganz so wie es nach den zugrundeliegenden Voraussetzungen auch sein sollte. Im Unterschied zu (87) resultiert das Maximum der Verteilung aus (89) bei

$$z_{\max} = -1 + e^2 \approx 6.4 \, . \qquad (\Pi = -1/3) \qquad (90)$$

Wie aus (87), (90) ersichtlich, ergibt sich also aus dem singularitätsfreien Linienelement (1) für alle $\Pi >$ -5/9 ein Maximum der Verteilung. Dies ganz im Unterschied zur Belegung $\Pi = $ -1, welche der oben zitierten Steady-State Theory entspricht (s. Abb. 3). Für die speziellen Werte $\Pi = $ -1/3, 0, +1/3, die einem stationären, einem druckfreien bzw. einem ultrarelativistischen Energie-Impuls-Tensor entsprechen, liegt dieses Maximum bei $z_{\max} = e^2-1$, 16/9 bzw. 1. Im ersten Fall also mit einem Zahlenwert von ca. 6.4 deutlich oberhalb des tatsächlich beobachteten Maximums bei $z_{\text{max-obs}} \approx 2.3$. In keinem Falle jedoch zeigt sich der daran anschließende, ebenfalls beobachtete steile Abfall der Verteilung auf nahezu Null im Bereich von etwa $2.3 < z < 4$.

Eine theoretische Verteilung gemäß (86) bzw. (89) wäre mit den Beobachtungstatsachen aber besser vereinbar, wenn man berücksichtigt, daß a) das Sternlicht in einem stationären Universum absorbiert werden muß, wobei eine mittlere Absorptionskonstante in der Größenordnung $c/H$ liegen sollte (ohne daß die Absorption unbedingt gleichmäßig sein müßte), und b) die technische Beobachtungswahrscheinlichkeit aller Objekte mit abnehmender scheinbarer Helligkeit jedenfalls so stark abnimmt, daß unterhalb einer gewissen technischen Nachweisgrenze überhaupt keine Objekte mehr registriert werden können[72]. Diese Möglichkeit wird zusätzlich gestützt durch ein anderes Phänomen (s. Abschn. 9/d).

**b. Zur geforderten Zeitabhängigkeit der relativen Photonenenergie**

Im stationären Fall $\Pi = $ -1/3 läßt sich anhand der Formel (44) für die Dichte $\bar{\mu}_0^*$ der im Mittel ruhenden kosmischen Materie auch eine plausible Überlegung zu einer relativistischen Ableitung der zeitlich veränderlichen Photonenenergie anstellen, die wegen der Rotverschiebung ansonsten – unter der Voraussetzung durchgängiger Gültigkeit – nur von der quantenmechanischen Energie-Frequenz-Beziehung $\varepsilon = hf$ gefordert wird.

Zwar ist zu beachten, daß Aussagen über eine homogen verteilte ruhende Materie nicht von vornherein auf die Energie von Photonen anwendbar sind. Doch andererseits stimmen die *Unterschiede* der Energie ruhender Atome vor und nach der Emission (Energiestufen) mit der Energie entsprechender Photonen am Ort und zur Zeit ihrer Entstehung überein. Nun folgt aus (44) für ein lokales Massenelement $\delta m_0 = \bar{\mu}_0^* \delta V$ mit Verwendung des aus (63) resultierenden Zusammenhangs $\delta V \approx e^{3Ht^*} \delta V^*$ und wegen der selbstverständlichen Beziehung $\delta m_0^* = \bar{\mu}_0^* \delta V^*$ zunächst für die Masse ruhender Atome $\delta m_0 \approx e^{Ht^*} \delta m_0^*$. Nimmt man außerdem an, daß sich die Energie eines Photons – ebenso wie die absolute Wellenlänge – während seiner Laufzeit nicht ändert (weil für ein Photon keine ,Eigenzeit' vergeht), so ergibt sich eine relative Energieabnahme im Vergleich zu später entstehenden Photonen, deren Energie wegen der inzwischen erfolgten Zunahme der Massen und Massenunterschiede von Atomen nun größer geworden sein muß. Demzufolge sollte also für die bei der Absorption am Beobachtungsort mit atomaren Energiestufen (d.h. auch mit neu entstehenden Photonen) verglichene Photonenenergie gelten $\varepsilon_0 = \varepsilon_E e^{-H\Delta t^*}$ (Bedeutung der Indizes wie in Abschn. 7). Neben dem Einfluß des Gravitationsfeldes auf spektrale Maßstäbe und Atomuhren könnte es also zusätzlich einen Einfluß auf Massen – und damit auch auf die dritte frei wählbare Basiseinheit – geben. Für ganz beliebige Teilchen schließlich könnte dann einheitlich gelten $\delta m \approx e^{H d\bar{s}^*/c} \delta m^*$, womit der (absoluten) Massenzunahme ruhender Teilchen einerseits und der (absoluten) Massenkonstanz von Photonen andererseits zugleich Rechnung getragen wäre[73].

---

[72]) Es soll an dieser Stelle noch einmal ausdrücklich betont werden, daß eine derartige Abschätzung über die Verteilung von *Inhomogenitäten* aus einer solchen Lösung der nicht-linearen EINSTEIN'schen Gleichungen, die wie alle bisherigen Versuche auf einer *homogenen* Materieverteilung basiert, jedenfalls zunächst grundsätzlich als fragwürdig gelten muß.

[73]) Eine relative zeitliche Abnahme der Photonenenergie steht natürlich im Widerspruch zu einer strikten Massenerhaltung der im kosmischen Bezugssystem nach Voraussetzung gleichmäßig verteilten, ruhenden Atome. Doch für den phänomenologischen kosmischen Energie-Impuls-Tensor $\bar{T}_{ik}^*$ (41) ist mit $\partial_k^*(\bar{\mu}_0^* \bar{u}^{*k}) \neq 0$ offensichtlich das Gesetz der lokalen Ruhmassenerhaltung nicht erfüllt (dessen Gültigkeit ansonsten vorausgesetzt werden muß, um aus der kovarianten Ableitung des lokalen kinetischen Energie-Impuls-Tensors die üblichen geodätischen Beschleunigungen zu erhalten, was aber gemäß Abschnitt 6 bei räumlicher Mittelung über kosmische Bereiche eben *nicht* der Fall zu sein braucht). Trotzdem aber ist die phänomenologische Energiedichte $\bar{T}_{00}^*$ konstant. – Davon abgesehen ist die Situation hier eine völlig andere als bei der Steady-State Theory, die bekanntlich eine *Materialisation* aus dem Nichts verlangt (s. z.B. [22]).

### c.  Zur scheinbaren Helligkeit als Funktion von z

Die scheinbare Helligkeit $I$ eines Objekts mit der absoluten Strahlungsleistung $L$ ist (bei einer bis zur Beobachtung vergangenen Laufzeit des Lichts $\Delta t^* > 0$) gegeben als

$$I = \frac{L}{4\pi l^{*2}} e^{-2H\Delta t^*}, \qquad (91)$$

wobei sich der Faktor $e^{-2H\Delta t^*}$ zunächst durch die kosmische Rotverschiebung (21) in Verbindung mit der Energie-Frequenz-Beziehung für Photonen $\varepsilon_{0/E} = h f_{0/E}$, d.h. also $\varepsilon_0 = \varepsilon_E (\delta\tau_E/\delta\tau_0)$, und dann aus der Eigenzeitabhängigkeit der in Bezug auf Atomuhren für ein Photon gemessenen Leistung $\varepsilon_{0/E}/\delta\tau_{0/E}$, insgesamt also schließlich aus $(\delta\tau_E/\delta\tau_0)^2$ ergibt (s. [22]). Einsetzen von $l_0^*$ aus (23) für $l^*$ in (91) führt zu dem Ergebnis

$$I(z) = \frac{LH^2(1+3\Pi)^2}{16\pi c^2 (1+z)^2 \left[1-(1+z)^{-\frac{1}{2}(1+3\Pi)}\right]^2}. \qquad (\Pi \neq -1/3) \qquad (92)$$

Mit $z \to \infty$ geht die scheinbare Helligkeit $I(z)$ für alle in (92) zugelassenen Werte von $\Pi$ gegen Null, wie dies vernünftigerweise zu erwarten ist. Auch der hier eigentlich ausgeschlossene Sonderfall $\Pi = -1/3$ bildet diesbezüglich keine Ausnahme. Für den stationären Fall ergibt sich nämlich

$$I(z) = \frac{L}{4\pi}\left[\frac{H}{c(1+z)\ln(1+z)}\right]^2. \qquad (\Pi = -1/3) \qquad (93)$$

Die Beziehungen (92), (93) sind insofern von besonderem Interesse, als sie jeweils einen unmittelbar überprüfbaren (theoretischen) Zusammenhang herstellen zwischen den direkt meßbaren Größen der scheinbaren Helligkeit $I$ und des Rotverschiebungs-Parameters $z$.

### d.  Zur mittleren kosmischen Strahlungsdichte

Für eine analoge Abschätzung der mittleren kosmischen Strahlungsdichte $I_K$ des Sternenlichts – allerdings unter denselben Vorbehalten wie in Abschnitt 9/a – machen wir den Ansatz

$$dI_K = I(z)\,dN. \qquad (94)$$

Zunächst für $\Pi \neq -1/3$ ergibt sich daraus mit $dN$ aus (86) und mit $I(z)$ aus (92) eine – bereits ohne Berücksichtigung jeder Absorption – *endliche* mittlere Intensität $I_K$ des Sternenlichts zu

$$I_K = \frac{2c\,n_S^*\,L_S}{H(5+3\Pi)}. \qquad (\Pi > -5/3) \qquad (95)$$

Hierbei haben wir das Produkt aus Objektzahldichte $n^*$ und absoluter Strahlungsleistung $L$ typischer Spiralnebel zur bequemeren Abschätzung durch das Produkt aus der mittleren Sternzahldichte $n_S^*$ und der absoluten Strahlungsleistung $L_S$ der Sonne (als eines in etwa typischen Sterns) ersetzt.

Bekanntlich ist nach STEFAN-BOLTZMANN die Gesamtintensität $I_{SB}$ einer schwarzen Strahlung der absoluten Temperatur $T_\vartheta$

$$I_{SB} = \frac{4\pi^5}{15}\frac{k^4 T_\vartheta^4}{c^2 h^3}. \qquad (96)$$

Ein Vergleich $I_K = I_{SB}$ beider Intensitäten liefert eine *effektive* Temperatur

$$T_{eff} = \frac{1}{k}\sqrt[4]{\frac{3}{2\pi^5(1+\frac{3}{5}\Pi)}\frac{c^3 h^3}{H}n_S^* L_S}. \qquad (97)$$

Mit der mittleren Sternzahldichte, für die gilt

$$n_S^* = \frac{\mu_S^*}{m_S}, \tag{98}$$

wobei die auf die Sonne bezogene Größe $m_S$ (analog zu $L_S$) für die mittlere Masse eines typischen Sterns stehen soll, folgt für den Fall, daß mit

$$\mu_S^* = q\rho_c = q\frac{3H^2}{8\pi G}, \tag{99}$$

– nicht zu verwechseln mit der gesamten mittleren Materiedichte $\bar{\mu}_0^* c^2$ (s. Abschn. 8/a) – ein Bruchteil $q$ der kritischen Dichte $\rho_c$ auf *baryonische* Materie entfällt, die insgesamt zur entsprechenden Strahlung beiträgt, also schließlich aus (97)

$$T_{\text{eff}} = \frac{1}{k}\sqrt[4]{\frac{9q}{16\pi^6\left(1+\tfrac{3}{5}\Pi\right)}\frac{c^3 h^3 H L_S}{G m_S}} \approx 5{,}4\,\text{K} \times \sqrt[4]{\frac{q}{\left(1+\tfrac{3}{5}\Pi\right)}}. \tag{100}$$

Da auch mit $\Pi = -1/3$ die Beziehung für die mittlere intergalaktische Intensität $I_K$ des Sternenlichts (95) unverändert gültig bleibt – wie eine eigene Berechnung bei Verwendung von (83), (89) und (93) anstelle von (23), (86) und (92) zeigt – gilt für die daraus abgeleitete effektive Temperatur (100) das gleiche.

Dies ergibt für ein stationäres Universum mit $\Pi = -1/3$ und einem plausiblen Schätzwert $q = 5\%$ (oder auch für den druckfreien Fall mit $\Pi = 0$, $q = 6\%$) numerisch gerade eine effektive Temperatur von ca. 3 K. Dies allerdings zunächst nur für eine *äquivalente* schwarze Strahlung, die der mittleren, hier abgeschätzten intergalaktischen Energiedichte des Sternenlichts vergleichbar ist. Doch selbst mit Blick auf die genannten Vorbehalte dürfte es sich bei dieser seltsamen Übereinstimmung wohl kaum um einen schieren Zufall handeln – ebensowenig wie bei dem fundamentalen Zusammenhang zwischen Rotverschiebung und mittlerer kosmischer Energiedichte.

Damit aber könnte man nun in der Beobachtungstatsache des steilen Abfalls der Galaxienverteilung auf nahezu Null im Bereich $2.3 < z < 4$ geradezu einen Hinweis auf die stationäre isotrope schwarze Wärmestrahlung bei etwa 3 K sehen: Weil ein gewisser Anteil der hier abgeschätzten räumlich und zeitlich konstanten mittleren kosmischen Strahlungsdichte bei einem beliebigen Beobachter offenbar *nicht* als einzelnen Objekten zuzuordnende Strahlung ankommt, sollte dieser ebenfalls stationäre Anteil (umgewandelt durch Absorption) zusätzlich als kosmische Wärme vorhanden sein. – Denkt man z.B. eine großräumig verteilte intergalaktische Substanz (beispielsweise ‚dark matter' und/oder ‚dark energy') mit (zumindest zeitweilig) *temperaturabhängigem* selektiven Emissions- bzw. Absorptionsverhalten [in (gegebenenfalls lokalen) Phasen von ‚re-combination' sowie ‚coupling' bzw. ‚decoupling'], so scheint eine Erwärmung bzw. Abkühlung dieser Substanz auf eine stationäre Temperatur von ca. 3 K jedenfalls nicht undenkbar[74]).

Das OLBERS'sche Paradoxon ist aufgrund der gemäß (95) endlichen mittleren Intensität $I_K$ des Sternenlichts – trotz einer im Falle $\Pi = -1/3$ unendlichen Anzahl von Sternen – jedenfalls für alle Werte $\Pi > -5/3$ des singularitätsfreien (1) bzw. stationären (38) Linienelements prinzipiell[75]) geklärt, ohne daß dies der Hypothese einer Expansion bedurft hätte.

---

[74]) Die Voraussage der kosmischen Hintergrundstrahlung (durch ALPHER, BETHE, GAMOW) wird seit ihrer Entdeckung als wohl wichtigster Stützpfeiler der ‚Big-Bang'-Kosmologie angesehen. Bei näherer Betrachtung aber stellt sich heraus, daß die Erklärung ihrer Natur als nahezu perfekte isotrope schwarze Strahlung zunächst einmal die ad-hoc Hypothese einer kosmischen ‚Inflations'-Phase verlangt, dann aber auch auf der Gültigkeit der Beziehung $R\,T_\vartheta$ = konstant beruht, was der Einführung einer zweiten *ad-hoc* Hypothese gleichkommt. Eine dritte *ad-hoc* Hypothese liegt schließlich darin, daß die Möglichkeit einer vollständigen ‚Entkopplung' der Wärmestrahlung von der Materie vorausgesetzt wird – und zwar über kosmische Zeiträume von einigen Milliarden Jahren (demgegenüber wäre es in einem stationären Universum eher umgekehrt: hier hat man grundsätzlich den Eindruck, *alles andere* als eine isotrope schwarze Strahlung wäre allein erklärungsbedürftig).

[75]) Hätte man in (91) den Korrekturfaktor außer acht gelassen, der sich aus der Energie-Frequenz-Beziehung für Photonen ergibt, so stünde in (95) statt der Zahl 5 die Zahl 3. Daß aber dieser zunächst externe quantenmechanische Korrekturfaktor auch eine Begründung innerhalb der allgemeinen Relativitätstheorie haben könnte, wurde in Abschnitt 9/b gezeigt.

### e. Zum Problem der ‚fehlenden kosmischen Energiedichte'

In Abschnitt 8/a haben wir festgestellt, daß im Beobachtungszeitpunkt $t^* = 0$ auf Basis des stationären Linienelements (38) – ohne den Beitrag des negativen Gravitationsdrucks – für die mittlere kosmische Materiedichte gelten sollte $\bar{\mu}_0^*(t^* = 0) = 2/3\rho_c$, wobei $\rho_c = \overline{T}_{00}^*/c^2 = 3H^2/(\kappa c^4)$ wie üblich für die kritische Dichte steht. Wenn wir nun davon ausgehen, daß tatsächlich nur diese Dichte $\bar{\mu}_0^*$ – und nicht die Komponente $\overline{T}_{00}^*$ – der astronomischen Beobachtung mit den üblichen Abschätzungsverfahren zugänglich ist, dann reduziert sich damit die Differenz zwischen dem in den letzten Jahren geschätzten Wert der (sowohl direkt als auch indirekt beobachteten) kosmischen Materiedichte zum theoretischen Wert (bisher $\rho_c$) sofort auf die Hälfte[76]).

## 10. Anmerkungen, Diskussion, Ausblick

In diesem Abschnitt seien zunächst einige Sätze zur gegenwärtigen Kosmologie erlaubt: Was wir *zweifelsfrei* als Grundlage haben, ist das Licht der Sterne an einem dunklen Nachthimmel[77]) mit dem Phänomen der Rotverschiebung sowie eine 3K-Hintergrundstrahlung. Die gesamte heutige Kosmologie beruht offensichtlich darauf, daß man glaubt, diese Beobachtungstatsachen nicht anders als mit einer ‚Expansion des Universums' erklären zu können.

Wenn nun gemeint wäre, daß die gesamte *Materie* des Universums seit einem Urknall auseinanderfliege, dann müßte der Ort der gemeinsamen Entstehung einen realen Mittelpunkt darstellen. In einem solchen Universum könnte das kosmologische Prinzip nicht einmal in seiner eingeschränkten Form gelten, demzufolge sich zur ‚gleichen Zeit' allen Beobachtern an jedem beliebigen Ort ein gleiches Bild bieten soll.

Wenn dagegen gesagt wird, es sei der *‚Raum'* selbst, der expandiere, dann fragen wir nach dem Koordinatensystem, das notwendig ist, um eine Expansion in Bezug auf dasselbe zu konstatieren – und nennen dieses den Repräsentanten des wahren Raums.

Das kosmologische Prinzip aber gilt unseres Erachtens entweder vollständig – ohne Expansion des Universums – oder es gilt überhaupt nicht[78]).

Doch wenn das kosmologische Prinzip tatsächlich Geltung hat – wovon wir ausgehen – dabei aber die kosmischen Entfernungen in Bezug auf eine gewisse Sorte von Maßstäben mit der Zeit anwachsen, dann kann dies nichts anderes bedeuten, als daß diese Sorte realer Maßstäbe gegenüber den unveränderlichen kosmischen Distanzen schrumpft. Diese Feststellung wäre allerdings eine Tautologie zur Behauptung einer Expansion, wenn man dabei nicht zu beachten hätte, daß reale Maßstäbe und Uhren notwendigerweise makroskopische Gebilde sind, die – im Unterschied zum Universum selbst – altern und vergehen, indem gleichzeitig immer wieder neue entstehen, sodaß alles in einem statistisch stationären Gleichgewicht[79]) bleibt.

Unsere Auffassung ist folgende: Der wahre Raum, in dem sich das kosmische Geschehen abspielt, und durch den insbesondere das Licht der Spiralnebel auf mathematisch nachvollziehbaren Wegen zu uns gelangt, ist der durch die euklidischen Systemkoordinaten beschriebene. In diesem aber entfernen sich die Spiralnebel (oder Galaxienhaufen, Supercluster, Quasare bzw. die möglicherweise von ‚Blasen' umgebenen originären Gravitationszentren im Sinne von Abschn. 1) *nicht* voneinander, sondern befinden sich – abgesehen von statistischen Abweichungen – im Zustand der Ruhe. Auch wenn andererseits mit Hilfe von Atomuhren durchgeführte, auf Lichtlaufzeiten beruhende Messungen gleichbleibender absoluter Entfernungen eine Zunahme anzeigen würden, so wäre damit nichts anderes bewiesen als eine zeitlich veränderliche Ganggeschwindigkeit der verwendeten Uhren (s. Abschn. 8/h).

---

[76]) entfällt

[77]) Dem wir wie der gesamten Natur in dem Bewußtsein gegenüberstehen, daß es niemals ein physikalisches Bild geben wird, das dieses Wunder vollständig fassen könnte (und aus dem allein sich eine Berechtigung für irreversible Eingriffe ableiten ließe, in den vergleichsweise verschwindend kleinen – doch angesichts seiner vielfältigen Möglichkeiten unbegreiflich großen und einzigen – Lebensraum Erde, der uns heute und in Zukunft zugänglich ist).

[78]) Eine Entstehung des Kosmos als Teil eines stationären Universums (s. auch Abschn. 1) wäre physikalisch nicht keineswegs inakzeptabel, wohl aber die geometrische Deutung einer korrespondierenden mathematischen Singularität als Beginn von ‚Raum und Zeit'.

[79]) Ein anderes Wort für ‚statistisch stationäres Gleichgewicht, aus dem alles immer wieder neu entsteht', wäre möglicherweise – mit allem gebotenen Respekt: *tohu w'a-bohu*.

Die Relativitätstheorie – und zwar die allgemeine in Verbindung mit der speziellen – verlangt offenbar, im Sinne einer Normalisierung innerhalb lokaler Inertialsysteme als Bezugspunkt für den jeweils ins Auge gefaßten Zeitraum immer wieder $t_0^* = 0$ zu setzen. Zu diesen Zeitpunkten stellen sich gewissermaßen alle absoluten kosmischen Einheiten, insbesondere die von Länge, Zeit und Masse, mit den entsprechenden atomaren Einheiten neu aufeinander ein[80]). Diese Auffassung ergibt sich aus der Gültigkeit der speziellen Relativitätstheorie im lokalen Inertialsystem und der Tatsache, daß die Eigenzeit ebenso wie die in Bezug auf spektrale Einheiten gemessene Länge in einem realistischen kosmischen Modell nicht streng integrabel sein können. Aus diesem Grund kann von einer Expansion des Universums nicht einmal in dem Sinne die Rede sein, daß sich ein spektral ausgemessener Kosmos *ständig* ausdehne. Was sich unseres Erachtens mit der absoluten kosmischen Zeit letztlich ändert, sind die atomaren Größen einschließlich der lokalen spektralen *Einheiten* von Zeit, Länge und Masse. Eine solche Änderung kann sich naturgemäß nur da bemerkbar machen, wo physikalische Größen miteinander verglichen werden, die aufgrund unterschiedlicher Entstehungszeiten unterschiedlich betroffen sind. – Als Beobachtungstatsachen, die sich aus dem stationären Linienelement (38) ergeben[81]), haben wir vor allem:

Elektromagnetische Strahlung erfährt zwischen ihrer Emission und Absorption eine Rotverschiebung in Form einer relativen Zunahme der Wellenlänge, und zwar mit der universellen Zeitkonstanten $H$. Damit verbunden ist eine entsprechende relative Abnahme von Frequenz und Energie der Photonen in Bezug auf atomare Einheiten. Weiterhin ergibt sich eine endliche, zeitlich konstante mittlere Dichte des Sternenlichts im intergalaktischen Raum, die in der Größenordnung der offenbar anteilig vorhandenen, ebenfalls stationären 3K-Strahlung eines schwarzen Körpers liegt. Für das stationäre Universum verlangt die allgemeine Relativitätstheorie den Wert $\Pi = -1/3$ des kosmischen Druckfaktors. Die beobachtete Rotverschiebung der statistisch in Ruhe befindlichen Spiralnebel stellt eine neue Form der *Gravitationsrotverschiebung* dar. Denn einerseits gibt es hier (statistisch) keinen realen linearen DOPPLER-Effekt, andererseits ist die kosmische Lichtgeschwindigkeit $c^*$ in diesem Fall notwendigerweise konstant gleich $c$, was sich bei 3-dimensionalem euklidischem Raum gemäß (28) sogar auch auf jedes nicht-stationäre kosmische Linienelement übertragen läßt.

Mit Blick auf die bei der RW-Form (F10) in der Quasi-Eigenzeit $t'$ auftretende Singularität – selbst bezüglich des mit $\Pi = -1/3$ in der kosmischen Zeit $t^*$ stationären Linienelements (38) – ist man versucht, diese etwa folgendermaßen zu interpretieren: Auch in einem insgesamt stationären Universum kann es keine Strukturen geben, die in ihrer (echten) ‚Eigenzeit' älter sind als größenordnungsmäßig $-\tau_H = 1/H$, dies beträfe insbesondere Uhren[82]), Sterne, Spiralnebel. Solch eine maximale Eigenzeit[83]) aber wäre wiederum zu verstehen als Konse-quenz einer prinzipiell nicht überschreitbaren Anwendbarkeitsgrenze der allgemeinen Relativitätstheorie – der Art nämlich, wie es bereits der SCHWARZSCHILD-Radius auf der anderen Seite zu sein scheint[84]). Gerade diese zuletzt genannte Grenze der physikalischen Beschreibung könnte es ja sein, hinter welcher sich diejenigen Prozesse abspielen, die bei jeweils lokaler Abnahme der Entropie für eine statistisch-stationäre Neubildung aller kosmischen Strukturen unverzichtbar sind. Eine solche Interpretation wäre auch insofern befriedigend, weil innerhalb der Physik unseres Erachtens einerseits kein Beginn des Universums, andererseits aber auch keine ewigen makroskopischen Strukturen vorstellbar sind.

---

[80]) Eine solche Interpretation erinnert deutlich an die bekannte quantenmechanische Reduktion des Wellenpakets, die hierzu eine gewisse Verwandtschaft zeigt.

[81]) Würde sich aber im Linienelement (1) eine andere Belegung als $\Pi = -1/3$, insbesondere z.B. der oben besprochene druckfreie Sonderfall $\Pi = 0$ als zutreffend erweisen, so blieben – abgesehen von der Möglichkeit, daß eine kosmologische Lösung der allgemeinen Relativitätstheorie auf Basis einer homogenen Materiedichte überhaupt untauglich wäre – die folgenden alternativen Erklärungsansätze: *Entweder* die rückblickende physikalische Beschreibung von Abläufen aus der Vergangenheit wäre im Rahmen der allgemeinen Relativitätstheorie nicht äquivalent mit der in die Zukunft vorausblickenden Beschreibung des gleichen Vorgangs. Die EINSTEIN'schen Gleichungen würden dann immer nur gelten mit vorübergehend angepaßten ‚heutigen' Werten. *Oder* der uns bekannte Kosmos hätte als Teil eines stationären Universums einen *gemeinsamen* zeitlichen Anfang vor ca. $10^{10}$ a – doch welche Objekte mögen es sein, die allesamt tatsächlich das gleiche Alter hätten? Sollte sich ein solcher Anfang wider Erwarten erweisen, dann bliebe nur die Möglichkeit, daß es sich dabei lediglich um den uns heute bekannten Teil eines statistisch-stationären Universums aus unendlich vielen solcher kosmischer Bereiche handelt, die ebenso wie Sterne und Spiralnebel immer wieder *entstehen* und *vergehen*. Im letzterem Fall aber gäbe es überhaupt kein stichhaltiges Kriterium mehr (außer eben $\Pi = -1/3$) für das richtige Linienelement eines stationären Universums – umgekehrt aber ließe sich aus keinem wie auch immer gearteten Linienelement auf einen zeitlichen Anfang schließen.

[82]) EINSTEIN hat im Zusammenhang mit seinen vielzitierten ‚natürlichen Uhren' immer wieder die scharfen Frequenzen der Spektrallinien angeführt. Doch gerade *scharfe* Frequenzen können für sich allein nicht als Uhren dienen, weil sie prinzipiell auf keinen gemeinsamen Zeitnullpunkt einstellbar sind. Uhren sind immer *makroskopische* Strukturen (bzw. abgegrenzte statistische Gesamtheiten) – und als solche vergänglich.

[83]) In Bezug auf quantitative Aussagen ist zu beachten, daß die Eigenzeit gemäß Abschnitt 8/c nicht streng integrabel ist. Deshalb ist die Existenz von Objekten, die älter wären als $-t_H'$, durch die Darstellungsmöglichkeit des stationären Linienelements im adaptierten (bzw. integrierten) Koordinatensystem (oder auch in der RW-Form) noch nicht unbedingt ausgeschlossen. Mit Blick auf die Unterscheidung von *nicht-integrabler* Eigenzeit und kosmischer Zeit bedürfen auch die verschiedenen Methoden der Altersbestimmung unseres Erachtens durchaus einer Überprüfung hinsichtlich ihrer Anwendbarkeit

Bereits in der vorausgegangenen Arbeit [13]/Anhang-f[85]) wurde aus der allgemeinen Relativitätstheorie gefolgert, daß nur um den Preis eines Verzichts auf eine vollständige deterministische Beschreibung an einem ausschließlichen Bezug auf spektrale Maßstäbe und Uhren festgehalten werden kann. Diese Auffassung wird bei kosmologischen Betrachtungen zusätzlich gestützt durch die Tatsache, daß das Produkt aus der Gravitationsenergie von Elementarteilchen mit der maximalen Eigenzeit größenordnungsmäßig in der Nähe der PLANCK'schen Konstanten liegt.

In der vorliegenden Arbeit konnte es zunächst nur darum gehen, denkbare Alternativen zur heutigen Urknall-Kosmologie aufzuzeigen und dazu das Bild eines stationären Universums – allerdings grob – zu skizzieren. Dabei haben wir vor allem festgehalten am Äquivalenzprinzip in seiner ursprünglichen Form und damit auch an der Gleichheit von träger und schwerer Masse. Dagegen haben wir *nicht* vorausgesetzt die Gleichberechtigung eines jeden beliebigen kosmischen Koordinatensystems oder eine Integrabilität von Eigenzeit und spektral gemessener Länge über kosmische Zeiträume bzw. kosmische Distanzen.

Obwohl dies mit EINSTEINs geometrischer Deutung eher unverträglich ist, beruht das Vorausgegangene doch sehr wohl auf den EINSTEIN'schen Gleichungen[86]).

Selbst vom Standpunkt der heutigen Kosmologie wäre die Behandlung der einschlägigen Fragen auf Basis der singularitätsfreien (1) bzw. skalarabhängigen (28) Lösungsschar nicht nur legitim, sondern unseres Erachtens auch sinnvoll, wobei sich jedes derartige Linienelement ja ganz unabhängig vom Druckfaktor $\Pi$ immer auch in die RW-Form (F10) transformieren läßt. *Warum aber sollte es zur Beschreibung des kosmischen Geschehens notwendig sein, zusätzlich zu den auf spektrale Einheiten bezogenen Längen einerseits abweichende räumliche Systemkoordinaten einzuführen, andererseits aber auf die Einführung einer angemessenen Systemzeit zusätzlich zur spektral gemessenen Eigenzeit von vornherein zu verzichten?*

Die Frage, welche Form des kosmischen Linienelements am Ende die richtige sei, läßt sich anhand der Beobachtungsdaten von Bewegungsabläufen mathematisch ebensowenig eindeutig entscheiden, wie die alte Frage, ob sich die Sonne um die Erde bewegt oder die Erde um die Sonne Gerade EINSTEINs allgemeine Relativitätstheorie lehrt doch, daß sich das eine Bild in das andere durch eine bloße Koordinatentransformation überführen läßt, die am eigentlichen Sachverhalt – in diesem Fall an den zu beschreibenden Bewegungsabläufen im Sonnensystem – überhaupt nichts ändert.

Trotzdem ist die erste Frage ebensowenig sinnlos wie die zweite. Denn die *physikalisch* richtige Antwort ist ohne jeden Zweifel im Sinne des KOPERNIKUS ausgefallen, weil sich allein dessen Sichtweise – ursprünglich herrührend von ARISTARCH – als fruchtbar und richtungsweisend erwiesen hat, indem sie über GALILEI und KEPLER zu NEWTON und der klassischen Physik führen konnte, ohne die auch die allgemeine Relativitätstheorie selbst undenkbar wäre. In einem ähnlichen Sinne (mit allem erdenklichen Vorbehalt) dürfen wir erwarten, daß

---

für kosmische Zeiträume der Größenordnung $1/H$. Auch würde eine insgesamt endliche vergangene ‚Eigenzeit' unter kosmologischen Gesichtspunkten im Rückblick lediglich die Anzeige einer zwar endlichen Anzahl dafür aber exponentiell anwachsender spektraler Zeiteinheiten (Takte) bedeuten, was gemessen mit astronomischen Planetenuhren jedenfalls wieder eine unendliche Vergangenheit ergibt.

[84]) Wir sprechen hier von den *makroskopischen* Anwendbarkeitsgrenzen der Relativitätstheorie, wohingegen die durch die PLANCK-Länge bezeichneten mikroskopischen Grenzen bei allen (quasi)stellaren Objekten im Vergleich zum SCHWARZSCHILD-Radius weit innerhalb liegen.

[85]) Gerade in den Ergebnissen dieser Arbeit lag die Motivation, die Möglichkeit eines stationären Universums in Verbindung mit der allgemeinen Relativitätstheorie noch einmal zu prüfen.

[86]) In Anhang B wird eine kosmische Einbettung lokaler Gravitationsfelder behandelt, die vollständig mit EINSTEINs Interpretation seiner allgemeinen Relativitätstheorie in Einklang steht, ohne daß dadurch die neuen Möglichkeiten einer relativistischen Kosmologie im Sinne eines stationären Universums ausgeschlossen würden.

sich das kosmische Linienelement der allgemeinen Relativitätstheorie in der Skalar-Form (28) als richtig und weiterführend erweisen wird – und zwar bei einem stationären Universum in Verbindung mit (29a)].

Gerade die immer besser zutreffende mathematische Beschreibung wiederkehrender Abläufe von Tages- und Jahreszeiten, Sonnen- und Mondfinsternissen hat im Laufe der Zeit für ganz gegensätzliche ‚Weltsysteme' als ‚Beweis' dienen müssen[87]). Warum sollte dies mit der Interpretation[88]) der vergleichsweise jungen kosmologischen Beobachtungstatsachen wie Rotverschiebung und 3K-Strahlung auf Anhieb anders sein?

### Anhang A: Die Erklärung einer scheinbaren Beschleunigung der Pioneer-Sonden 10/11

In diesem Anhang der Version v1 wurde detailliert aufgezeigt, wie ein realer, nicht-konventioneller Pioneer-Effekt [28] aus der Relativitätstheorie folgen könnte, allerdings nur bei Übereinstimmung der kosmischen Zeit mit der Ephemeridenzeit. Da diese Möglichkeit aufgrund handfester Beobachtungstatsachen jedoch ausgeschlossen werden muß, bleibt nur die abschließende Feststellung, daß ein nicht-konventioneller Pioneer-Effekt mit der Relativitätstheorie selbst bei Berücksichtigung möglicher kosmischer Effekte unvereinbar wäre [bzgl. Klärung dieser Frage s. die Version v1 der vorliegenden Arbeit sowie die diesbezüglichen, weiteren Arbeiten d. Verf. „*Relativity Theory and a Real Pioneer Effect*" – e-print <http://arXiv.org/abs/gr-qc/0212004> v1 (2002), v2, v3 (2003)]. – Bei Berücksichtigung einer unabhängigen Untersuchung [29] steht diese Folgerung im Einklang mit der bereits vorher vertretenen Position [30], [31], daß es sich bei dem genannten Effekt um eine tatsächliche Beschleunigung aufgrund anisotroper Energieabstrahlung handeln sollte.

### Anhang B: Die lokale Einbettung in den kosmischen Hintergrund

In diesem Anhang B wollen wir nun prüfen, inwieweit sich eine u.a. durch die Meßdaten des binären Pulsars bestätigte *gleiche* Ganggeschwindigkeit von Atomuhr und Planetenuhr mit einer relativistischen Kosmologie in Einklang bringen läßt. Denn mit der Atomzeit, der Ephemeridenzeit lokaler Planetensysteme und der kosmischen Zeit gibt es (mindestens) drei fundamentale Zeitskalen, die sich – obwohl die beiden erstgenannten identisch sein sollten – prinzipiell unabhängig voneinander definieren lassen.

*Das folgende stellt also die Möglichkeit einer lokalen Einbettung im Sinne eines stationären Universums dar. Und zwar entspricht diese Möglichkeit durchgängig dem bisherigen Verständnis der allgemeinen Relativitätstheorie, indem sie davon ausgeht, daß es ein einziges Linienelement sein soll, aus dem sich Energie-Impuls-Tensor, Bewegungsgesetz und die Anzeige von Atomuhren ableiten lassen.*

Wir beschränken uns darauf, lediglich die hauptsächlichen Ergänzungen zum vorausgegangenen darzustellen. Nach dem folgenden, mathematisch widerspruchsfreien Modell kann es einen Pioneer-Effekt nicht geben.

Das kosmische Linienelement (38) bietet nun eine außerordentlich einfache denkbare Möglichkeit zur Berücksichtigung lokaler Inhomogenitäten der Materieverteilung geradezu an, und zwar in Form des – hier zunächst stationär – *eingebetteten* Linienelements:

---

[87]) In seiner Geschichte der Physik bemerkt MAX V. LAUE: „*Und wenn dann die Theorien wechseln, so wird aus einem schlagenden Beweise für die eine leicht ein ebenso starkes Argument für eine ganz entgegengesetzte.*"

[88]) Die Hochachtung des Verfassers vor dem unvergleichlichen Physiker ALBERT EINSTEIN läßt sich nicht deutlicher zum Ausdruck bringen als durch die abschließende Feststellung, daß die vorliegende Arbeit auf inzwischen bereits jahrzehntelanger (fast immer stillschweigender) Auseinandersetzung mit dessen faszinierendem Werk beruht, das im Wirken vieler bedeutender Naturwissenschaftler und Mathematiker seine Wurzeln hat, von denen hier stellvertretend neben H. A. LORENTZ vor allem auch HENRI POINCARÉ genannt sein soll (wobei das Privileg dieser Auseinandersetzung durchwegs den Charakter eines Lernprozesses hatte). Nicht zuletzt im Sinne NEWTONs sei trotzdem – oder gerade deshalb – die subjektive Bemerkung erlaubt, daß es sich bei EINSTEINs kosmologischem Konzept eines durch ‚Krümmung' geschlossenen dreidimensionalen Raumes um eine wohl eher romantische (um nicht zu sagen unmögliche) Vorstellung gehandelt hat, die allein erst die spätere *Akzeptanz* eines – im Rahmen des physikalisch Beschreibbaren schlechterdings unmöglichen – Beginns von ‚Raum und Zeit' verständlich werden läßt (nach meiner privaten Einschätzung könnte EINSTEINs Bezeichnung der kosmologischen Konstanten als ‚größte Eselei' seines Lebens möglicherweise unbewußt auch auf die daraus abgeleitete ‚Krümmung des Raums' gemünzt gewesen sein). Doch tatsächlich ist es allein EINSTEINs – im lokalen Gravitationsfeld ausnahmslos experimentell bestätigte – allgemeine Relativitätstheorie, auf deren Grundlage sich überhaupt quantitative Aussagen zur Kosmologie ableiten lassen, die in erstaunlicher Weise den konkreten Beobachtungstatsachen nahe kommen.

$$\mathrm{d}\widetilde{s}^{*2} \;=\; \widetilde{g}_{ik}^{*}\,\mathrm{d}x^{*i}\mathrm{d}x^{*k} \;=\; \mathrm{e}^{2Ht^{*}}\,\mathrm{d}s_{\mathrm{isoliert}}^{*2} \;=\; \mathrm{e}^{2Ht^{*}} g_{ik}^{*}\,\mathrm{d}x^{*i}\mathrm{d}x^{*k}. \tag{B1}$$

In diesem Anhang B unterscheiden wir die Ephemeridenzeit $t$ des Planetensystems K ausdrücklich von der kosmischen Zeit $t^*$, in der sich das Universum als stationär beschreiben läßt ($x^{*\alpha}$ sind die entsprechenden räumlichen Koordinaten des kosmischen Bezugssystems K*). Dabei stehen die Bezeichnungen $g^*_{ik}$ und $ds^*_{\text{isoliert}}$ nach Umbenennung der Koordinaten ($x^i := x^{*i}$) für denjenigen gewöhnlichen Fundamentaltensor bzw. für dasjenige entsprechende Linienelement der allgemeinen Relativitätstheorie, die sich bei Vernachlässigung des kosmischen Hintergrunds ergeben, insbesondere also bei Lösung der Vakuum-Feldgleichungen $R_{ik}^* = 0$ für den Raum außerhalb der gewöhnlichen Materie. Hier ist allerdings zu beachten, daß jedes in der ursprünglichen Fundamentalform $g_{ik}$ des isoliert betrachteten Linienelements auftretende NEWTON'sche Potential $c^2\Phi$ gemäß

$$c^2\Phi \equiv -\frac{GM}{r} \approx -\frac{GM}{r^*}\mathrm{e}^{-Ht^*} \equiv c^2\Phi^* \tag{B2}$$

durch $c^2\Phi^*$ zu ersetzen wäre. Dies deshalb, weil es sich bei dem Abstand $r$ im Ausdruck des wie üblich geschriebenen NEWTON'schen Potentials ($c^2\Phi = -GM/r$) im Sinne dieses Anhangs B um eine auf *spektrale* Einheiten bezogene Entfernung handeln sollte, welche – entsprechend der zweiten Beziehung (63) ergibt sich für kleine Abstände $r \approx r^* \cdot \mathrm{e}^{Ht^*}$ – hier durch die absolute Entfernung $r^*$ auszudrücken wäre[89]. Denn im Linienelement der allgemeinen Relativitätstheorie stehen bekanntlich keine spektralen, sondern immer nur absolute Systemkoordinaten.

So wäre für $ds^*_{\text{isoliert}}$ in (B1) insbesondere die wohlbekannte SCHWARZSCHILD-Lösung mit dem NEWTON'schen Potential in der Form $c^2\Phi^*$ einzusetzen, wenn es – bei Vernachlässigung der durch die Planeten hervorgerufenen Störungen – um Bewegungsabläufe von Massenpunkten im Gravitationsfeld der Sonne geht:

$$\mathrm{d}\widetilde{s}_{(S)}^{*2} \;=\; \mathrm{e}^{2Ht^*}\cdot\left\{(1+2\Phi^*)c^2\mathrm{d}t^{*2} - \frac{\mathrm{d}r^{*2}}{1+2\Phi^*} - r^{*2}\left(\sin^2\vartheta^*\,\mathrm{d}\varphi^{*2} + \mathrm{d}\vartheta^{*2}\right)\right\}. \tag{B3}$$

Wir haben bereits in (68) die vollständig ausgeschriebene Form des Variationsprinzips gewählt, weil sich (B3) formal auch verstehen läßt als $\delta\!\int m_0^*c\,\mathrm{d}s = 0$ mit $m_0^* = m_0\zeta^*$, $(GM)^* = (GM)/\zeta^*$ bei zeitlich konstanten Werten von $m_0$ und $(GM)$. Dazu wäre es konsequenterweise allerdings erforderlich, Photonen eine – wenn auch beliebig kleine – Ruhemasse zuzuschreiben.

Bei der – durch den Index '(S)' gekennzeichneten – phänomenologischen Einbettung der SCHWARZSCHILD-Lösung gemäß (B3) ergibt sich als entsprechender EINSTEIN-Tensor

$$\widetilde{E}^{*k}_{i(S)} \;=\; \frac{H^2\mathrm{e}^{-2Ht^*}}{c^2(1+2\Phi^*)^2}\cdot\begin{pmatrix} 3+8\Phi^* & 0 & 0 & 0 \\ 0 & 1+4\Phi^* & 0 & 0 \\ 0 & 0 & \dfrac{1+7\Phi^*+14\Phi^{*2}}{1+2\Phi^*} & 0 \\ 0 & 0 & 0 & \dfrac{1+7\Phi^*+14\Phi^{*2}}{1+2\Phi^*} \end{pmatrix} \;=\; \kappa\,\widetilde{T}^k_i, \tag{B4}$$

wobei dieses für den stationären Fall berechnete Ergebnis gemäß (34) auch für die anderen Belegungen von $\Pi$ wieder als brauchbare erste Näherung in $Ht^*$ gelten kann. Wie man sieht, entspricht die Energiedichte außerhalb einer Kugel praktisch überall der mittleren kosmischen Energiedichte, so daß man tatsächlich von einer Einbettung in den kosmischen Hintergrund sprechen könnte[90]. Selbst bei Annäherung auf einen Abstand $r^* = 2r_G$ vom

---

[89] In der Regel läßt sich dies formal auch dadurch bewerkstelligen, daß man die Gravitationskonstante $G$ beim Übergang zum eingebetteten Linienelement einfach durch $G\mathrm{e}^{-Ht^*}$ (bzw. $G/\zeta^*$) ersetzt.

[90] Hätte man übrigens – im Sinne einer 2. Alternative – $\Phi^*$ in (B3) ohne den oben begründeten reziproken kosmischen Zeitfaktor $\mathrm{e}^{-Ht^*}$ verwendet, so wäre im ansonsten annähernd gleich gebliebenen Energie-Impuls-Tensor (B4) zusätzlich eine inakzeptable negative radiale Energieströmung $S_r = -Hr_G/[\kappa r^{*2}(1-r_G/r^*)]$ aufgetreten. Integriert man diese nämlich über eine Kugelfläche in hinreichend großer Entfernung $R \gg r_G$, so findet man als Energiezuwachs pro Zeit näherungsweise $HMc^2$ für das Innere der Kugel. Eine damit verbundene zeitliche Massenzunahme eingebetteter Objekte – näherungsweise der Form $M \approx M_0\mathrm{e}^{Ht^*}$ – würde allerdings z.B. bei der Sonne mit ca. $4\cdot 10^{29}$ W/$c^2$ den Massenverlust durch Abstrahlung *um etwa das Tausendfache* übersteigen. Gäbe es eine solche zeitliche Massenzunahme tatsächlich, so wäre auch eine daraus resultierende Zu-

Mittelpunkt einer hinreichend kompakten Kugel (Gravitationsradius $r_G = 2GM/c^2$) wäre die Energiedichte dort im Vergleich zum mittleren kosmischen Wert nur viermal so groß (wohingegen sie bei weiterer Annäherung an $r_G$ räumlich divergiert). Hier wäre zu beachten, daß der EINSTEIN-Tensor – wie bereits weiter in Abschnitt 3 erörtert – nicht unbedingt die ‚mikroskopisch' wahre, sondern gegebenenfalls eine phänomenologische (durchschnittliche) Materiedichte repräsentiert[91]).

In Analogie zum stationären Fall (B1) bietet sich aber auch für ganz beliebige Werte des Druckfaktors $\Pi$ die entsprechende Möglichkeit an, jedes lokal isolierte Linienelement, das bisher unter Vernachlässigung des kosmischen Hintergrunds gewonnen wurde, phänomenologisch in eine kosmische Lösung einzubetten, und zwar in der allgemeinen Form

$$d\widetilde{s}^{(c)2} = \widetilde{g}_{ik}^{(c)} dx^i dx^k = \zeta^{(c)2} ds_{\text{isoliert}}^2 = \zeta^{(c)2} g_{ik} dx^i dx^k. \tag{B5}$$

Dabei möge $\zeta^{(c)} = \zeta^{(c)}[t^{(c)}]$ mit $d\zeta^{(c)}/dt^{(c)} > 0$ wieder den kosmischen Zeitskalar bezeichnen. Jedes in der Fundamentalform $g^{(c)}_{ik}$ des isoliert betrachteten Linienelements auftretende NEWTON'sche Potential $c^2\Phi$ wäre analog zu dem oben gesagten nun durch $c^2\Phi^{(c)} = -\zeta^{(c)-1} GM/r^{(c)}$ zu ersetzen.

Die skalarabhängige Einbettung (B1), (B5) des isolierten konventionellen Linienelements zeichnet sich dadurch aus, daß sie einerseits im Sinne dieses Anhangs B dem kosmischen Hintergrundpotential Rechnung trägt[92]), ohne andererseits aber den gegebenenfalls lokal orts- und zeitabhängigen Koordinatenwert der Lichtgeschwindigkeit $c^{(c)}_{\text{isoliert}}$ des jeweils ins Auge gefaßten Teilsystems spürbar zu beeinflussen. Wegen der demzufolge bereits in relativ geringen Entfernungen von den lokalen Quellen praktisch in die Konstante $c$ übergehenden Lichtgeschwindigkeit $c^{(c)}$ ist (B5) sehr gut geeignet für die Behandlung aller astronomischen Bewegungsprobleme, bei denen Bahndaten mit Hilfe elektromagnetischer Signale – sei es nun in Form von Laufzeiten oder DOPPLER-Verschiebungen – gemessen werden.

Bei stationärer Einbettung der SCHWARZSCHILD-Lösung gemäß (B3) erhält man z.B. (mit den Abkürzungen ‚rad', ‚tan' für radial bzw. tangential)

$$\begin{matrix} c^*_{\text{rad}} \\ c^*_{\text{tan}} \end{matrix} = c^*_{\text{isoliert}} + O(Ht^* \cdot \Phi^*_{\text{isoliert}}) = \begin{matrix} c(1+2\Phi^*) = c\left(1 - \dfrac{2GM}{r^*}e^{-Ht^*}\right) \\ c\sqrt{1+2\Phi^*} \approx c\left(1 - \dfrac{GM}{r^*}e^{-Ht^*}\right) \end{matrix}, \tag{B6}$$

wobei hier anstelle des Faktors $e^{-Ht^*}$ im allgemeinen Falle $\zeta^{(c)-1}$ zu stehen hätte. Mit zunehmender Entfernung $r^*$ geht $c^*$ über in $c$, was wieder eine konstante *kosmische* Lichtgeschwindigkeit bedeutet. Bei Verwendung des durch die spezielle Form des skalar eingebetteten Linienelements (B5) eindeutig festgelegten Bezugssystems wollen wir hier der Kürze halber einfach von $c$-Koordinaten sprechen. Wenn man nun bedenkt, daß die von SHAPIRO gemessene Laufzeitverzögerung von Radarsignalen (s. z.B. [22]) gemäß (B6) in 100 Jahren lediglich eine relative Änderung von ca. $10^{-8}$ des ursprünglich zu erwartenden Wertes erfahren sollte, dann ist daraus ersichtlich, daß man bei Verwendung von $c$-Koordinaten innerhalb des Sonnensystems auch in diesem Anhang praktisch immer mit $c^* = c^*_{\text{isoliert}}$ rechnen kann.

Das Linienelement (B3) erlaubt es nun weiter – ebenso wie das skalar eingebettete Linienelement (B5) – den daraus resultierenden dynamischen Einfluß des kosmischen Hintergrunds im Vergleich zu den durch lokale Gravitationsquellen verursachten Beschleunigungen leicht abzuschätzen. Anstatt aber unmittelbar eine Lösung der entsprechenden Bewegungsgleichungen auf Basis des stationär eingebetteten SCHWARZSCHILD-Elements

---

nahme der Schwerebeschleunigung der Sonne von etwa $a_M \approx Ht^* a_{\text{NEWTON}}$ zu berücksichtigen, die z.B. beim Merkur bereits über einen Zeitraum von 10 Jahren in der Größenordnung von einigen Prozenten derjenigen relativistischen Korrekturen liegen müßte, die für seine Periheldrehung verantwortlich sind. Das aber wäre der Beobachtung wohl nicht entgangen. Auch die Frage, wie sich dies gerade mit der Energiebilanz des binären Pulsars PSR 1913+16 vereinbaren lassen sollte, zeigt mehr als deutlich, welch schwerwiegende Widersprüche einer solchen Massenveränderlichkeit entgegen stehen.

[91]) Im Unterschied hierzu wird aber im Hauptteil dieser Arbeit davon ausgegangen, daß zwar sowohl die lokale als auch die kosmisch gemittelte Materieverteilung phänomenologisch richtig beschrieben werden, nicht aber beide zugleich.

[92]) Auch aus dem zum Linienelement (B5) gehörenden EINSTEIN-Tensor erhält man natürlich außerhalb der lokalen Quelle des Gravitationsfeldes nicht mehr die Energiedichte Null, sondern eine solche, die im Unendlichen in die mittlere kosmische Dichte übergeht.

(B3) anzustreben, gehen wir folgendermaßen vor. Wir transformieren vorübergehend die absoluten Koordinaten $t^*$, $x^{*\alpha}$ auf die Koordinaten $t''$, $x''^{\alpha}$ des *adaptierten* Systems (s. Abschn. 8/d) durch die der Transformation (65) entsprechenden Substitutionen

$$t^* = \frac{1}{H}\ln(1+Ht'') - \frac{Hr''^2}{2c^2} \qquad \Leftrightarrow \qquad t'' = \frac{1}{H}\left[e^{H(t^*+Hr^2/2c^2)} - 1\right],$$
$$x^{*\alpha} = \frac{x''^{\alpha}}{1+Ht''} \qquad \qquad x''^{\alpha} = x^{*\alpha} e^{H(t^*+Hr''^2/2c^2)}. \tag{B7}$$

Bei den beiden Relationen rechter Hand läßt sich das Abstandsquadrat $r''^2 = x''^2+y''^2+z''^2$ im Falle $Ht^*$, $Hr^*/c \ll 1$ bis auf Korrekturen 2. Ordnung näherungsweise durch $r^{*2} = x^{*2}+y^{*2}+z^{*2}$ ersetzen, sodaß sie dementsprechend als Umkehrtransformationen verwendet werden können[93]. Die stationär eingebettete SCHWARZSCHILD-Form (B3) geht nach Transformation und Entwicklung bis zur 2. Ordnung in $Ht''$, $Hr''/c$ sowie anschließender Vernachlässigung von Faktoren $H^2\Phi''$ gegenüber $H^2$ und zugleich von $H\Phi''^2$ gegenüber $H\Phi''$ schließlich über in

$$d\tilde{s}''^2 \approx \left(B'' - \frac{2H^2r''^2}{c^2}\right)c^2 dt''^2 - 4H^2 t'' r'' dt'' dr'' - \left(\frac{1}{B''} - \frac{2H^2r''^2}{c^2}\right)dr''^2 - \left(1 - \frac{H^2r''^2}{c^2}\right)d\Sigma''^2, \tag{B8}$$

wobei als Abkürzung $B'' = 1+2\Phi''$ verwendet ist. Bei zusätzlicher Vernachlässigung auch aller Terme in $H^2$ aber – was zur Behandlung von Bewegungsabläufen im Sonnensystem durchaus genügen würde – geht die Entwicklung dieses Linienelements unmittelbar über in die klassische SCHWARZSCHILD-Lösung [das Element $d\Sigma''$ ist wieder wie in (66) als das einer gewöhnlichen Kugelfläche zu verstehen]. Dies aber zeigt:

*Auf Basis des eingebetteten Linienelements werden Planeten, Monde, Satelliten sowie auch Partner in Doppelsternsystemen bezüglich der Anzeige einer im Schwerpunkt des Systems ruhenden Atomuhr als lokal adaptierte Systemzeit – und bei Verwendung spektraler Längeneinheiten – über charakteristische Beobachtungszeiträume $t'' \approx 100$ a keine meßbaren Abweichungen erfahren (im Vergleich zu den erwarteten Bahnen des im Rahmen der allgemeinen Relativitätstheorie wie üblich isoliert betrachteten lokalen Gravitationsfelds).*

Umgekehrt aber bedeutet dies, daß sich durch bloße Rücktransformation mittels (B7) z.B. aus den Daten der gewöhnlichen Schwarzschild-Bahnen die geodätische Bewegung makroskopischer Körper in *absoluten* Koordinaten für das Sonnensystem gewinnen lassen.

Beispielsweise liege die relativistisch berechnete Bahn eines makroskopischen Körpers ($v/c \ll 1$) als $R^{\alpha} = R^{\alpha}(t)$ für das jeweilige zunächst isoliert betrachtete System bereits vor. Dann sollte diese Bahn aufgrund des vorher gesagten in ausreichender Näherung übereinstimmen mit der Bahn in adaptierten Koordinaten. Daraus wiederum ergibt sich die gleiche Bahn schließlich in *absoluten* Koordinaten durch Resubstitution gemäß (B7) zu

$$R^{*\alpha}(t^*) = R^{\alpha}\left(t := \frac{e^{Ht^*}-1}{H}\right) \cdot e^{-Ht^*}. \tag{B9a}$$

Nun impliziert (B7) für $r \ll c/H$ gemäß den Voraussetzungen dieses Anhangs B natürlich

$$dt \approx dt^* e^{Ht^*}, \tag{B10}$$

sodaß für kleine Beobachtungszeiten $t, t^* \ll 1/H$ näherungsweise folgen müßte

$$R^{*\alpha}(t^*) = R^{\alpha}\left(t \approx t^* + \frac{1}{2}Ht^{*2}\right) \cdot e^{-Ht^*}. \tag{B9b}$$

Aus (B9b) dürfte also etwa gefolgert werden, daß sich die kosmische Einbettung in sehr wichtigen Fällen als geschwindigkeitsabhängige Reibungskraft $F_K^* = -\mu_K v^*$ mit dem kosmischen ‚Reibungskoeffizienten' $\mu_K = mH$ ($m$ = Masse des Körpers) bzw. als Bremsbeschleunigung $-Hv^*$ auswirken sollte. Dieses würde insbesondere – anstelle GALILEIs klassischem Trägheitssatz – für die Bewegung weitab von lokalen Gravitationsquellen gelten (ansonsten nur näherungsweise für die rein radiale Bewegung im Schwerefeld eines Zentralkörpers).

---

[93]) Die exakten Umkehrtransformationen lassen sich mit Hilfe der LAMBERT-Funktion darstellen.

Betrachten wir nun aber den Fall idealisierter Planetenbahnen: Für diese würde aus (B9b) eine Korrektur der absoluten Bahnradien resultieren, die mit einer Zeitabhängigkeit der absoluten Winkelgeschwindigkeiten einherginge. Den zeitlich abnehmenden Abstand $R_{(S)}{}^*$ der – bezüglich spektraler Einheiten auf unveränderlichen Kreisbahnen mit gleichbleibendem Radius $R_{(S)}$ – umlaufenden Objekte finden wir für das stationär eingebettete SCHWARZSCHILD-Element zu

$$R^*_{(S)} \approx R_{(S)} \mathbf{e}^{-Ht^*}, \tag{B11}$$

was in absoluten Koordinaten eine allmähliche spiralförmige Annäherung bedeuten würde. Auch die absoluten Umlaufzeiten $T_{(S)}{}^*$ würden sich gleichermaßen ändern, nämlich gemäß

$$T^*_{(S)} \approx T_{(S)} \mathbf{e}^{-Ht^*}. \tag{B12}$$

*Umgekehrt aber sind die unmittelbar mit geeignet plazierten Atomuhren gemessenen Umlaufzeiten $T_{(S)}$ sowie die auf spektrale Längeneinheiten bezogenen großen Halbachsen $a_{(S)}$ auf Basis des eingebetteten Linienelements im Sonnensystem konstant!*

Die aus dem NEWTON'schen Gravitationsgesetz resultierenden unveränderlichen Werte der großen Halbachsen der Planetenbahnen sind also mit hinreichender Genauigkeit über große Zeiträume als Grenzfall in (B1) enthalten, wenn man sich in diesem Fall auf Basis des eingebetteten Linienelements unmittelbar auf die Anzeige entsprechend positionierter Atomuhren und die Verwendung spektraler Längeneinheiten bezieht. Dieses eingebettete Linienelement zeichnet sich gerade dadurch aus, daß es auch bei Berücksichtigung des kosmischen Hintergrunds eine – im Rahmen der herkömmlichen Auffassung der allgemeinen Relativitätstheorie ganz selbstverständlich erwartete – gleiche Ganggeschwindigkeit für alle natürliche Uhren, d.h. insbesondere für Atomuhr *und* Planetenuhr ergibt. Was also könnte außer dem Pioneer-Effekt gegen das in diesem Anhang B skizzierte Modell sprechen?

Ein grundsätzlicher Einwand besteht darin, daß im eingebetteten Linienelement zwei Betrachtungsweisen möglicherweise unzulässig vermischt werden, nämlich die lokale (Planetensystem) mit der kosmischen (homogene Materieverteilung). Dies wird besonders deutlich, wenn man die Dimensionen des eingebetteten Gravitationsfelds erweitert, bis es zunächst galaktische Abmessungen erreicht. Dehnt man nämlich die Dimensionen des betrachteten Bereichs noch weiter aus, so findet man schließlich die mittlere kosmische Materiedichte im eingebetteten Energie-Impuls-Tensor *doppelt* vertreten, was natürlich widersinnig wäre. Es sei denn man beschränkt dieses ganze Modell ausdrücklich auf eine *lokale* Einbettung. Zwar taucht dann sofort die Frage auf, warum die Einbettung nicht in die mittlere Materiedichte der *Milchstraße* erfolgen sollte, doch läßt sich dies dahingehend beantworten, daß die Einbettung eben nur in den bezüglich ultra-galaktischer Maßstäbe *homogenen* (und *isotropen*) Hintergrund erfolgen darf.

Unabhängig davon aber, ob sich die zur Behandlung lokaler Bewegungsabläufe in einem stationären Universum hier aufgezeigte Möglichkeit einmal unmittelbar als richtig erweisen wird, so dürfte jedenfalls mit dem im Vergleich zur herkömmlichen Form leicht zu behandelnden Linienelement (28) der theoretische Boden zur endgültigen Entscheidung dieser Frage bereitet sein – und damit auch für einen experimentellen Zugang zur Kosmologie, der über die reine Beobachtung hinausgeht.

---

box@peter-ostermann.de

## Literatur

[1] A. EINSTEIN: *Kosmologische Betrachtungen zur allgemeinen Relativitätstheorie*, Sitz.ber. Preuß. Akad. Wiss. (1917), S. 142

[2] E. P. HUBBLE: *A RELATION BETWEEN DISTANCE AND RADIAL VELOCITY AMONG EXTRA-GALACTIC NEBULAE*, Proc. N. Acad. Sci. **15** (1929), S. 168

[3] W. DE SITTER: *On the relativity of inertia. Remarks concerning EINSTEIN'S latest hypothesis*, Proc. Kkl. Akad. Amsterdam **XIX** (1917), S. 1217; *On the curvature of space*, ebendort **XX** (1917), S. 229; *On Einstein's Theory of Gravitation , and its Astronomical Consequences*, M. Not. Roy. Astron. Soc. **LXXVIII** (1917), S. 3

[4] A. FRIEDMAN(N): *Über die Krümmung des Raumes*, Zeitschr. f. Physik **10** (1922), S. 377; *Über die Möglichkeit einer Welt mit konstanter negativer Krümmung des Raumes*, ebendort **21** (1924), S. 326

[5] G. LEMAÎTRE: *UN UNIVERS HOMOGÈNE DE MASSE CONSTANTE ET DE RAYON CROISSANT, RENDANT COMPTE DE LA VITESSE RADIALE DES NÉBULEUSES EXTRA-GALACTIQUES*, Ann. Soc. Sci. Bruxelles **XLVII** (1927), S. 49; *A Homogeneous Universe of Constant Mass and Increasing Radius accounting for the Radial Velocity of Extra-Galactic Nebulae*, M. Not. Roy. Astron. Soc. **XCI** (1931), S. 483; *The Expanding Universe*, ebendort, S. 490, 703

[6] W. H. BONDI, T. GOLD: *The Steady-State Theory of the Expanding Universe*, M. Not. R. Astr. Soc. **108** (1948), S. 252

[7] F. HOYLE: *A New Model for the Expanding Universe*, M. Not. Roy. Astron. Soc. **108** (1948), S. 372; *On the Cosmological Problem*, ebendort **109** (1949), S. 365

[8] H. W. M. OLBERS: *Ueber die Durchsichtigkeit des Weltraums*, (Bodes) Astron. Jahrbuch f. 1826, **51** (1823), S. 110

[9] P. EHRENFEST: *Gleichförmige Rotation starrer Körper und Relativitätstheorie*, Phys. Zeitschr. **10** (1909), S. 918

[10] Th. KALUZA: *Zur Relativitätstheorie*, Phys. Zeitschr. **11** (1910), S. 977

[11] A. EINSTEIN: *Die Grundlage der allgemeinen Relativitätstheorie*, Ann. d. Phys. **49** (1916), S. 769

[12] A. EINSTEIN: *Zur Elektrodynamik bewegter Körper*, Ann. d. Phys. **17** (1905), S. 891

[13] P. OSTERMANN: *Die Einweg-Lichtgeschwindigkeit auf der rotierenden Erde und die Definition des Meters*, e-Print in <http://arXiv.org/abs/gr-qc/0208056>, v1 (2002), S. 1-22

[14] K. SCHWARZSCHILD: *Über das Gravitationsfeld eines Massenpunktes nach der EINSTEINschen Theorie*, Sitz. d. Kgl. Preuß. Akad. d. Wiss. **XIX** (1916), S. 189; *... einer Kugel aus inkompressibler Flüssigkeit ...* , ebendort S. 424

[15] C. W. MISNER, K. S. THORNE, J. A. WHEELER: *Gravitation*, New York 1970-71-73

[16] H. P. ROBERTSON: *KINEMATICS AND WORLD-STRUCTURE*, Astrophys. J. **82** (1935), S. 284; ebendort **83** (1936), S. 187, S. 257

[17] A. G. WALKER: *ON MILNE'S THEORY OF WORLD-STRUCTURE*, Proc. London Math. Soc. **42** (1936), S. 90

[18] L.D. LANDAU, E.M. LIFSCHITZ: *Lehrbuch d. theor. Physik*, Bd. II, Klassische Feldtheorie, 12. Aufl., Berlin 1992

[19] W. PAULI: *Encyklopädie der mathematischen Wissenschaften*, Bd. V Teil 2, S. 539, Leipzig 1921

[20] N. ROSEN: *Flat-Space Metric in General Relativity Theory*, Ann. of Physics **22** (1963), S. 1

[21] M. V. LAUE: *Die Relativitätstheorie* Bd. I/II, 7. durchges. Aufl., Braunschweig 1961

[22] S. WEINBERG: *Gravitation and Cosmology*, New York 1972

[23] G. NORDSTRÖM: *Relativitätsprinzip und Gravitation*, Zeitschr. f. Phys. **13** (1912), S. 1126; *Träge und schwere Masse in der Relativitätsmechanik*, Ann. d. Phys. **40** (1913), S. 856; *Zur Theorie der Gravitation vom Standpunkt des Relativitätsprinzips*, ebendort **42** (1913), S. 533

[24] A. EINSTEIN, A. D. FOKKER: *Die Nordströmsche Gravitationstheorie vom Standpunkt des absoluten Differentialkalküls*, Ann. d. Phys. **44** (1914), S. 321

[25] H. WEYL: *Raum - Zeit - Materie*, 8. Aufl., Hg. u. erg. v. J. Ehlers, Berlin Heidelberg New York 1993; *Gravitation und Elektrizität*, Sitz.ber. Preuß. Akad. d. Wiss. (1918), S. 465; *Eine neue Erweiterung der Relativitätstheorie*, Ann. d. Phys. **59** (1919), S. 101

[26] R. A. HULSE, J. H. TAYLOR: *Discovery of a Pulsar in a Binary System*, Astrophys. J. **195** (1975), L51

[27] C. M. WILL: *Theory and experiment in gravitational physics*, Rev. Ed., Cambridge 1993

[28] J. D. ANDERSON, PH. A. LAING, E. L. LAU, A. S. LIU, M. M. NIETO, S. G. TURYSHEV: *Study of the anomalous acceleration of Pioneer 10 and 11,* Phys. Rev. D **65** (2002), 082004, e-Print in <http://arXiv.org/abs/gr-qc/0104064>, v4 (2002), S. 1-54; *Indication, from Pioneer 10/11, Galileo, and Ulysses Data, of an Apparent Anomalous, Weak, Long-Range Acceleration*, Phys. Rev. Lett. **81** (1998), S. 2858

[29] C. B. MARKWARDT: *Independent Confirmation of the of Pioneer 10 Anomalous Acceleration,* e-Print in <http://arXiv.org/abs/gr-qc/0208046>, v1 (2002), S. 1-21

[30] J. I. KATZ: *Comment on "Indication, from Pioneer 10/11, Galileo, and Ulysses Data of an ... Acceleration",* Phys. Rev. Lett. **83** (1999), S. 1892, e-Print in <http://arXiv.org/abs/gr-qc/9809070>, v3 (1998)

[31] E. M. MURPHY: *A Prosaic Explanation for the Anomalous Accelerations Seen in Distant Spacecraft*, Phys. Rev. Lett. **83** (1999), S. 1890*;* e-Print in <http://arXiv.org/abs/gr-qc/9810015> (1998)